\documentclass[10pt]{article}
\usepackage{epsfig}

\renewcommand\baselinestretch{1.2}
\large\normalsize

\begin{document}

\title{Dynamic probes of quantum spin chains 
       \protect\\
       with the Dzyaloshinskii-Moriya interaction}

\author{Oleg Derzhko$^{1,2}$, Taras Verkholyak$^1$, Taras Krokhmalskii$^1$, and Helmut B\"{u}ttner$^3$
  \protect\\
  $^1$Institute for Condensed Matter Physics, National Academy of Sciences of Ukraine,
  \protect\\
  1 Svientsitskii Street, L'viv-11, 79011, Ukraine
  \protect\\
  $^2$National University ``Lvivska Politechnika'',
  12 S.~Bandera Street, L'viv, 79013, Ukraine
  \protect\\
  $^3$Theoretische Physik I, Universit\"{a}t Bayreuth, Bayreuth, D-95440, Germany}

\maketitle

\begin{abstract}
We consider the spin-$\frac{1}{2}$ anisotropic $XY$ chain 
in a transverse ($z$) field
with the Dzyaloshinskii-Moriya interaction
directed along $z$-axis in spin space 
to examine the effect of the Dzyaloshinskii-Moriya interaction 
on the $zz$, $xx$ and $yy$ dynamic structure factors.
Using the Jordan-Wigner fermionization approach 
we analytically calculate 
the dynamic transverse spin structure factor.
It is governed by a two-fermion excitation continuum.
We analyze the effect of the Dzyaloshinskii-Moriya interaction 
on the two-fermion excitation continuum.
Other dynamic structure factors 
which are governed by many-fermion excitations 
are calculated numerically. 
We discuss 
how the Dzyaloshinskii-Moriya interaction manifests itself 
in the dynamic properties of the quantum spin chain
at various fields and temperatures.
\end{abstract}

\renewcommand\baselinestretch{1.45}
\large\normalsize

\section{Introduction. Jordan-Wigner representation}

\setcounter{equation}{0}

Recently the dynamic properties of quasi-one-dimensional quantum spin systems 
have become an intense area of research.
On the one hand, 
a variety of quasi-one-dimensional materials 
were discovered during last decades.
On the other hand,
many exactly solvable statistical mechanics models 
refer to one spatial dimension. 
Very often these models permit to make a detailed analysis 
not only of their thermodynamic properties 
but also of their dynamic properties. 
Such theoretical investigations are necessary for interpretation 
of the experimental data 
observable in the scattering or resonance experiments 
on quasi-one-dimensional compounds \cite{01}.

In this paper we analyze the effect of the Dzyaloshinskii-Moriya interaction 
on the dynamic properties of quantum spin chains.
The Dzyaloshinskii-Moriya interaction is present in many low-dimensional materials 
\cite{02,03,04,05,06,07,08,09,10,11,12,13,14,15}.
Such antisymmetric exchange interaction,
${\bf D}_{{\bf{n}}{\bf{m}}}\cdot\left[{\bf{s}}_{\bf{n}}\times{\bf{s}}_{\bf{m}}\right]$,
takes place,
if allowed by crystal symmetry,
due to spin-orbit coupling \cite{16} 
and is weaker than the symmetric Heisenberg superexchange interaction 
$J_{{\bf{n}}{\bf{m}}}\left({\bf{s}}_{\bf{n}}\cdot{\bf{s}}_{\bf{m}}\right)$.
In spite of this,
this interaction has a number of important consequences 
and may cause a number of unconventional phenomena. 
Interestingly, 
the Dzyaloshinskii-Moriya interaction may appear 
while analyzing the nonequilibrium steady states of quantum spin chains with currents
\cite{17,18,19}.
It may also appear in the quantum spin representation 
in the stochastic kinetics 
of adsorption-desorption processes \cite{20}
(see also \cite{21}).

The analysis of the effect of the Dzyaloshinskii-Moriya interaction 
on the dynamic quantities of quantum spin chains 
was reported in several papers \cite{22,23}.
In particular in Refs. \cite{22,23} 
such analysis was performed using the symmetry arguments and field-theoretical methods
for the isotropic Heisenberg ($XXX$) chain.

In our study we restrict ourselves to a simpler model,
i.e. anisotropic $XY$ chain,
the dynamic properties of which are amenable to detailed analytical and numerical analysis.
We should also note 
that Cs$_2$CoCl$_4$ provides a new example 
of spin-$\frac{1}{2}$ $XY$ chain \cite{24}
that may renew interest in the calculation of observable quantities 
for such a chain \cite{25,26,27}.

To  be specific,
we consider $N$ spins one-half 
governed by the following Hamiltonian
\begin{eqnarray}
\label{1.01}
H=\sum_n\left(J^xs_n^xs_{n+1}^x+J^ys_n^ys_{n+1}^y\right)
+\sum_nD\left(s_n^xs_{n+1}^y-s_n^ys_{n+1}^x\right)
+\sum_n\Omega s_n^z.
\end{eqnarray} 
Here $J^x$, $J^y$ are the anisotropic $XY$ exchange interaction constants,
$D$ is the $z$-component of the Dzya\-lo\-shin\-skii-Moriya interaction
and
$\Omega$ is the transverse magnetic field.
Such model was introduced in Refs. \cite{28,29}.
Some of its dynamic properties were examined in Refs. \cite{30,31,32,33,34}.
In particular,
the transverse dynamic susceptibility $\chi_{zz}(\kappa,\omega)$ 
of the model (\ref{1.01}) 
was derived explicitly 
for $\kappa=0$ \cite{30}
and $\kappa\ne 0$ \cite{32}.
Moreover,
the case of isotropic $XY$ interaction $J^x=J^y$ 
(in this case the Dzyaloshinskii-Moriya interaction can be eliminated from the Hamiltonian 
by a spin axes rotation)
was analyzed in some detail \cite{34,35}.

The performed analysis is based on the Jordan-Wigner transformation,
\begin{eqnarray}
\label{1.02}
c_1=s_1^-,
\;\;\;
c_n=\left(-2s_1^z\right)\left(-2s_2^z\right)\ldots\left(-2s_{n-1}^z\right)s_n^-,
\;n=2,\ldots,N,
\end{eqnarray}
which transforms (\ref{1.01}) into the following bilinear Fermi form
\begin{eqnarray}
\label{1.03}
H=\sum_n\left(\frac{J+{\mbox{i}}D}{2}c_n^+c_{n+1}
-\frac{J-{\mbox{i}}D}{2}c_nc_{n+1}^+
+\frac{\gamma}{2}\left(c_n^+c_{n+1}^+-c_nc_{n+1}\right)
+\Omega \left(c_n^+c_n-\frac{1}{2}\right)\right)
\end{eqnarray}
with
$J=\frac{1}{2}\left(J^x+J^y\right)$
and
$\gamma=\frac{1}{2}\left(J^x-J^y\right)$.

In our calculations we consider both periodic and open boundary conditions.
Of course, 
in the thermodynamic limit the results for bulk characteristics should be insensitive 
to the boundary conditions implied.
In the former case, 
i.e. when $s_{N+1}^\alpha=s_1^\alpha$, 
the bilinear Fermi form (\ref{1.03}) is periodic or antiperiodic 
depending on whether the number of fermions is odd or even.
After the Fourier transformation,
\begin{eqnarray}
\label{1.04}
c_\kappa=\frac{1}{\sqrt{N}}\sum_n\exp\left({\mbox{i}}\kappa n\right)c_n,
\;\;\;
c_n=\frac{1}{\sqrt{N}}\sum_\kappa\exp\left(-{\mbox{i}}\kappa n\right)c_\kappa
\end{eqnarray}
(here $\kappa$ takes $N$ values 
$\frac{2\pi}{N}n$ for periodic boundary conditions
or 
$\frac{2\pi}{N}\left(n+\frac{1}{2}\right)$ for antiperiodic boundary conditions;
$n=-\frac{N}{2},-\frac{N}{2}+1,\ldots,\frac{N}{2}-1$ for $N$ even
and
$n=-\frac{N-1}{2},-\frac{N-1}{2}+1,\ldots,\frac{N-1}{2}$ for $N$ odd), 
and the Bogolyubov transformation with Fermi-operators $\beta$
\begin{eqnarray}
\label{1.05}
\beta_\kappa={\mbox{i}}u_\kappa c_\kappa+v_\kappa c^+_{-\kappa},
\;\;\;
c_{\kappa}=-{\mbox{i}}u_\kappa \beta_\kappa +v_\kappa \beta^+_{-\kappa}
\end{eqnarray}
with 
\begin{eqnarray}
\label{1.06}
u_\kappa={\mbox{sgn}}\left(\gamma\sin\kappa\right)
\frac{1}{\sqrt{2}}\sqrt{1+\frac{\Omega+J\cos\kappa}{\lambda_\kappa}},
\;\;\;
v_\kappa=\frac{1}{\sqrt{2}}\sqrt{1-\frac{\Omega+J\cos\kappa}{\lambda_\kappa}},
\nonumber\\
\lambda_\kappa
=\sqrt{\left(\Omega+J\cos\kappa\right)^2+\gamma^2\sin^2\kappa}
\end{eqnarray}
the Hamiltonian (\ref{1.03}) becomes
\begin{eqnarray}
\label{1.07}
H=\sum_\kappa\Lambda_\kappa\left(\beta^+_{\kappa}\beta_\kappa-\frac{1}{2}\right),
\end{eqnarray}
\begin{eqnarray}
\label{1.08}
\Lambda_\kappa=D\sin\kappa+\lambda_\kappa.
\end{eqnarray}
It should be noted here 
that only the energy spectrum $\Lambda_{\kappa}$ (\ref{1.08}) 
but not the coefficients of the Bogolyubov transformation $u_\kappa$, $v_\kappa$ (\ref{1.06})
depends on $D$.
Using (\ref{1.08}) one immediately finds 
that the energy spectrum is gapless
when $\gamma^2\le D^2$ and $\Omega^2\le J^2+D^2-\gamma^2$
or
when $\gamma^2>D^2$ and $\Omega^2=J^2$.

In our numerical calculations we use open boundary conditions \cite{36}.
The bilinear Fermi form (\ref{1.03}) with open boundary conditions
can be brought into the form (\ref{1.07}) 
after a linear transformation
\begin{eqnarray}
\label{1.09}
\eta_k=\sum_{j=1}^N\left(g_{kj}c_j+h_{kj}c_j^+\right),
\;\;\;
c_j=\sum_{k=1}^N\left(g_{kj}^*\eta_k+h_{kj}\eta_k^+\right)
\end{eqnarray}
where 
\begin{eqnarray}
\label{1.10}
\sum_{j=1}^N\left(g_{kj}A_{jm}-h_{kj}B_{jm}^*\right)
=\Lambda_kg_{km},
\;\;\;
\sum_{j=1}^N\left(g_{kj}B_{jm}-h_{kj}A_{jm}^*\right)
=\Lambda_kh_{km},
\nonumber\\
\sum_{j=1}^N\left(g_{kj}g_{qj}^*+h_{kj}h_{qj}^*\right)=\delta_{kq},
\;\;\;
\sum_{j=1}^N\left(g_{kj}g_{km}^*+h_{kj}^*h_{km}\right)=\delta_{jm},
\nonumber\\
\sum_{j=1}^N\left(g_{kj}h_{qj}+h_{kj}g_{qj}\right)
=\sum_{k=1}^N\left(g_{kj}h_{km}^*+h_{kj}^*g_{km}\right)=0
\end{eqnarray}
with 
$A_{nm}=\Omega\delta_{nm}
+\frac{1}{2}\left(J+{\mbox{i}}D\right)\delta_{m,n+1}
+\frac{1}{2}\left(J-{\mbox{i}}D\right)\delta_{m,n-1}$,
$B_{nm}=\frac{1}{2}\gamma\left(\delta_{m,n+1}-\delta_{m,n-1}\right)$.
We are not aware of a general analytical solution for this problem.
For the two particular cases,
namely,
the anisotropic $XY$ chain without field
and the Ising chain in a transverse field 
such solutions can be found e.g. in Ref. \cite{37}.
In our study we solve Eqs. (\ref{1.10}) numerically for chains of about a few hundred sites.

The relation between the spin model (\ref{1.01}) 
and the noninteracting Jordan-Wigner fermions (\ref{1.07}) 
is a key step 
in the statistical mechanics calculations 
for one-dimensional spin-$\frac{1}{2}$ $XY$ systems.

In our study of the dynamic properties 
we focus on the dynamic spin structure factors
\begin{eqnarray}
\label{1.11}
S_{\alpha\alpha}(\kappa,\omega)
=
\frac{1}{N}\sum_{j=1}^N
\sum_{n=1}^N\exp\left({\mbox{i}}\kappa n\right)
\int_{-\infty}^{\infty}{\mbox{d}}t\exp\left({\mbox{i}}\omega t\right)
\left(\langle s_j^\alpha(t)s_{j+n}^\alpha\rangle
-\langle s_j^\alpha\rangle\langle s_{j+n}^\alpha\rangle\right),
\;\;\;
\alpha=x,y,z.
\end{eqnarray}
The dynamic structure factors are of considerable importance 
since they are directly comparable with inelastic neutron scattering experiments 
of some quasi-one-dimensional substances. 
The dynamic transverse spin structure factor $S_{zz}(\kappa,\omega)$ 
can be easily evaluated analytically (Section 2).
The transverse dynamics is governed exclusively by a two-fermion excitation continuum 
the properties of which 
in the case of $XY$ chain without the Dzyaloshinskii-Moriya interaction 
were discussed earlier \cite{38}.
Another quantity 
which is also governed by the two-fermion excitation continuum 
is the dynamic dimer structure factor \cite{39,40,41}. 
Therefore,
the effect of the Dzyaloshinskii-Moriya interaction 
on the two-fermion excitation continuum 
deserves a separate discussion (Section 3).
The $xx$ and $yy$ dynamic structure factors are computed numerically (Section 4).
We compare and contrast the properties of different dynamic structure factors
at different values of the transverse field and temperature
emphasizing the effect of the Dzyaloshinskii-Moriya interactions.
We end up with conclusions (Section 5).

Some preliminary results of this study 
were announced in the conference paper \cite{42}.

\section{Dynamics of transverse spin correlations}

\setcounter{equation}{0}

We start with the $zz$ dynamic structure factor of the model (\ref{1.01}).
For analytical calculation of this quantity 
one can consider only periodic
(or only antiperiodic) 
boundary conditions for the bilinear Fermi form (\ref{1.03}).
Using the relation between spin and Fermi operators (\ref{1.02})
and the transformations (\ref{1.04}) and (\ref{1.05}), (\ref{1.06}) 
after standard calculations 
using the Wick-Bloch-de Dominicis theorem 
we arrive at
\begin{eqnarray}
\label{2.01}
S_{zz}(\kappa,\omega)
=\int_{-\pi}^{\pi}
{\mbox{d}}\kappa_1
\left(
\left(
u_{\kappa_1}^2u_{\kappa_1-\kappa}^2-u_{\kappa_1}u_{\kappa_1-\kappa}v_{\kappa_1}v_{\kappa_1-\kappa}
\right)n_{\kappa_1}\left(1-n_{\kappa_1-\kappa}\right)
\delta\left(\omega+\Lambda_{\kappa_1}-\Lambda_{\kappa_1-\kappa}\right)
\right.
\nonumber\\
\left.
+
\left(
u_{\kappa_1}^2v_{\kappa_1-\kappa}^2+u_{\kappa_1}u_{\kappa_1-\kappa}v_{\kappa_1}v_{\kappa_1-\kappa}
\right)n_{\kappa_1}n_{-\kappa_1+\kappa}
\delta\left(\omega+\Lambda_{\kappa_1}+\Lambda_{-\kappa_1+\kappa}\right)
\right.
\nonumber\\
\left.
+
\left(
u_{\kappa_1-\kappa}^2v_{\kappa_1}^2+u_{\kappa_1}u_{\kappa_1-\kappa}v_{\kappa_1}v_{\kappa_1-\kappa}
\right)\left(1-n_{-\kappa_1}\right)\left(1-n_{\kappa_1-\kappa}\right)
\delta\left(\omega-\Lambda_{-\kappa_1}-\Lambda_{\kappa_1-\kappa}\right)
\right.
\nonumber\\
\left.
+
\left(
v_{\kappa_1}^2v_{\kappa_1-\kappa}^2-u_{\kappa_1}u_{\kappa_1-\kappa}v_{\kappa_1}v_{\kappa_1-\kappa}
\right)\left(1-n_{-\kappa_1}\right)n_{-\kappa_1+\kappa}
\delta\left(\omega-\Lambda_{-\kappa_1}+\Lambda_{-\kappa_1+\kappa}\right)
\right)
\end{eqnarray} 
where 
$n_\kappa=\left(\exp\left(\beta\Lambda_\kappa\right)+1\right)^{-1}$
is the Fermi function.
This result agrees with the corresponding formula 
for the transverse dynamic susceptibility $\chi_{zz}(\kappa,\omega)$ derived in Ref. \cite{32}.

Introducing the function
\begin{eqnarray}
\label{2.02}
f\left(\kappa_1,\kappa\right)
=\frac{\left(\Omega+J\cos\left(\kappa_1-\frac{\kappa}{2}\right)\right)
\left(\Omega+J\cos\left(\kappa_1+\frac{\kappa}{2}\right)\right)
-\gamma^2\sin\left(\kappa_1-\frac{\kappa}{2}\right)\sin\left(\kappa_1+\frac{\kappa}{2}\right)}
{\lambda_{\kappa_1-\frac{\kappa}{2}}\lambda_{\kappa_1+\frac{\kappa}{2}}}
\end{eqnarray} 
the dynamic structure factor 
$S_{zz}(\kappa,\omega)$ (\ref{2.01}) 
can be expressed as follows
\begin{eqnarray}
\label{2.03}
S_{zz}(\kappa,\omega)
=\int_{-\pi}^{\pi}
{\mbox{d}}\kappa_1
\left(
\frac{1+f\left(\kappa_1,\kappa\right)}{2}
\left(1-n_{\kappa_1-\frac{\kappa}{2}}\right)n_{\kappa_1+\frac{\kappa}{2}}
\delta\left(\omega-\Lambda_{\kappa_1-\frac{\kappa}{2}}+\Lambda_{\kappa_1+\frac{\kappa}{2}}\right)
\right.
\nonumber\\
\left.
+\frac{1-f\left(\kappa_1,\kappa\right)}{4}
\left(
\left(1-n_{\kappa_1-\frac{\kappa}{2}}\right)
\left(1-n_{-\kappa_1-\frac{\kappa}{2}}\right)
\delta\left(\omega-\Lambda_{\kappa_1-\frac{\kappa}{2}}-\Lambda_{-\kappa_1-\frac{\kappa}{2}}\right)
\right.
\right.
\nonumber\\
\left.
\left.
+
n_{\kappa_1+\frac{\kappa}{2}}
n_{-\kappa_1+\frac{\kappa}{2}}
\delta\left(\omega+\Lambda_{\kappa_1+\frac{\kappa}{2}}+\Lambda_{-\kappa_1+\frac{\kappa}{2}}\right)
\right)
\right).
\end{eqnarray} 
In the limit of isotropic $XY$ interaction $\gamma=0$
Eq. (\ref{2.03}) yields the result obtained earlier \cite{34}. 
In the limit $D=0$ 
(when $\Lambda_\kappa=\lambda_\kappa=\lambda_{-\kappa}\ge 0$)
and $T=0$ 
($\beta\to\infty$)
Eq. (\ref{2.03}) becomes
\begin{eqnarray}
\label{2.04}
S_{zz}(\kappa,\omega)
=\int_{-\pi}^{\pi}
{\mbox{d}}\kappa_1
\frac{1-f\left(\kappa_1,\kappa\right)}{4}
\delta\left(\omega-\Lambda_{\kappa_1-\frac{\kappa}{2}}-\Lambda_{\kappa_1+\frac{\kappa}{2}}\right).
\end{eqnarray} 
This coincides with the expression obtained earlier in Ref. \cite{38}.
We notice that 
Eq. (\ref{2.04}) is more generally valid for the case 
$\gamma^2>D^2$ (when $\Lambda_\kappa>0$) and $T=0$.
In the case $D^2>\gamma^2$ and $T=0$
or in the most general case of arbitrary $D$
and $T>0$ 
(as well as in the case $D=0$ but $T>0$ which was not considered in Ref. \cite{38})
the $zz$ dynamic structure factor
exhibits new qualitative features 
in comparison with the analysis reported in Ref. \cite{38}.
Again the $zz$ dynamic structure factor 
is governed exclusively by two-fermion excitations 
as can be seen from Eq. (\ref{2.03}),
however,
for $D^2>\gamma^2$, $T=0$ or for $T>0$ 
all three $\delta$-functions in Eq. (\ref{2.03}) may come into play.

In Fig. 1  
we show the gray-scale plots for $S_{zz}(\kappa,\omega)$ 
at low temperature ($\beta J=50$) 
for several typical sets of parameters
($J=1$, $\gamma=0.5$, $D=0,\;0.5,\;1$, 
$\Omega=0,\;0.5,\;1,\;\frac{\sqrt{7}}{2}$).
For $\gamma^2>D^2$ (right column in Fig. 1)
at such a low temperature only one two-fermion excitation continuum is relevant 
(the one which arises owing to 
$\delta\left(\omega-\Lambda_{\kappa_1-\frac{\kappa}{2}}-\Lambda_{-\kappa_1-\frac{\kappa}{2}}\right)$
in (\ref{2.03}))
whereas in the opposite case $D^2>\gamma^2$ (left column in Fig. 1) 
all three two-fermion excitation continua contribute to transverse dynamics.
In Fig. 2 
we demonstrate typical low-temperature frequency profiles of 
$S_{zz}(0,\omega)$,
$S_{zz}(\frac{\pi}{2},\omega)$,
$S_{zz}(\pi,\omega)$
for a chain with $J=1$, $\gamma=0.5$, $\Omega=0.5$
and different values of the Dzyaloshinskii-Moriya interaction
$D=0,\;0.5,\;1$.
In Fig. 3
we show the gray-scale plots for $S_{zz}(\kappa,\omega)$ 
at intermediate and high temperatures ($\beta J=10,\;1,\;0.1$)
for a chain with $J=1$, $\gamma=0.5$, $D=1$, $\Omega=0.5$.  
All these data are used herein  
to discuss generic properties of a two-fermion dynamic structure factor 
(Section 3)
and 
to compare different dynamic structure factors 
(Section 4).

\section{Two-fermion excitation continua}

\setcounter{equation}{0}

The $zz$ dynamic structure factor (\ref{2.03}) 
can be rewritten in the form
\begin{eqnarray}
\label{3.01}
S_{zz}(\kappa,\omega)
=\sum_{j=1}^3 S_{zz}^{(j)}(\kappa,\omega),
\;\;\;
S_{zz}^{(j)}(\kappa,\omega)
=\int_{-\pi}^{\pi}{\mbox{d}}\kappa_1
B^{(j)}(\kappa_1,\kappa)C^{(j)}(\kappa_1,\kappa)
\delta\left(\omega-E^{(j)}(\kappa_1,\kappa)\right)
\end{eqnarray}
with
\begin{eqnarray}
\label{3.02}
B^{(1)}(\kappa_1,\kappa)=B^{(3)}(\kappa_1,\kappa)=\frac{1-f(\kappa_1,\kappa)}{4},
\nonumber\\
B^{(2)}(\kappa_1,\kappa)=\frac{1+f(\kappa_1,\kappa)}{2},
\end{eqnarray}
\begin{eqnarray}
\label{3.03}
C^{(1)}(\kappa_1,\kappa)
=\left(1-n_{\kappa_1-\frac{\kappa}{2}}\right)\left(1-n_{-\kappa_1-\frac{\kappa}{2}}\right),
\nonumber\\
C^{(2)}(\kappa_1,\kappa)
=\left(1-n_{\kappa_1-\frac{\kappa}{2}}\right)n_{\kappa_1+\frac{\kappa}{2}},
\nonumber\\
C^{(3)}(\kappa_1,\kappa)
=n_{\kappa_1+\frac{\kappa}{2}}n_{-\kappa_1+\frac{\kappa}{2}},
\end{eqnarray}
\begin{eqnarray}
\label{3.04}
E^{(1)}(\kappa_1,\kappa)
=\Lambda_{\kappa_1-\frac{\kappa}{2}}+\Lambda_{-\kappa_1-\frac{\kappa}{2}},
\nonumber\\
E^{(2)}(\kappa_1,\kappa)
=\Lambda_{\kappa_1-\frac{\kappa}{2}}-\Lambda_{\kappa_1+\frac{\kappa}{2}},
\nonumber\\
E^{(3)}(\kappa_1,\kappa)
=-\Lambda_{\kappa_1+\frac{\kappa}{2}}-\Lambda_{-\kappa_1+\frac{\kappa}{2}}.
\end{eqnarray}
In accordance with (\ref{3.01})
we distinguish three two-fermion excitation continua
(they correspond to $j=1,2,3$ in Eq. (\ref{3.01}))
which govern $S_{zz}(\kappa,\omega)$.
The gray-scale plots of $S_{zz}^{(j)}(\kappa,\omega)$, $j=1,2,3$
for a typical set of parameters, 
$J=1$, $\gamma=0.5$, $D=1$, $\Omega=0.5$, $\beta=50$,
are shown in Fig. 4.
As can be seen from (\ref{3.01}) and (\ref{3.02}), (\ref{3.03}), (\ref{3.04})
the specific features of the considered two-fermion dynamic structure factor
which describes the dynamics of the transverse spin fluctuations 
are controlled by the $B^{(j)}$-functions.
Therefore,
in Fig. 5 we also plot  
$S_{zz}^{(j)}(\kappa,\omega)$ (\ref{3.01})
for the same set of parameters as in Fig. 4,
although,
with $B^{(j)}(\kappa_1,\kappa)=1$.
By comparing Figs. 4 and 5 
one can distinguish 
the specific features  
(coming from the $B^{(j)}$-functions (\ref{3.02}))
and 
the generic features  
(coming from the $C^{(j)}$-functions (\ref{3.03})
and the $E^{(j)}$-functions (\ref{3.04})) 
of the dynamic quantity considered  
(see discussion below).

We remark
that although we were not able to find a simple analytical form 
of the most important lines in the $\kappa$--$\omega$ plane 
characterizing the two-fermion excitation continua
for a general case of the spin chain (\ref{1.01})
it is easy to determine these functions numerically 
for any set of parameters 
using MAPLE or/and FORTRAN codes.
We also recall 
that in the simplest case of the isotropic $XY$ model in a transverse field 
($\gamma=0$, $D=0$)
the corresponding results have been derived analytically
\cite{43}.
However,
in the case $\gamma\ne 0$, $D=0$
the analytical results 
have been reported only 
in the limiting cases
$\Omega=0$
or
$\gamma=1$ \cite{38}.
Of course, 
the case $\gamma\ne 0$, $D\ne 0$ considered in the present paper 
is even more complicated.
In what follows 
we take a typical set of parameters 
$J=1$, $\gamma=0.5$, $D=1$, $\Omega=0.5$.

We begin with the case of infinite temperature $T\to\infty$ ($\beta=0$).
The two-fermion dynamic structure factor may have nonzero values 
in the plane wave-vector $\kappa$ -- frequency $\omega$
if the equation 
\begin{eqnarray}
\label{3.05}
\omega-E^{(j)}(\kappa_1,\kappa)=0
\end{eqnarray}
has at least one solution $\kappa_1^{\star}$,
$-\pi\le\kappa_1^{\star}<\pi$.
In Fig. 6 
(panels a ($j=1$), b ($j=2$), c ($j=3$))
we show the regions in the $\kappa$--$\omega$ plane 
in which equation (\ref{3.05}) has 
four solutions (dark-gray regions), 
two solutions (gray regions) 
or has no solutions (white regions).
The bounding lines 
of the regions in which equation (\ref{3.05}) has solutions 
constitute 
the upper ($\omega^{(j)}_u(\kappa)$) 
and 
the lower ($\omega^{(j)}_l(\kappa)\le\omega^{(j)}_u(\kappa)$) 
boundaries 
of the two-fermion excitation continuum, 
respectively;
moreover,
for some regions of the wave-vector $\kappa$
the lower boundaries 
$\omega^{(j)}_l(\kappa)$ may be equal to zero.
Alternatively,
we may find the upper and the lower boundaries of the two-fermion excitation continuum 
seeking for the maximal and minimal values of $E^{(j)}(\kappa_1,\kappa)$ 
while $\kappa_1$ varies from $-\pi$ to $\pi$,
i.e.,
\begin{eqnarray}
\label{3.06}
\omega^{(j)}_u(\kappa)=\max_{-\pi\le\kappa_1<\pi}\left\{E^{(j)}(\kappa_1,\kappa)\right\},
\;\;\;
\omega^{(j)}_l(\kappa)=\min_{-\pi\le\kappa_1<\pi}\left\{0,\;E^{(j)}(\kappa_1,\kappa)\right\}.
\end{eqnarray}
We have found that $\omega^{(j)}_u(\kappa)$ and $\omega^{(j)}_l(\kappa)\ne 0$ 
occur for the values of $\kappa_1$,
$\kappa_1^{\star}$, 
which satisfy the equation
\begin{eqnarray}
\label{3.07}
\left.\frac{\partial}{\partial\kappa_1}E^{(j)}(\kappa_1,\kappa)\right\vert_{\kappa_1=\kappa_1^{\star}}=0.
\end{eqnarray}
Moreover,
Eq. (\ref{3.07}) also holds along the boundary $\omega(\kappa)$ 
between the gray and the dark-gray regions 
in the upper ($j=1$) and the middle ($j=2$) panels in the left column in Fig. 6.
On the other hand,
the quantities  
\begin{eqnarray}
\label{3.08}
S^{(j)}(\kappa,\omega)
=\int_{-\pi}^{\pi}{\mbox{d}}\kappa_1 
{\cal{S}}^{(j)}(\kappa_1,\kappa)
\delta\left(\omega-E^{(j)}(\kappa_1,\kappa)\right)
=\sum_{\left\{\kappa_{1}^{\star}\right\}}
\left.
\frac{{\cal{S}}^{(j)}(\kappa_1,\kappa)}{\left\vert\frac{\partial}{\partial\kappa_1}E^{(j)}(\kappa_1,\kappa)\right\vert}
\right\vert_{\kappa_1=\kappa_1^\star},
\end{eqnarray}
where $\left\{\kappa_{1}^{\star}\right\}$
are solutions of Eq. (\ref{3.05}),
may exhibit a van Hove singularity
along the line 
$\omega^{(j)}_s(\kappa)=E^{(j)}(\kappa_{1}^{\star},\kappa)$,
where $\kappa_{1}^{\star}$ satisfies Eq. (\ref{3.07}).
Thus,
the mentioned boundary lines in the panels in the left column in Fig. 6 
are the lines of van Hove singularities 
akin to the density of states effect in one dimension.
Further,
we have found that for almost all cases 
$\frac{\partial^2}{{\partial\kappa_1}^2}E^{(j)}(\kappa_1,\kappa)\ne 0$
for the values of $\kappa_1$ which satisfy 
(\ref{3.05}) with $\omega=\omega^{(j)}_s(\kappa)$ 
that obviously implies a familiar square-root divergence 
\begin{eqnarray}
\label{3.09}
S^{(j)}(\kappa,\omega)
\propto \epsilon^{-\frac{1}{2}}
\end{eqnarray}
when $\omega$ approaches $\omega^{(j)}_s(\kappa)$,
$\epsilon=\left\vert\omega-\omega^{(j)}_s(\kappa)\right\vert$.
However,
for $j=2$ and $\kappa\approx 1.07844531$
(see panel b in Fig. 6)
Eq. (\ref{3.07}) holds for $\kappa_1\approx 2.16480069$ 
and at this point  
$\frac{\partial^2}{{\partial\kappa_1}^2}E^{(2)}(\kappa_1,\kappa)=0$ 
but
$\frac{\partial^3}{{\partial\kappa_1}^3}E^{(2)}(\kappa_1,\kappa)
\approx 2.96548741 \ne 0$.
As a result, 
if $\omega$ is in the $\epsilon$-vicinity of
$\omega^{(2)}_s(\kappa\approx 1.07844531)\approx 0.78594502$ 
the quantity $S^{(2)}(\kappa,\omega)$ (\ref{3.08}) 
exhibits singularity with another exponent
\begin{eqnarray}
\label{3.10}
S^{(2)}(\kappa\approx 1.07844531,\omega\approx 0.78594502\pm\epsilon)
\propto \epsilon^{-\frac{2}{3}}.
\end{eqnarray}
We illustrate different types of singularities in Fig. 7.
In particular,
in Fig. 7a one can see the square-root divergencies (\ref{3.09}), 
whereas in Fig. 7b aside from the square-root divergences (\ref{3.09}) 
(solid and dotted curves)
one can also see the dependence (\ref{3.10})
(dashed curve).
We notice
that the $\epsilon^{-\frac{2}{3}}$ singularity remains for other values of $D$
and is also present when $D\to 0$.
For $J=1$, $\gamma=0.5$, $D=0$, $\Omega=0.5$ 
it occurs at $\kappa\approx 1.68213734$
while $\omega$ 
approaches $\omega_s^{(2)}(\kappa\approx 1.68213734)=0$.
Interestingly,
this fact could not be detected in the earlier study 
on the $zz$ dynamics in the anisotropic $XY$ chain in a transverse field 
without the Dzyaloshinskii-Moriya interaction \cite{38}
since that study refers to the zero-temperature case 
when only the continuum $j=1$ manifests itself in the transverse spin dynamics
(see discussion after Eq. (\ref{2.04})).
We also notice
that the observation of the $\epsilon^{-\frac{2}{3}}$ singularity 
may be difficult because of the fact 
that this peculiarity takes place only at one specific value of $\kappa$
(in contrast to the $\epsilon^{-\frac{1}{2}}$ singularity).
However,
for the values of wave-vector in the vicinity of this specific value 
one easily distinguishes a reminiscence of the $\epsilon^{-\frac{2}{3}}$ singularity
(see the dotted curve in Fig. 7b). 
Finally we note 
that for some of the lines characterizing the two-fermion excitation continua 
we can give simple analytical expressions.
Thus, for $j=1$ 
the maximum/minimum of $E^{(j)}(\kappa_1,\kappa)$ occurs at $\kappa_1=0$ and $\kappa_1=-\pi$ 
and hence the corresponding boundary lines are given by
$E^{(1)}(0,\kappa)$
and 
$E^{(1)}(-\pi,\kappa)$.
We did not find simple analytical expressions 
for the boundary lines between the white and the dark-gray regions 
and 
for the nonzero lower boundary
(see panel a in Fig. 6).
For $j=3$ the upper boundary is given by 
$E^{(3)}(0,\kappa)$
and 
$E^{(3)}(-\pi,\kappa)$.

Next we turn to the zero-temperature case $T=0$ ($\beta\to\infty$) 
for which the effect of the Fermi functions involved in $C^{(j)}$-functions becomes important. 
The Fermi functions contract a region of possible values of $\kappa_1$
in Eq. (\ref{3.05});
now $\kappa_1$ varies only within a part of the region between $-\pi$ and $\pi$ 
where $C^{(j)}(\kappa_1,\kappa)\ne 0$. 
In Fig. 6 
(panels d ($j=1$), e ($j=2$), f ($j=3$))
we show the regions in the $\kappa$--$\omega$ plane
in which Eq. (\ref{3.05}) taking into account the condition $C^{(j)}(\kappa_1,\kappa)\ne 0$ 
has four solutions (dark-gray regions in panel d),
two solutions (gray regions),
one solution (light-gray regions in panel e) 
or has no solutions (white regions).
The upper and the lower boundaries may be also found according to Eq. (\ref{3.06}),
although 
with $\kappa_1$ varying within a part of the region between $-\pi$ and $\pi$ 
where $C^{(j)}(\kappa_1,\kappa)\ne 0$.
The values of $\kappa$ at which $\omega^{(j)}_l(\kappa)=0$ 
correspond to the potential soft modes $\kappa_c$
(see the right panels in Fig. 6).
(Note  
that the soft modes may not occur owing to $B^{(j)}$-functions 
(compare Figs. 4 and 5).)
The two-fermion dynamic quantities may exhibit the above discussed van Hove singularities
if their occurrence is not prohibited by the Fermi functions
(and $B^{(j)}$-functions,
compare Figs. 4 and 5).
Moreover,
along the bounding lines between the regions of $\kappa$--$\omega$ plane 
corresponding to different numbers of solution of Eq. (\ref{3.05}) 
the two-fermion dynamic quantities may abruptly change their values
exhibiting a finite jump 
(for example, 
we refer to the boundary line between the gray and light-gray regions in Fig. 6e; 
see also the panels b in Figs. 4 and 5).
We can write down the analytical expressions 
for some characteristic lines seen in Figs. 6d, 6e, 6f.
We introduce 
\begin{eqnarray}
\label{3.11}
\cos\kappa_{\pm}
=
\frac{-\Omega J\pm\sqrt{\left(D^2-\gamma^2\right)\left(J^2+D^2-\gamma^2-\Omega^2\right)}}
{J^2+D^2-\gamma^2},
\;\;\;
D\sin\kappa_{\pm}<0;
\end{eqnarray}
$\kappa_{\pm}$ obey the equation $\Lambda_{\kappa_{\pm}}=0$.
For the chosen set of parameters 
$J=1$, $\gamma=0.5$, $D=1$, $\Omega=0.5$ 
we have 
$\kappa_+\approx -1.24466864$,
$\kappa_-\approx -2.67211739$.
The lines $\Lambda_{-\kappa_{\pm}-\kappa}$ 
form the lower boundary of the continuum $j=1$;
the lines $\Lambda_{\kappa_{\pm}-\kappa}$, $-\Lambda_{\kappa_{\pm}+\kappa}$  
form the lower boundary of the continuum $j=2$;
the lines $-\Lambda_{\kappa-\kappa_{\pm}}$
form the lower boundary of the continuum $j=3$.
The soft modes may occur at the values of wave-vector
$-2\kappa_{\pm},\;-\kappa_+-\kappa_-$
($j=1$),
$0,\;\pm\left(\kappa_+-\kappa_-\right)$
($j=2$),
$2\kappa_{\pm},\;\kappa_++\kappa_-$
($j=3$).

Finally,
we emphasize a role of 
$B^{(j)}$-functions (\ref{3.02}) 
which are responsible for the specific features 
of the dynamic transverse spin structure factor $S_{zz}(\kappa,\omega)$.
The functions $B^{(j)}(\kappa_1,\kappa)$ 
modify and add some additional structure to $S_{zz}(\kappa,\omega)$ 
in the $\kappa$--$\omega$ plane
(compare Fig. 4 and Figs. 5, 6d, 6e, 6f 
referring to the low-temperature limit
and Fig. 3c and Figs. 6a, 6b, 6c 
referring to the high-temperature limit).
In particular,
the function $B^{(2)}(\kappa_1,\kappa)$ removes 
the soft modes at $\kappa=\pm\left(\kappa_+-\kappa_-\right)$ 
but not at $\kappa=0$ 
from $S_{zz}(\kappa,\omega)$
(see Fig. 4b).
Furthermore,
comparing Figs. 4a, 4c and 5a, 5c 
one sees that van Hove singularities 
along the lines 
$\omega=E^{(1)}(0,\kappa)$, $\omega=E^{(1)}(-\pi,\kappa)$ (panels a)
and
along the lines 
$\omega=E^{(3)}(0,\kappa)$, $\omega=E^{(3)}(-\pi,\kappa)$ (panels c)
disappear
since
$B^{(1)}(0,\kappa)=B^{(3)}(0,\kappa)=B^{(1)}(-\pi,\kappa)=B^{(3)}(-\pi,\kappa)=0$.

To summarize this Section,
the two-fermion dynamic structure factors 
have a nonzero value 
only in a restricted area of the $\kappa$--$\omega$ plane
(two-fermion excitation continua)
and may exhibit the van Hove singularities 
not only with exponent $\frac{1}{2}$ but also with exponent $\frac{2}{3}$. 
Moreover,
at zero temperature 
the two-fermion dynamic structure factors
may exhibit jumps at which their values abruptly increase by a finite value.

\section{$xx$ and $yy$ dynamic structure factors}

\setcounter{equation}{0}

In the present Section 
we calculate the $xx$ and $yy$ dynamic structure factors.
For numerical calculations it is convenient to rewrite (\ref{1.11}) 
in the following form
\begin{eqnarray}
\label{4.01}
S_{\alpha\alpha}(\kappa,\omega)
=\int_{-\infty}^{\infty}{\mbox{d}}t\exp\left({\mbox{i}}\omega t\right)
\left(\langle s_j^{\alpha}(t)s_{j}^{\alpha}\rangle
-\langle s^\alpha\rangle^2\right)
\nonumber\\
+2\sum_{n=1,2,3,\ldots}
\int_{-\infty}^{\infty}{\mbox{d}}t
\;{\mbox{Re}}
\left(
\exp\left({\mbox{i}}\kappa n\right)
\exp\left({\mbox{i}}\omega t\right)
\left(\langle s_j^{\alpha}(t)s_{j+n}^{\alpha}\rangle
-\langle s^\alpha\rangle^2\right)
\right)
\end{eqnarray}
(we have omitted $\frac{1}{N}\sum_{j=1}^N$ in (\ref{1.11})
and
have used the relation
$\langle s_j^{\alpha}(-t)s_{j+n}^{\alpha}\rangle
=
\langle s_j^{\alpha}(t)s_{j-n}^{\alpha}\rangle^\star$).
As a result,
to get $S_{\alpha\alpha}(\kappa,\omega)$ 
according to (\ref{4.01})
we have to compute 
the time-dependent spin correlation functions 
$\langle s_j^{\alpha}(t)s_{j+n}^{\alpha}\rangle$, $n=0,1,2,\ldots$ 
with $t$ varying from $-\infty$ to $\infty$.
We compute 
$\left\langle s_j^{\alpha}(t)s_{j+n}^{\alpha}\right\rangle$
$\alpha=x,y$
numerically.
For this purpose we express 
the spin operators entering the time-dependent correlation functions 
in terms of auxiliary operators 
$\varphi_l^{\pm}=c_l^+\pm c_l$
which according to Eq. (\ref{1.09}) are linear combinations of operators $\eta_k^+$, $\eta_k$.
Since the $x$ and $y$ spin components at each site 
are essentially nonlocal objects in terms of Jordan-Wigner fermions 
(see (\ref{1.02}))
the resulting expressions for the time-dependent spin correlation functions 
$\langle s_j^{\alpha}(t)s_{j+n}^{\alpha}\rangle$,
$\alpha=x,y$
are complicated averages of products 
of a large number of Fermi operators
attached not only the sites $j$ and $j+n$ 
but to two strings of sites extending to the boundary of the chain. 
After applying the Wick-Bloch-de Dominicis theorem 
one gets an intricate result which can be compactly written 
as the Pfaffian of the $2(2j+n-1)\times 2(2j+n-1)$ antisymmetric matrix
constructed from elementary contractions 
$\langle \varphi_l^+(t)\varphi^+_m\rangle$,
$\langle \varphi_l^+(t)\varphi^-_m\rangle$,
$\langle \varphi_l^-(t)\varphi^+_m\rangle$,
$\langle \varphi_l^-(t)\varphi^-_m\rangle$.
The elementary contractions are easily expressed in terms of $\Lambda_k$, $g_{kj}$, $h_{kj}$ (\ref{1.09}), (\ref{1.10}).
Finally we numerically evaluate the Pfaffians.
Typically we take $N=400$,
assume $j=31,\;71,\;131$ 
and calculate the correlation functions with $n$ up to $50,\;70,\;140$
in the time range 
$-50\div 50,\;-100\div 100,\;-300\div 300,\;-600\div 600$.
To remove the effect of a finite time cut-off 
we multiply the integrands in (\ref{4.01}) 
by $\exp\left(-{\rm{i}}\epsilon\vert t\vert\right)$ 
with $\epsilon=0.01,\;0.02$.
To be sure that our results pertain to the thermodynamic limit 
we examine in detail different types of finite size effects
(for further details see Refs. \cite{36}).

In Fig. 8 (Fig. 9) we plot $S_{xx}(\kappa,\omega)$
($S_{yy}(\kappa,\omega)$)
at low temperature ($\beta J=50$)
for several typical sets of parameters 
($J=1$, $\gamma=0.5$, 
$D=0,\;0.5,\;1$, 
$\Omega=0,\;0.25,\;0.5,\;1$).
(For larger values of the transverse field $\Omega$ 
there are no qualitative changes in these dynamic structure factors:
they have a simple single-mode structure.)
In Fig. 10 (Fig. 11) we show typical low-temperature frequency profiles 
of $S_{xx}(\kappa,\omega)$ ($S_{yy}(\kappa,\omega)$)
at $\kappa=0$, $\kappa=\frac{\pi}{2}$, $\kappa=\pi$ 
for a chain with $J=1$, $\gamma=0.5$, $\Omega=0.5$ 
and different values of the Dzyaloshinskii-Moriya interaction 
$D=0,\;0.5,\;1$.
In Fig. 12 (Fig. 13) we plot $S_{xx}(\kappa,\omega)$ ($S_{yy}(\kappa,\omega)$) 
at intermediate and high temperatures 
($\beta J=10,\;1,\;0.1$)
for a chain with $J=1$, $\gamma=0.5$, $D=1$, $\Omega=0.5$.

In contrast to transverse dynamic structure factor, 
the $xx$ and $yy$ dynamic structure factors are essentially more complicated quantities 
within the Jordan-Wigner method.
Really,
owing to a nonlocal relation between the $x$ and $y$ spin components  
and Fermi operators (\ref{1.02}) 
the $xx$ and $yy$ time-dependent spin correlation functions 
are expressed through 
many-particle correlation functions   
of noninteracting Jordan-Wigner fermions.
Let us now discuss the obtained numerical results.
First of all we note 
that both dynamic structure factors  
$S_{xx}(\kappa,\omega)$ and $S_{yy}(\kappa,\omega)$
show similar behavior for the taken value of $\gamma=0.5$; 
obviously they become identical in the isotropic limit $\gamma=0$. 
We start with the dynamic structure factors at low temperatures.
As can be seen in Figs. 8 and 9
(and Figs. 12a and 13a) 
these dynamic structure factors show several washed-out excitation branches 
which are roughly in correspondence with the characteristic lines 
of the two-fermion excitation continua
(compare three dynamic structure factors 
in the panel a and the panels b and c in Fig. 14;
note that these quantities are shown for $\kappa$ that varies from $-\pi$ to $3\pi$).
Thus,
although $S_{xx}(\kappa,\omega)$ and $S_{yy}(\kappa,\omega)$ are many-particle quantities 
within the Jordan-Wigner picture 
and hence they are not restricted to some region in the $\kappa$--$\omega$ plane,
their values outside the two-fermion continua are rather small.
This observation 
(i.e. two-particle features dominate many-particle dynamic quantities 
$S_{xx}(\kappa,\omega)$ and $S_{yy}(\kappa,\omega)$ at low temperatures)
agrees with our previous studies on isotropic $XY$ chains \cite{36,34}
(see also Ref. \cite{38}).
The constant frequency scans for several values of the wave-vector displayed in Figs. 10, 11 
show the redistribution of spectral weight 
$S_{xx}(\kappa,\omega)$ and $S_{yy}(\kappa,\omega)$ 
as the Dzyaloshinskii-Moriya interaction $D$ increases.
We note that these frequency profiles exhibit one or several peaks 
that may be relatively sharp or broad.
The Dzyaloshinskii-Moriya interaction affects 
the positions of the peaks, their shapes and even their number
(see, for example, 
the dependences
$S_{xx}(0,\omega)$ vs $\omega$
and 
$S_{yy}(0,\omega)$ vs $\omega$
displayed in Figs. 10a and 11a).
Constant frequency (and wave-vector) scans 
can be obtained for quasi-one-dimensional compounds 
by neutron scattering or resonance techniques 
and our findings may be useful in explaining the experimental data 
for the corresponding materials. 
As can be seen from our results,
the Dzyaloshinskii-Moriya interaction clearly manifests itself  
in the frequency or wave-vector profiles of the dynamic structure factors 
that can be used in determining the magnitude of this interaction.

It should be remarked 
that the dynamic structure factors of quantum spin chains 
are often examined within the framework of a bosonization approach \cite{44,23}.
Note, however, 
that field-theoretical approaches 
do not apply to small length scales and short time scales 
when the discreteness of the lattice becomes important.
Therefore,
since these methods can describe only the low-energy physics,
the high-frequency features nicely seen in Figs. 8 and 9 
cannot be reproduced by these theories.

As temperature increases,  
the low-temperature structure gradually disappears 
and the dynamic structure factors
$S_{xx}(\kappa,\omega)$ and $S_{yy}(\kappa,\omega)$
become $\kappa$-independent in the high-temperature limit
(see Figs. 12 and 13).
This is in agreement with earlier studies in the infinite temperature limit 
\cite{31}.

To summarize,
$xx$ and $yy$ dynamic structure factors at low temperatures 
are not restricted to the two-fermion excitation continua 
and have (small) nonzero values outside these continua.
They exhibit several washed-out excitations 
concentrated along the characteristic lines of the two-fermion excitation continua.
The Dzyaloshinskii-Moriya interaction manifests
itself in the constant frequency/wave-vector scans
influencing the detailed structure of such profiles.
In the high-temperature limit 
$xx$ and $yy$ dynamic structure factors become $\kappa$-independent.

\section{Conclusions}

\setcounter{equation}{0}

In this paper,
we have obtained the detailed dynamic structure factors 
$S_{\alpha\alpha}(\kappa,\omega)$, $\alpha=x,y,z$
of the spin-$\frac{1}{2}$ anisotropic $XY$ chain in a transverse field 
with the Dzyaloshinskii-Moriya interaction. 
The Dzyaloshinskii-Moriya interaction leads to nontrivial changes 
in the dynamic quantities.
The two-fermion excitations which exclusively govern the $zz$ dynamic structure factor, 
form three excitation continua
and all of them manifest themselves even at zero temperature 
for sufficiently large strength of the Dzyaloshinskii-Moriya interaction.
The two-fermion dynamic quantities have nonzero values in a restricted region of the $\kappa$--$\omega$ plane;
they may exhibit van Hove singularities 
(not only with exponent $\frac{1}{2}$ but also with $\frac{2}{3}$); 
moreover,
they may exhibit finite jumps at zero temperature.
The $xx$ and $yy$ dynamic structure factors involve many-fermion excitations.
However,
the two-fermion excitations dominate their low-temperature behavior:
at low temperatures these quantities show several washed-out excitation branches 
which correspond to specific lines of the two-fermion excitation continua.

The Dzyaloshinskii-Moriya interaction clearly manifests itself 
in the frequency/wave-vector profiles 
which makes it possible to determine the magnitude of this interaction by measuring the dynamic structure factors.
Dynamical structure factors can be measured by neutron scattering.
Another experimental technique 
which yields dynamic quantities is electron spin resonance (ESR).
If a static magnetic field along $z$ axis 
and the electromagnetic wave polarized in $\alpha\perp z$ direction 
are applied to the spin-$\frac{1}{2}$ anisotropic $XY$ chain 
with the Dzyaloshinskii-Moriya interaction, 
the experimentally measurable absorption intensity is given by
\begin{eqnarray}
\label{5.01}
I(\omega)
\propto
\omega\frac{1-\exp\left(-\beta\omega\right)}{2}
S_{\alpha\alpha}(0,\omega).
\end{eqnarray}
Thus,
the theoretical results for the dynamic structure factors 
presented in Section 4 
are directly related to the ESR absorption intensity $I(\omega)$ (\ref{5.01}).
Similar study for the spin-$\frac{1}{2}$ isotropic $XY$ chain 
has been recently reported in Ref. \cite{45}.
A detailed analysis of the Dzyaloshinskii-Moriya interaction effect 
on ESR experiments on the materials,
which can be modeled 
by the spin-$\frac{1}{2}$ anisotropic $XY$ chain,
seems to be an interesting issue.
The work in this direction is in progress.

\section*{Acknowledgments}

O. D. acknowledges the kind hospitality 
of the University of Magdeburg in the end of 2005 and in the beginning of 2006.
T. V. acknowledges the kind hospitality 
of the University of Bayreuth in the spring and autumn of 2004.
The paper was presented 
at the International Conference
on Strongly Correlated Electron Systems 
(Vienna, July 26th - 30th, 2005).
O. D., T. V. and T. K. thank the Organizing Committee
for the support for attending the conference.

\renewcommand\baselinestretch{1.20}
\large\normalsize

\vspace{10mm}

{\bf {FIGURE CAPTIONS}}
\\

FIG. 1.
Gray-scale plot
for the $zz$ dynamic structure factor $S_{zz}(\kappa,\omega)$ (\ref{2.03}) 
for the spin chain (\ref{1.01})
with $J=1$, $\gamma=0.5$
at low temperature, $\beta =50$,
for different strengths of the Dzyaloshinskii-Moriya interaction,
$D=0$ (right column),
$D=0.5$ (middle column),
$D=1$ (left column),
and different fields,
$\Omega=0,\;0.5,\;1,\;\frac{\sqrt{7}}{2}$
(from bottom to top).
\\

FIG. 2.
Frequency profiles of $S_{zz}(\kappa,\omega)$
at different wave-vectors $\kappa=0$ (left panel),
$\kappa=\frac{\pi}{2}$ (middle panel)
and $\kappa=\pi$ (right panel)
for the spin chain (\ref{1.01})
with $J=1$, $\gamma=0.5$, 
$D=0$ (dotted lines),
$D=0.5$ (dashed lines),
$D=1$ (solid lines),
$\Omega=0.5$
at low temperature $\beta =50$.
\\

FIG. 3.
The $zz$ dynamic structure factor (\ref{2.03}) 
for the spin chain (\ref{1.01})
with $J=1$, $\gamma=0.5$, $D=1$, $\Omega=0.5$
at different temperatures 
$\beta=10$ (a),
$\beta=1$ (b)
and
$\beta=0.1$ (c).
\\

FIG. 4.
The gray-scale plots of 
$S_{zz}^{(1)}(\kappa,\omega)$ (a),
$S_{zz}^{(2)}(\kappa,\omega)$ (b)
and
$S_{zz}^{(3)}(\kappa,\omega)$ (c)
for the set of parameters  
$J=1$, $\gamma=0.5$, $D=1$, $\Omega=0.5$, $\beta=50$.
\\

FIG. 5.
For comparison 
$S_{zz}^{(j)}(\kappa,\omega)$ 
(\ref{3.01}) 
with $B^{(j)}(\kappa_1,\kappa)=1$
for the same set of parameters as in Fig. 4.
a: $j=1$,
b: $j=2$,
c: $j=3$.
\\

FIG. 6.
Towards the properties of the two-fermion excitation continua;
$J=1$, $\gamma=0.5$, $D=1$, $\Omega=0.5$;
left panels -- infinite temperature limit $T\to\infty$ ($\beta=0$),
right panels -- zero temperature limit $T=0$ ($\beta\to\infty$);
$j=1$ (panels a and d),
$j=2$ (panels b and e),
$j=3$ (panels c and f).
\\

FIG. 7.
Frequency profiles (\ref{3.08}) (with ${\cal{S}}^{(j)}(\kappa_1,\kappa)=1$)
for the chain (\ref{1.01}) 
with $J=1$, $\gamma=0.5$, $D=1$, $\Omega=0.5$, 
$\beta=0$
which illustrate van Hove singularities with different exponents
$\frac{1}{2}$ and $\frac{2}{3}$.
Panel a:
$S^{(1)}(\kappa,\omega)$ (\ref{3.08}) vs $\omega$
at 
$\kappa=0.5$ (solid line),
$\kappa=2$ (dashed line),
$\kappa=3$ (dotted line);
panel b:
$S^{(2)}(\kappa,\omega)$ (\ref{3.08}) vs $\omega$
at 
$\kappa=0.9$ (solid line),
$\kappa\approx 1.07844531$ 
(dashed line; 
the low-frequency singularity has power-law exponent $\frac{2}{3}$),
$\kappa=1.2$ (dotted line).
\\

FIG. 8.
The $xx$ dynamic structure factor $S_{xx}(\kappa,\omega)$ 
for the spin chain (\ref{1.01})
with $J=1$, $\gamma=0.5$
at low temperature $\beta=50$.
$D=0$ (right column),
$D=0.5$ (middle column),
$D=1$ (left column),
$\Omega=0,\;0.25,\;0.5,\;1$
(from bottom to top).
\\

FIG. 9.
The $yy$ dynamic structure factor $S_{yy}(\kappa,\omega)$ 
for the spin chain (\ref{1.01})
with $J=1$, $\gamma=0.5$
at low temperature $\beta=50$.
$D=0$ (right column),
$D=0.5$ (middle column),
$D=1$ (left column),
$\Omega=0,\;0.25,\;0.5,\;1$
(from bottom to top).
\\

FIG. 10.
Frequency profiles of $S_{xx}(\kappa,\omega)$
at $\kappa=0$ (left panel),
$\kappa=\frac{\pi}{2}$ (middle panel)
and $\kappa=\pi$ (right panel)
for the spin chain (\ref{1.01})
with $J=1$, $\gamma=0.5$, 
$D=0$ (dotted lines),
$D=0.5$ (dashed lines),
$D=1$ (solid lines),
$\Omega=0.5$
at low temperature $\beta =50$.
\\

FIG. 11.
Frequency profiles of $S_{yy}(\kappa,\omega)$
at $\kappa=0$ (left panel),
$\kappa=\frac{\pi}{2}$ (middle panel)
and $\kappa=\pi$ (right panel)
for the spin chain (\ref{1.01})
with $J=1$, $\gamma=0.5$, 
$D=0$ (dotted lines),
$D=0.5$ (dashed lines),
$D=1$ (solid lines),
$\Omega=0.5$
at low temperature $\beta =50$.
\\

FIG. 12.
The $xx$ dynamic structure factor $S_{xx}(\kappa,\omega)$ 
for the spin chain (\ref{1.01})
with $J=1$, $\gamma=0.5$, $D=1$, $\Omega=0.5$
at different temperatures: 
$\beta=10$ (left panel),
$\beta=1$ (middle panel)
and
$\beta=0.1$ (right panel).
\\

FIG. 13.
The $yy$ dynamic structure factor $S_{yy}(\kappa,\omega)$ 
for the spin chain (\ref{1.01})
with $J=1$, $\gamma=0.5$, $D=1$, $\Omega=0.5$
at different temperatures: 
$\beta=10$ (left panel),
$\beta=1$ (middle panel)
and
$\beta=0.1$ (right panel).
\\

FIG. 14.
$S_{zz}(\kappa,\omega)$,
$S_{xx}(\kappa,\omega)$
and
$S_{yy}(\kappa,\omega)$
(from top to bottom)
for the spin chain (\ref{1.01})
with $J=1$, $\gamma=0.5$, $D=1$, $\Omega=0.5$
at low temperature $\beta=50$. 

\newpage

\begin{figure}
\epsfig{file = 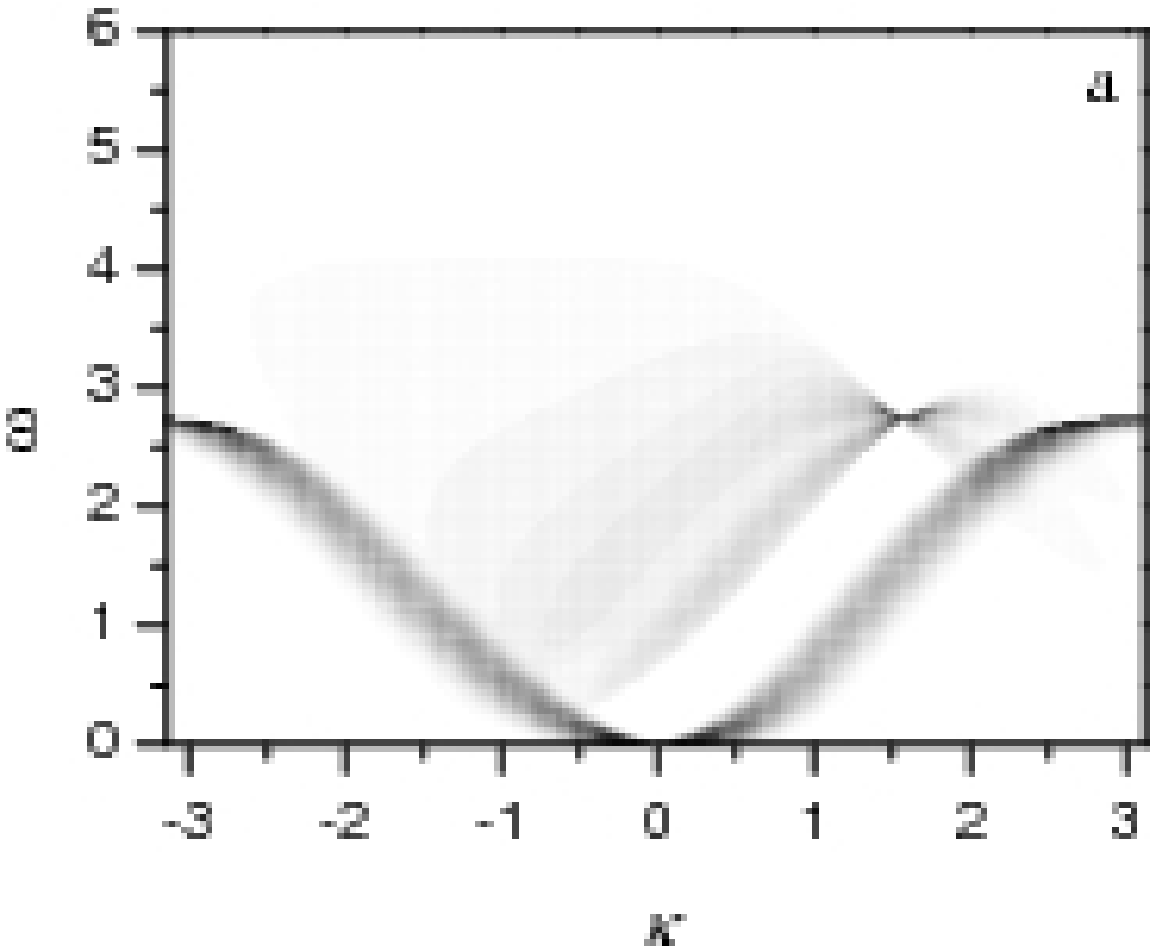, height = 0.25\linewidth}
\epsfig{file = 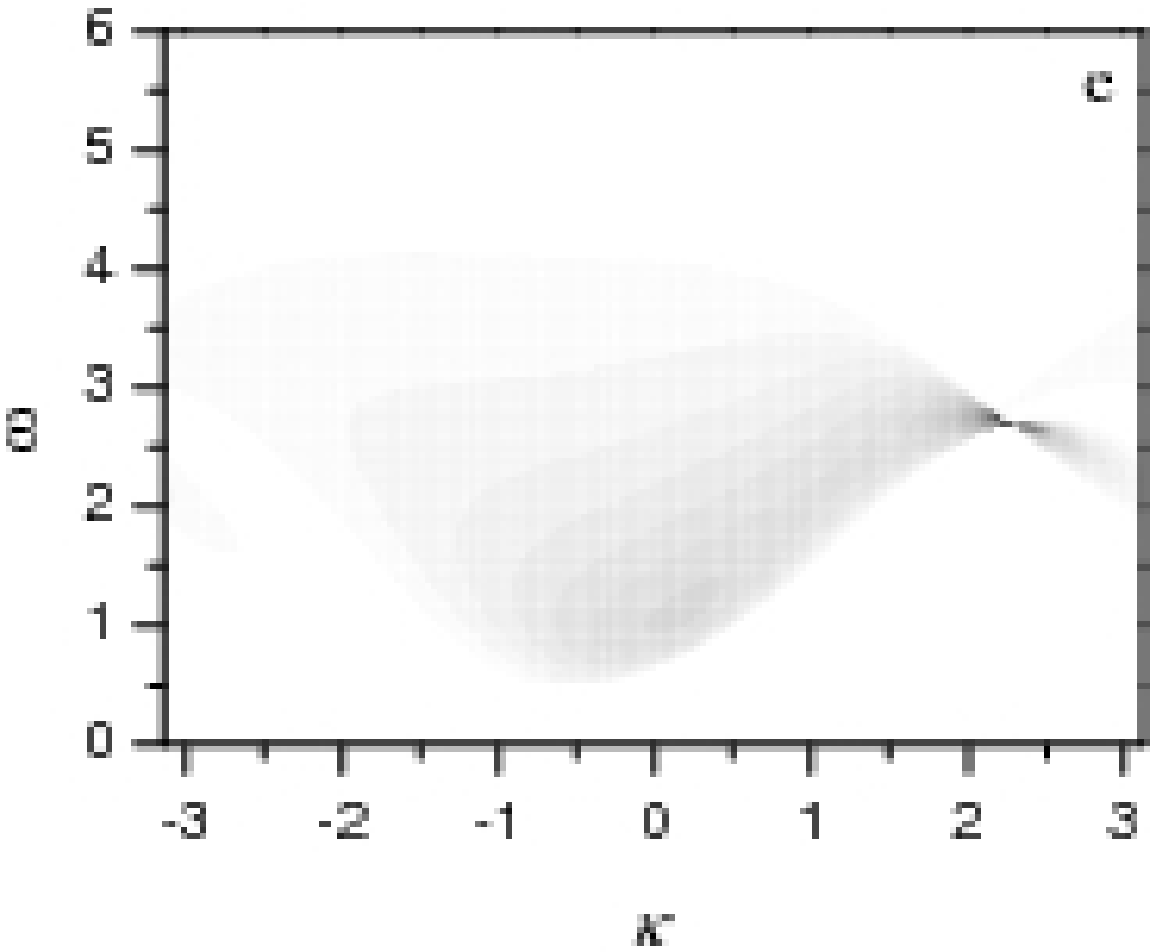, height = 0.25\linewidth}
\epsfig{file = 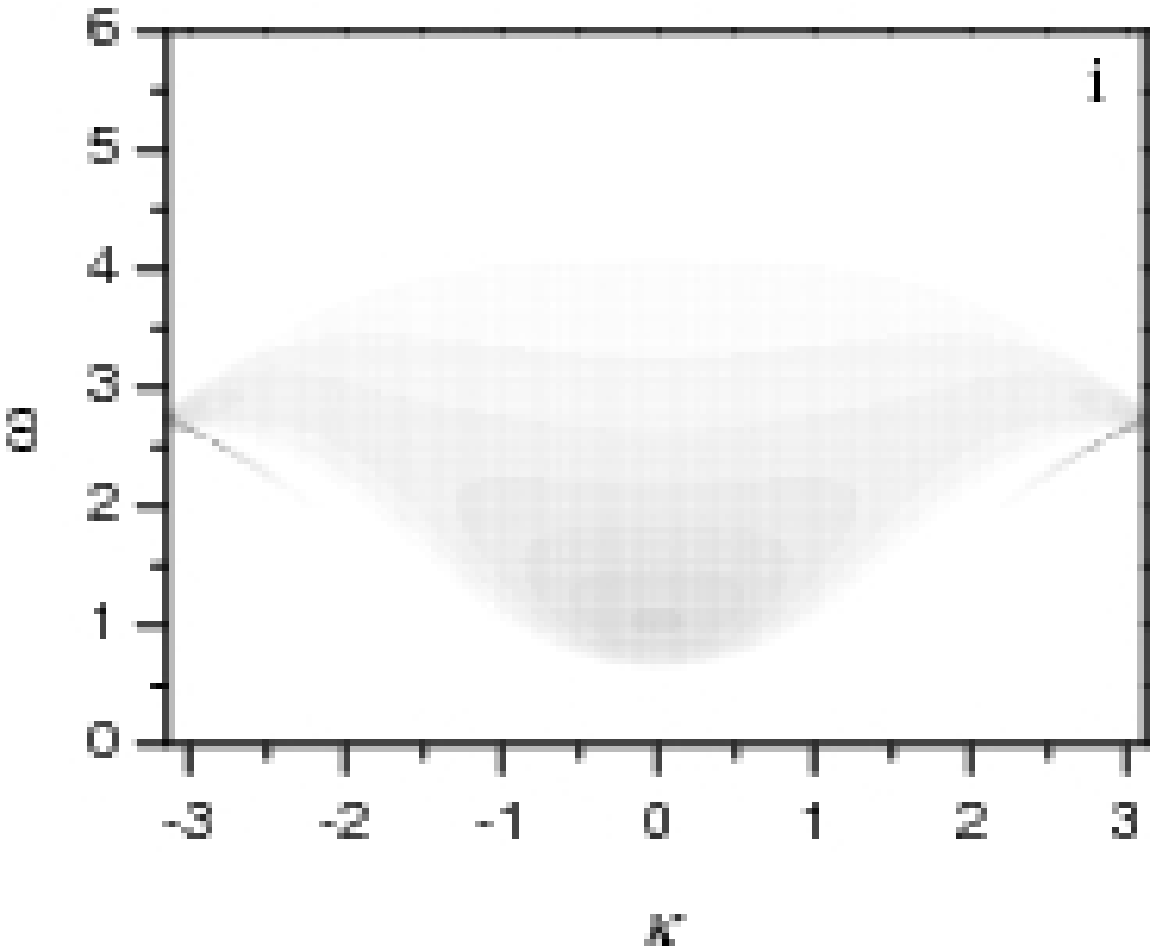, height = 0.25\linewidth}\\
\epsfig{file = 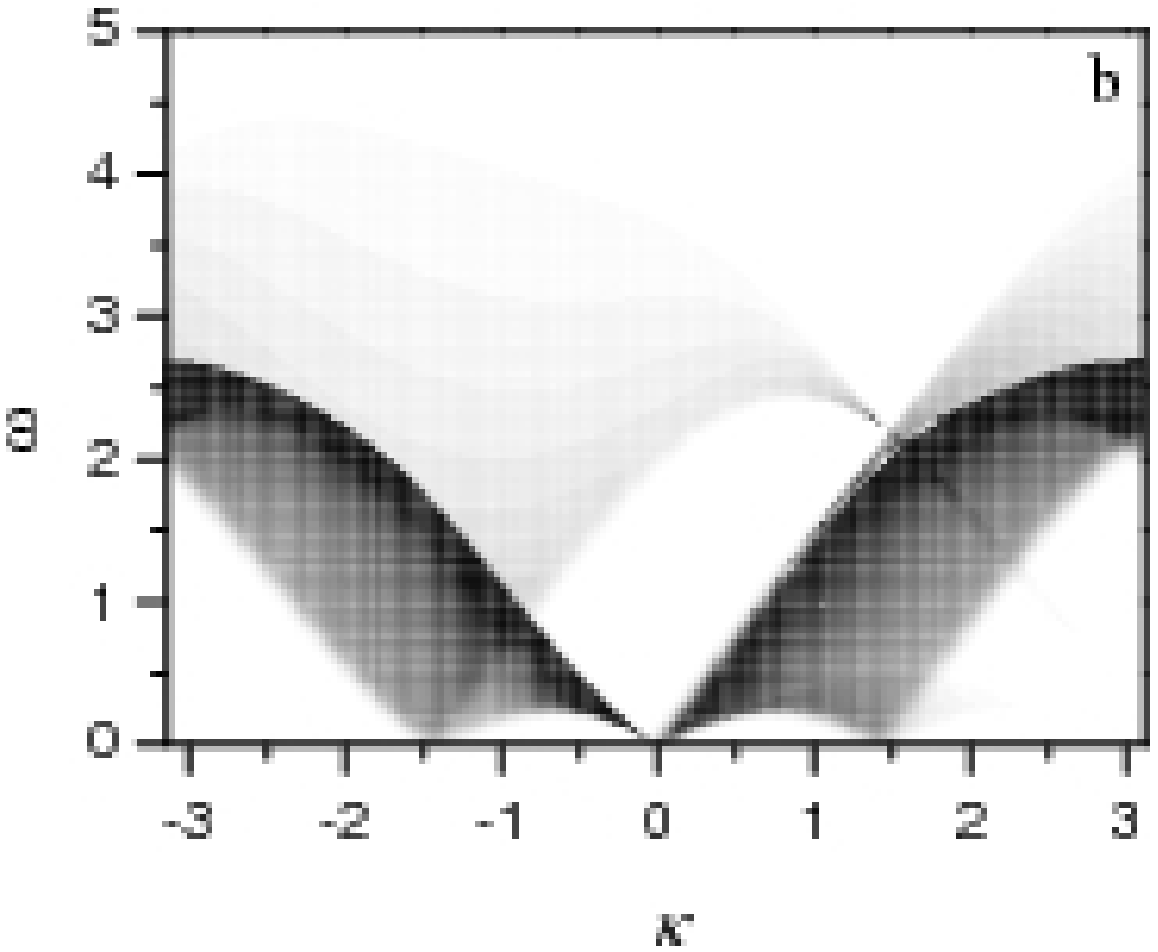, height = 0.25\linewidth}
\epsfig{file = 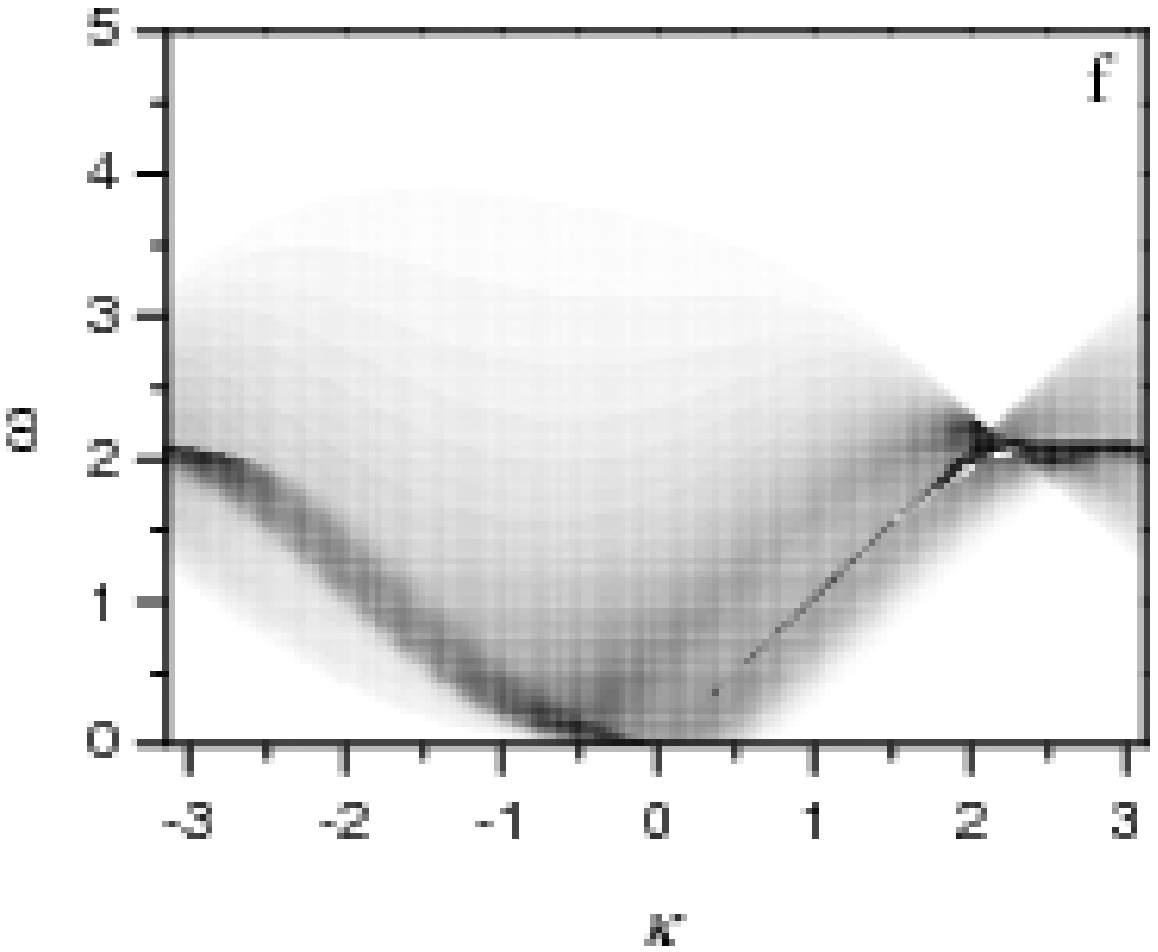, height = 0.25\linewidth}
\epsfig{file = 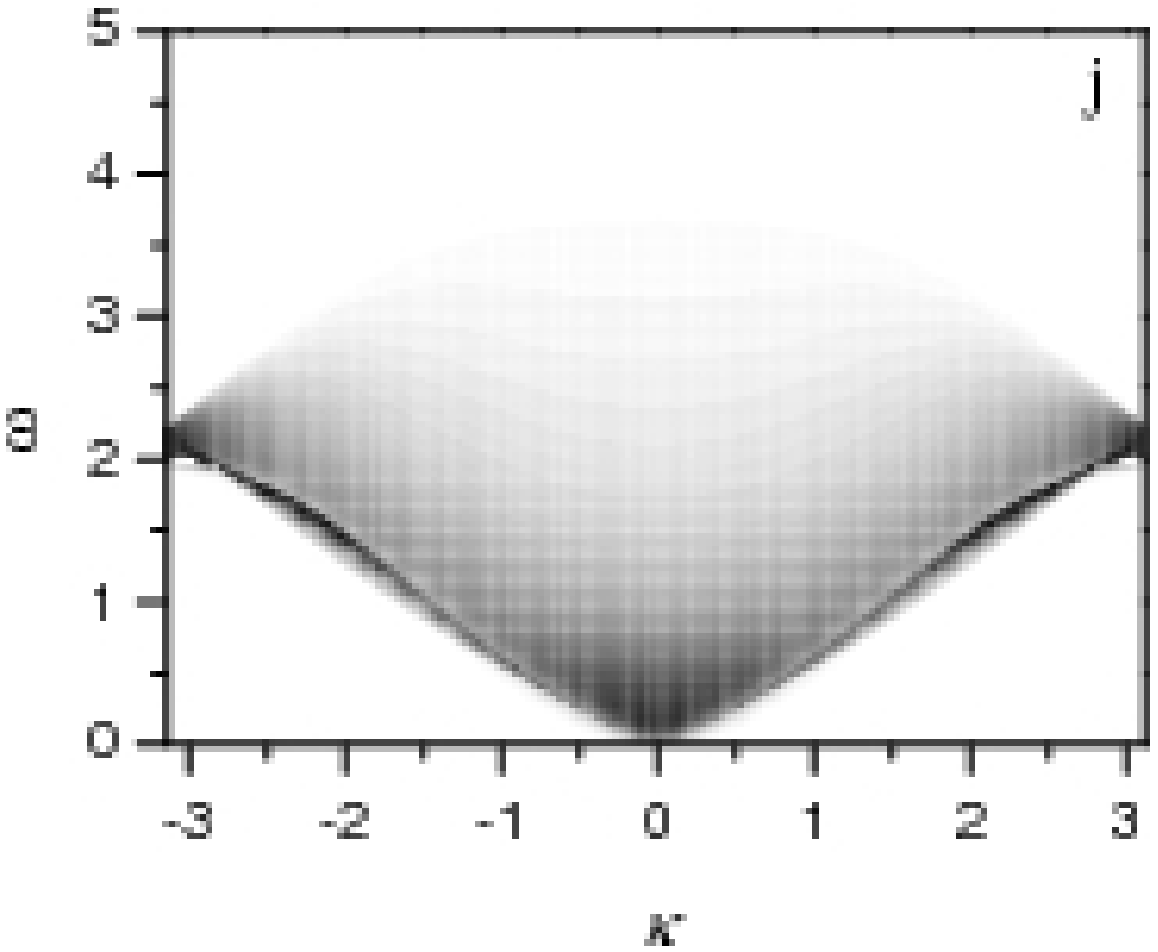, height = 0.25\linewidth}\\
\epsfig{file = 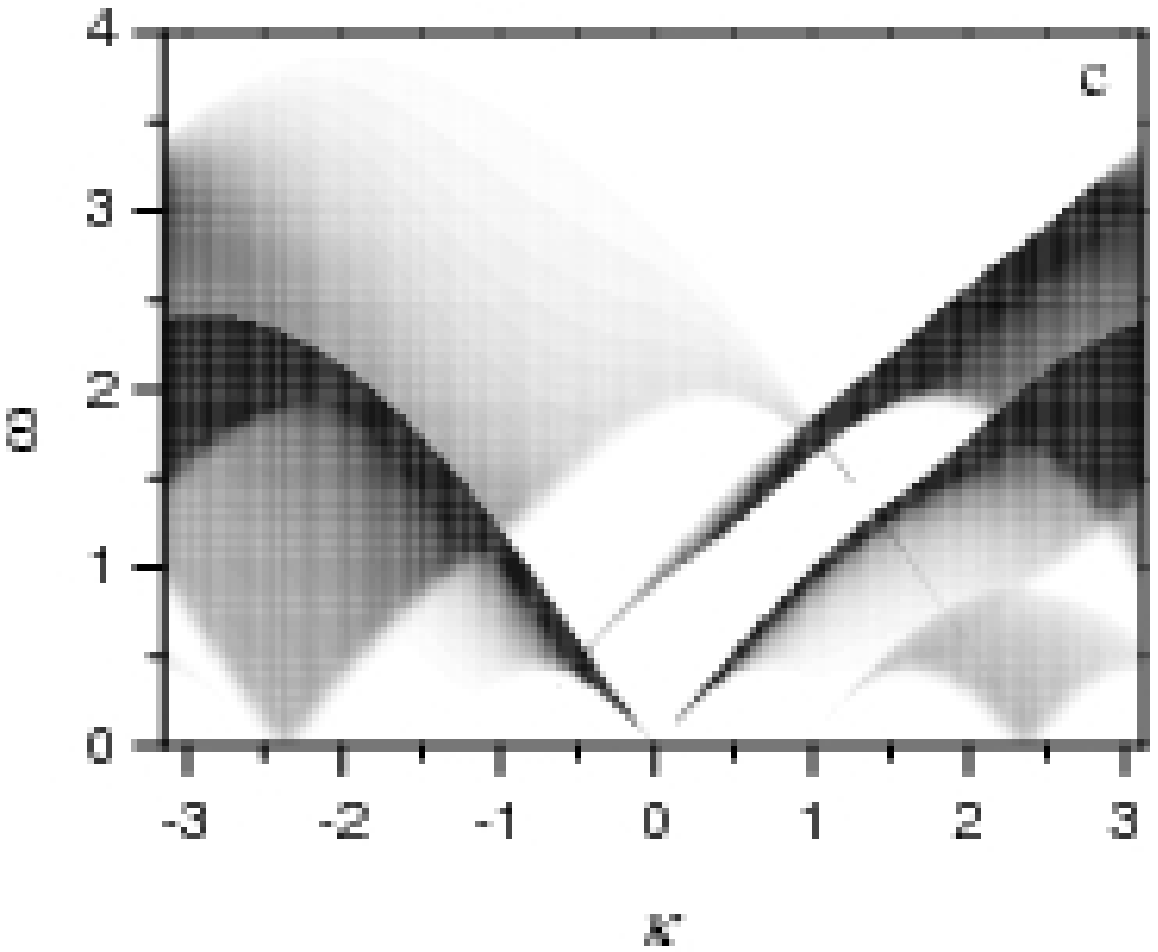, height = 0.25\linewidth}
\epsfig{file = 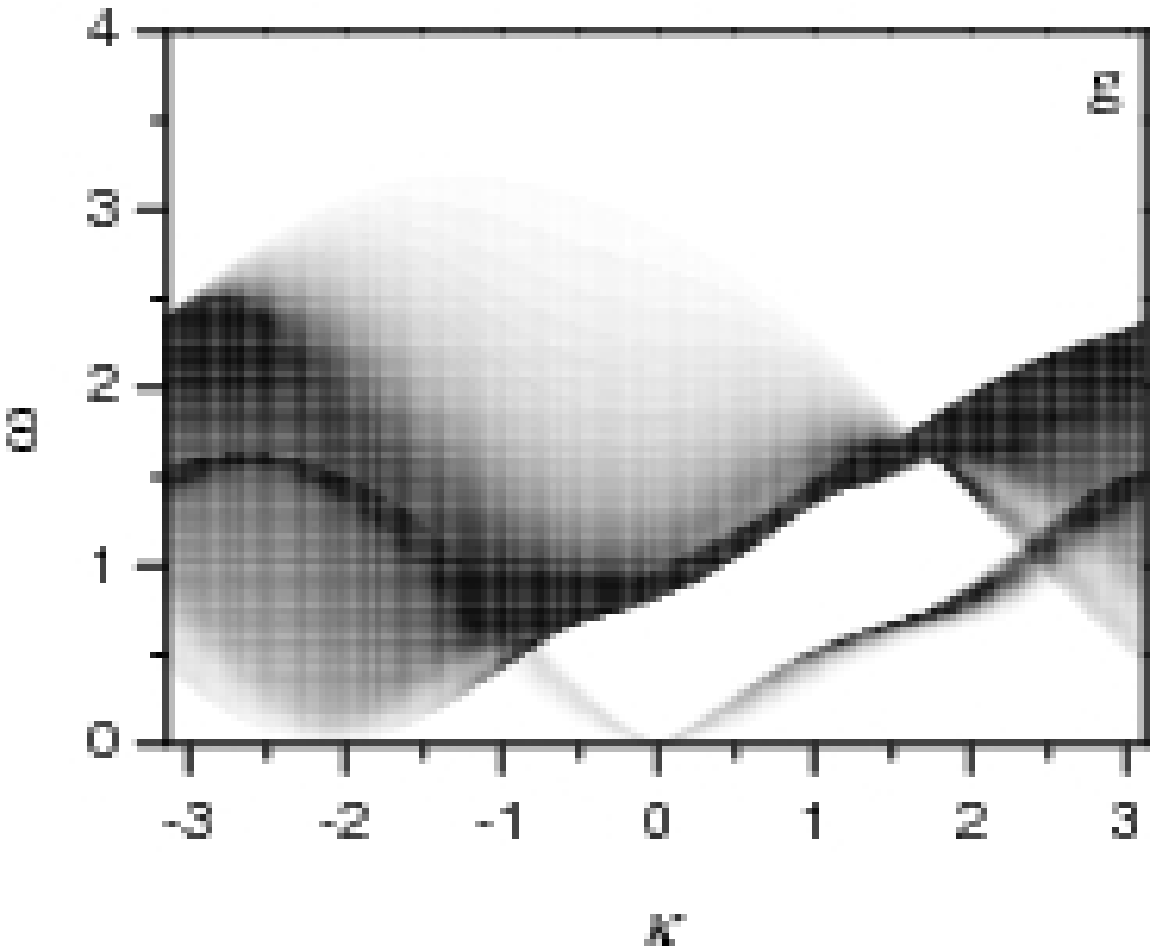, height = 0.25\linewidth}
\epsfig{file = 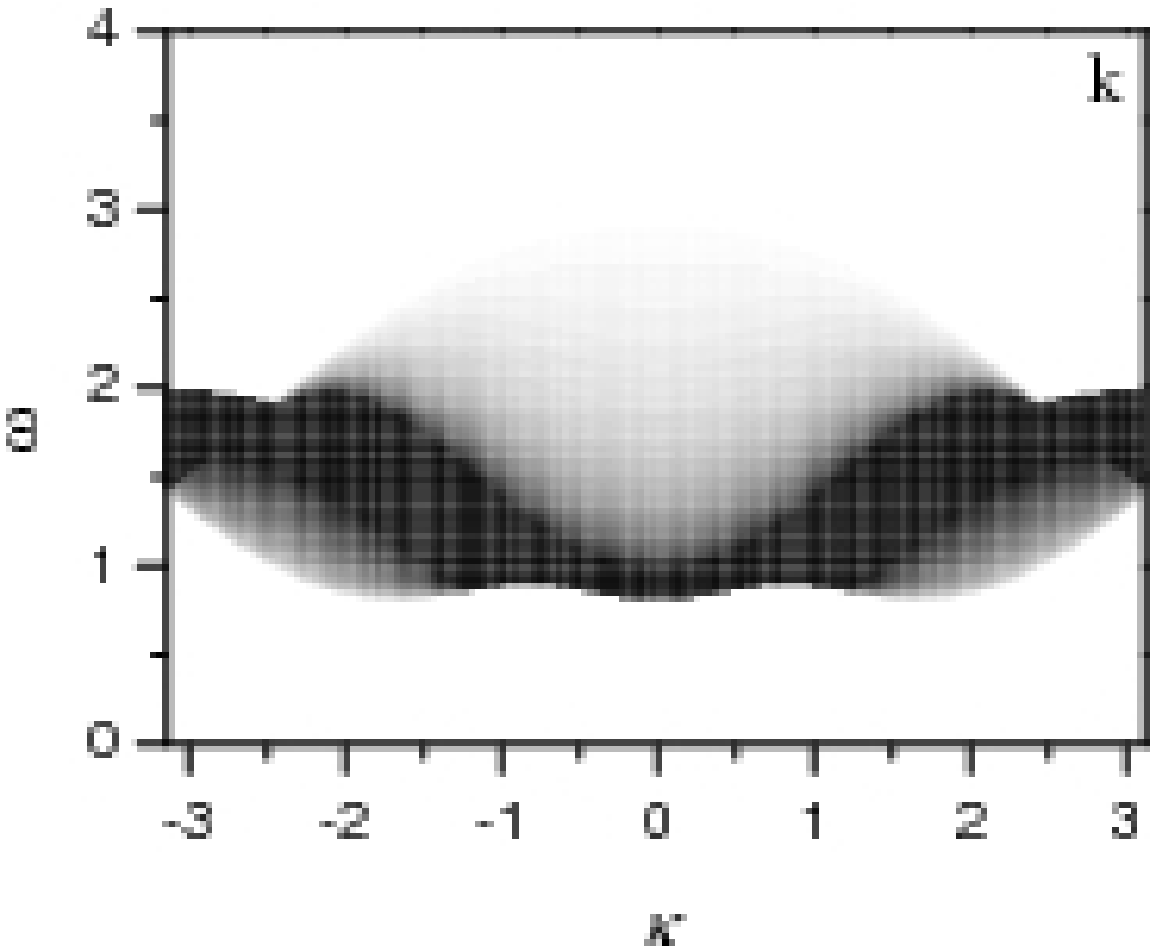, height = 0.25\linewidth}\\
\epsfig{file = 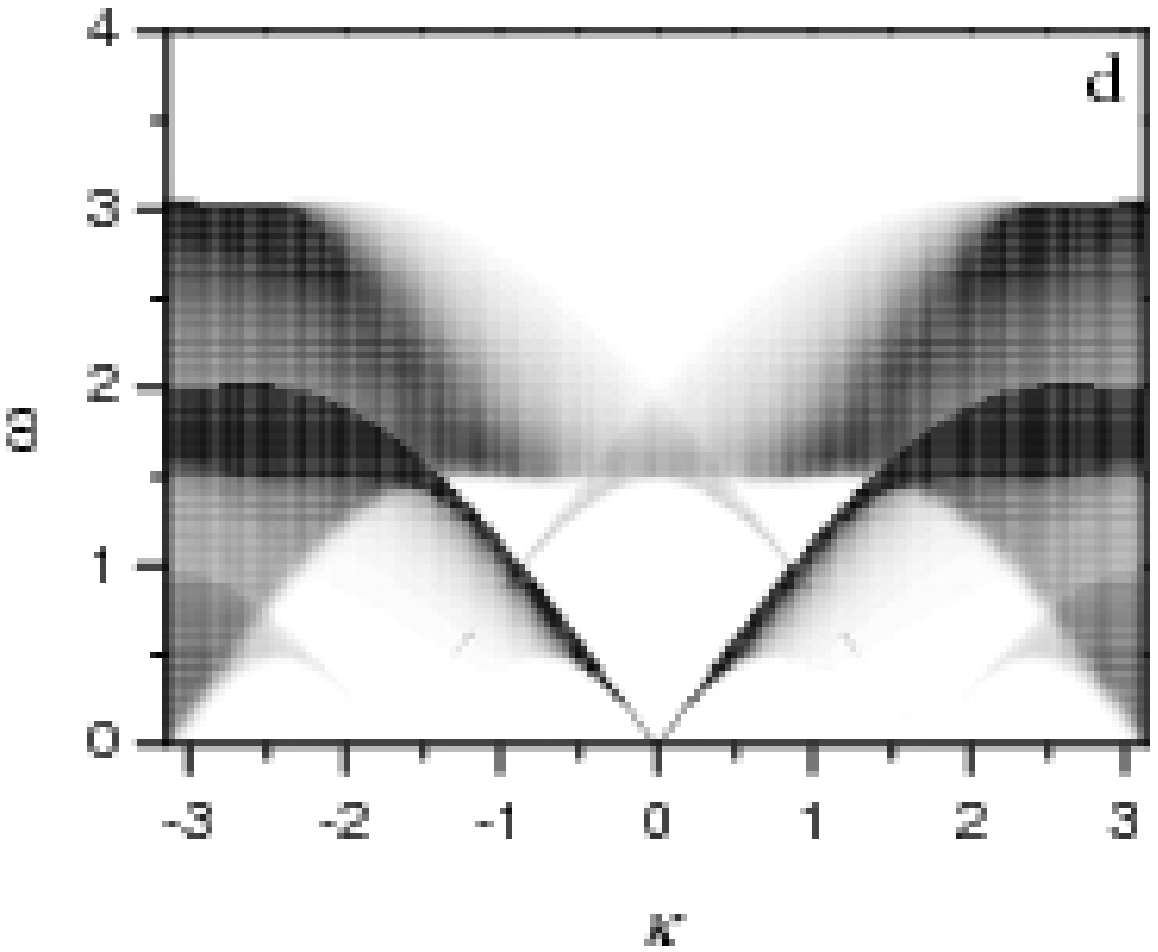, height = 0.25\linewidth}
\epsfig{file = 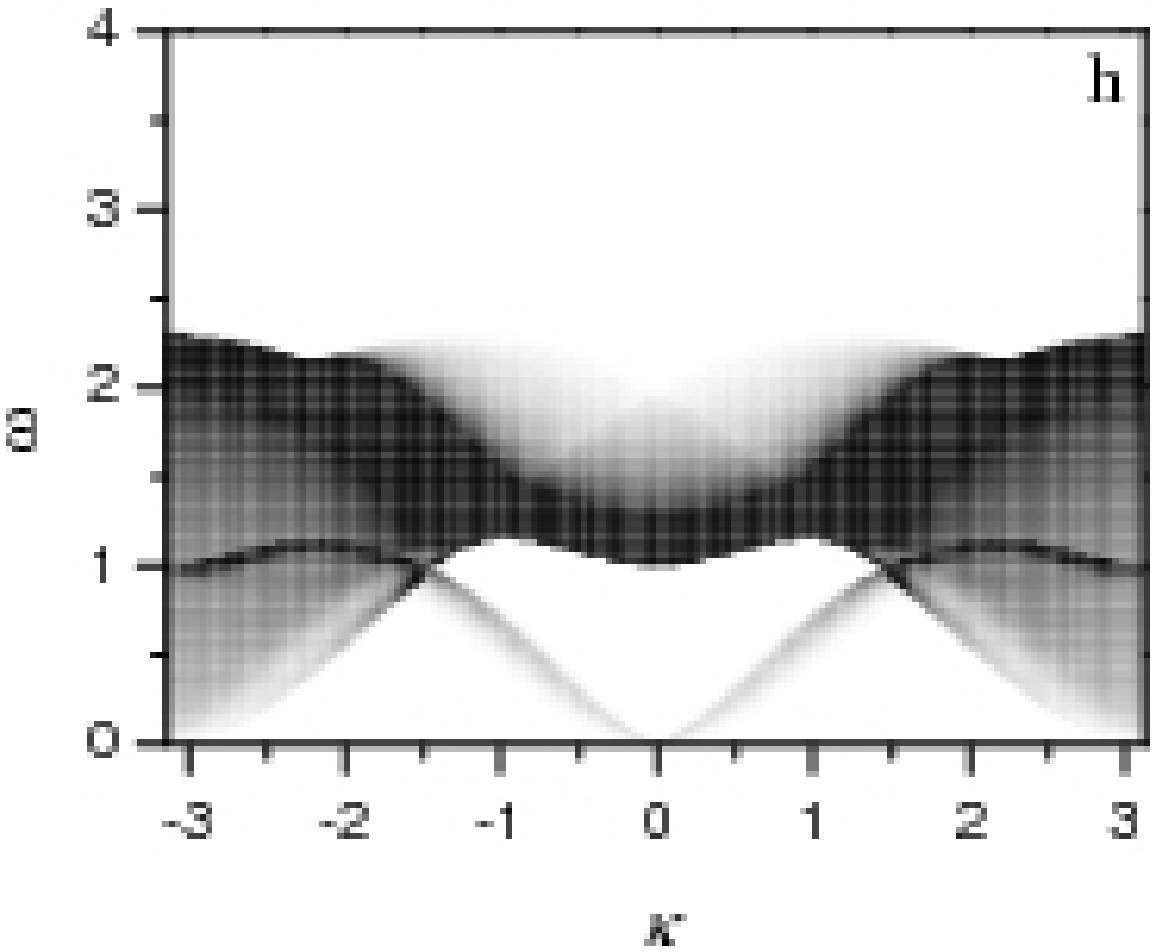, height = 0.25\linewidth}
\epsfig{file = 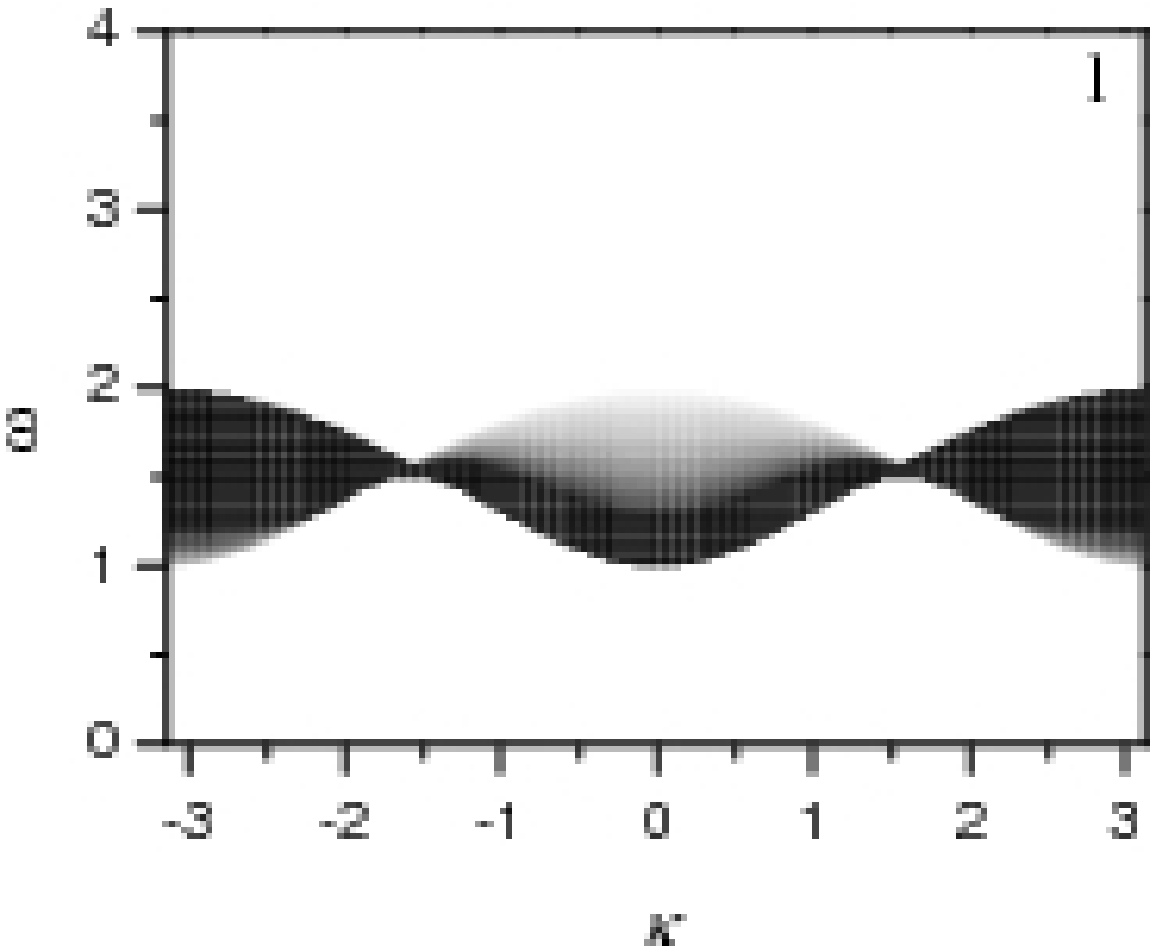, height = 0.25\linewidth}
\caption{}
\end{figure}

\newpage

\begin{figure}
\epsfig{file = 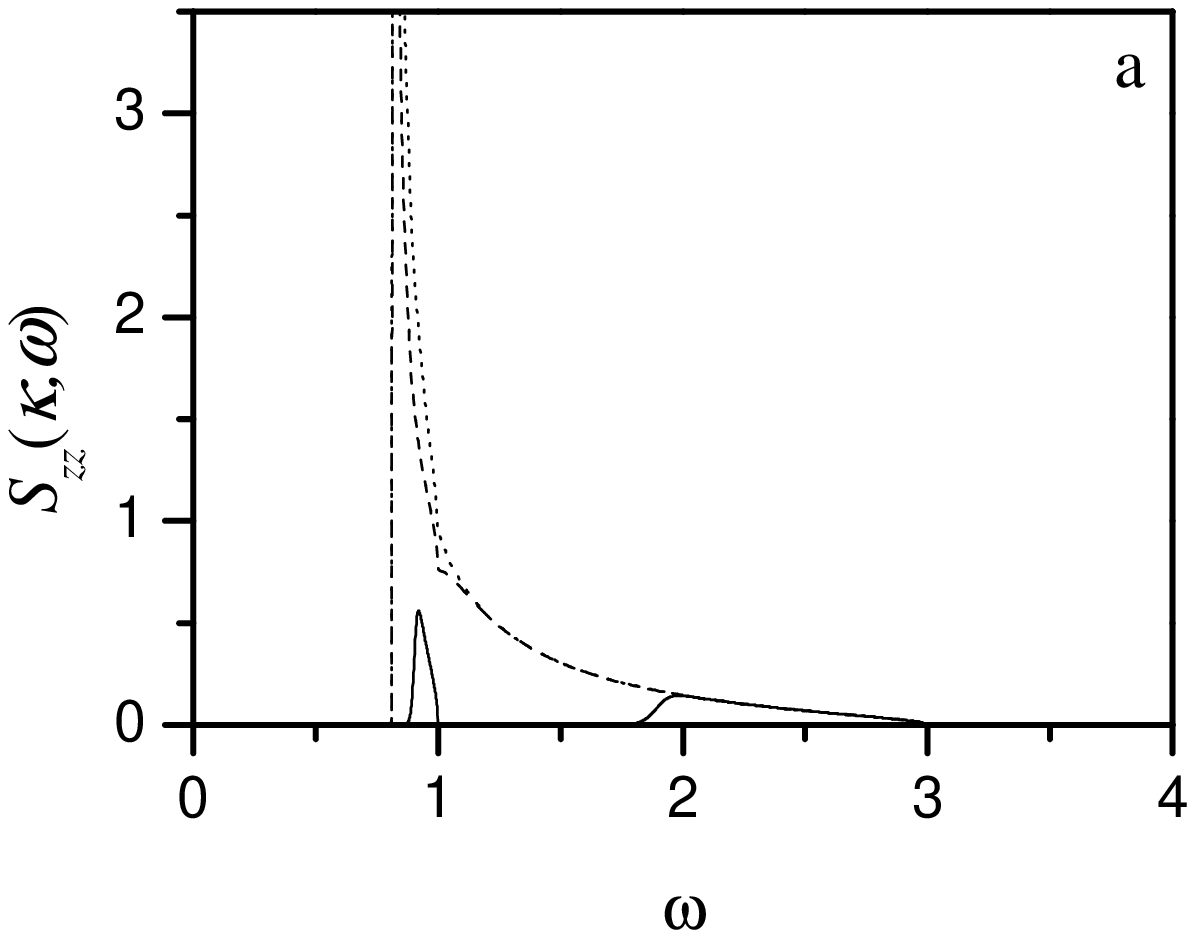, height = 0.25\linewidth}
\epsfig{file = 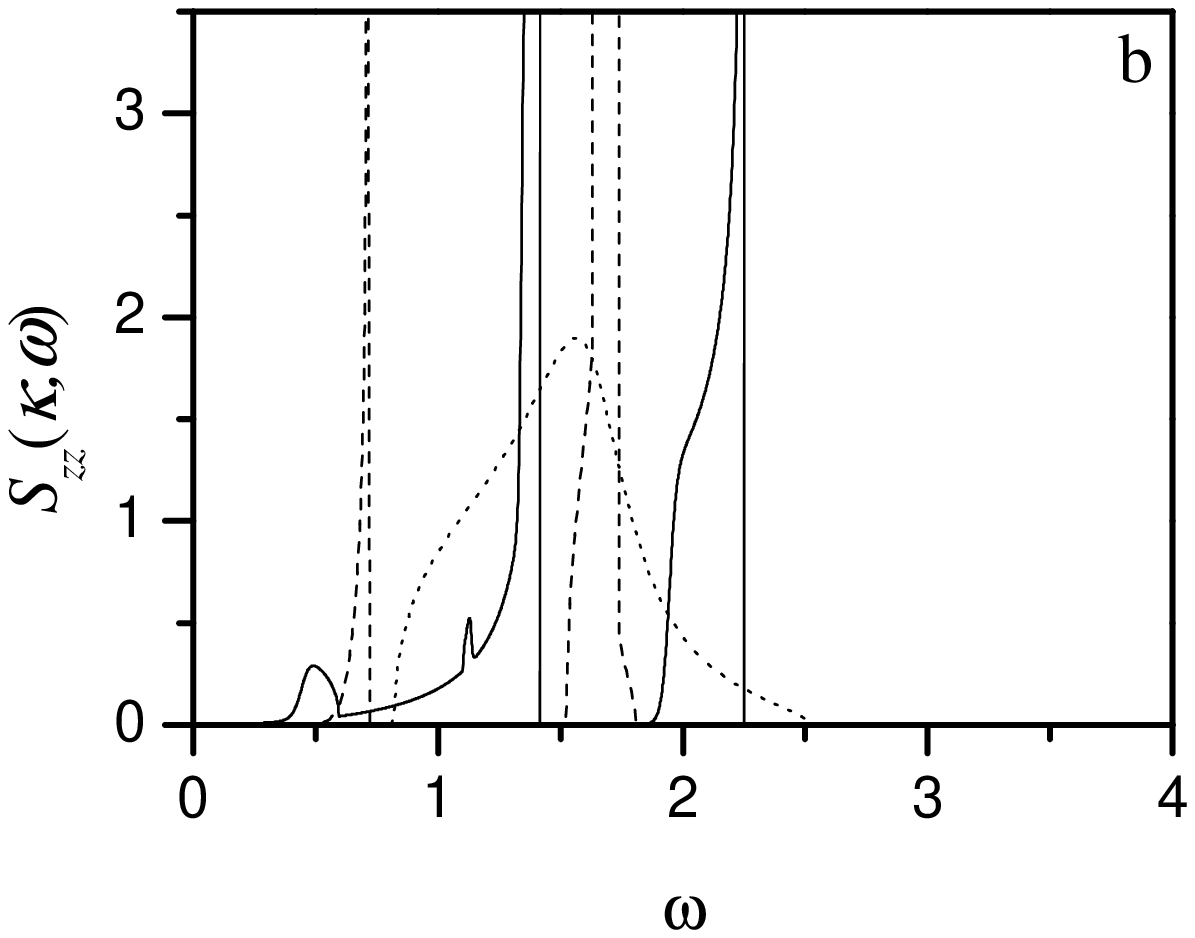, height = 0.25\linewidth}
\epsfig{file = 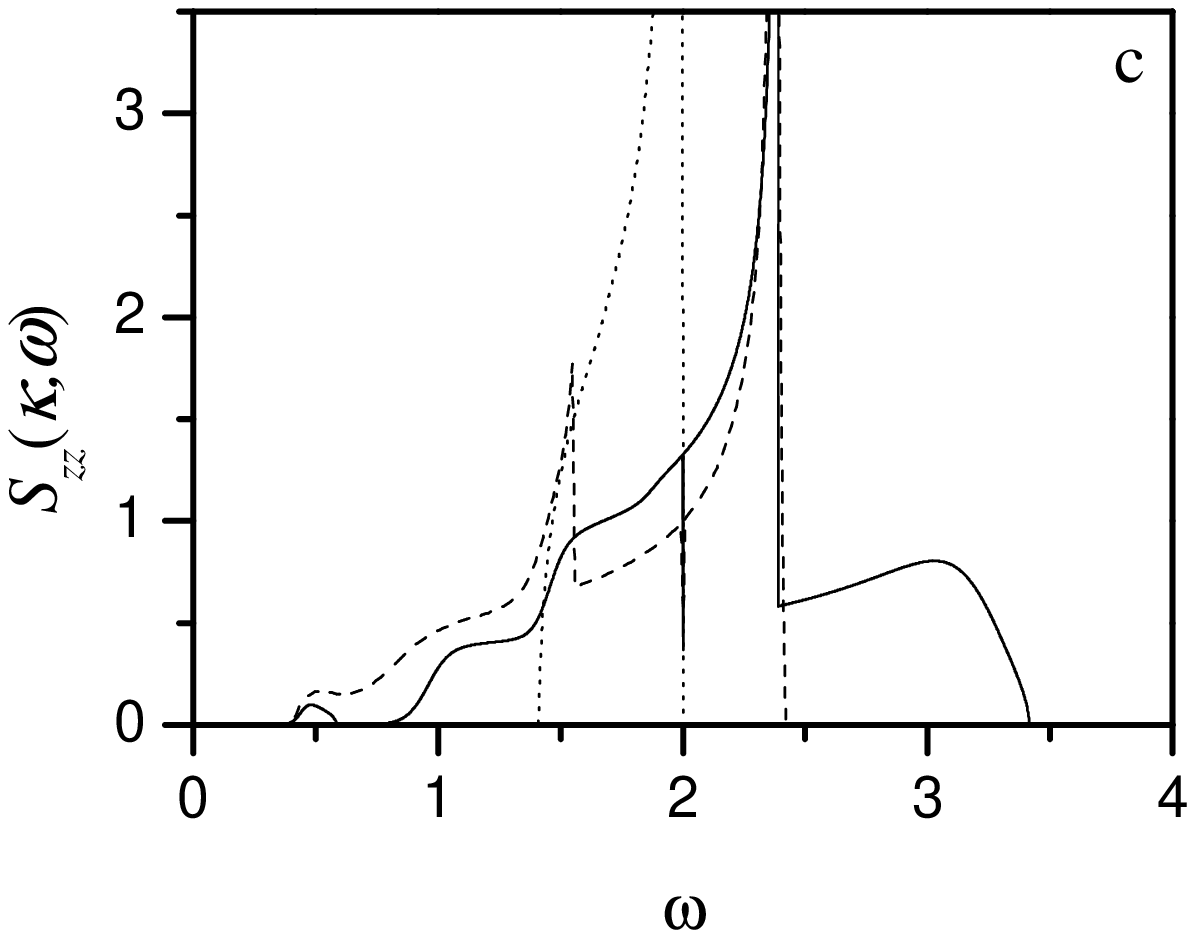, height = 0.25\linewidth}
\caption{}
\end{figure}

\clearpage

\begin{figure}
\epsfig{file = 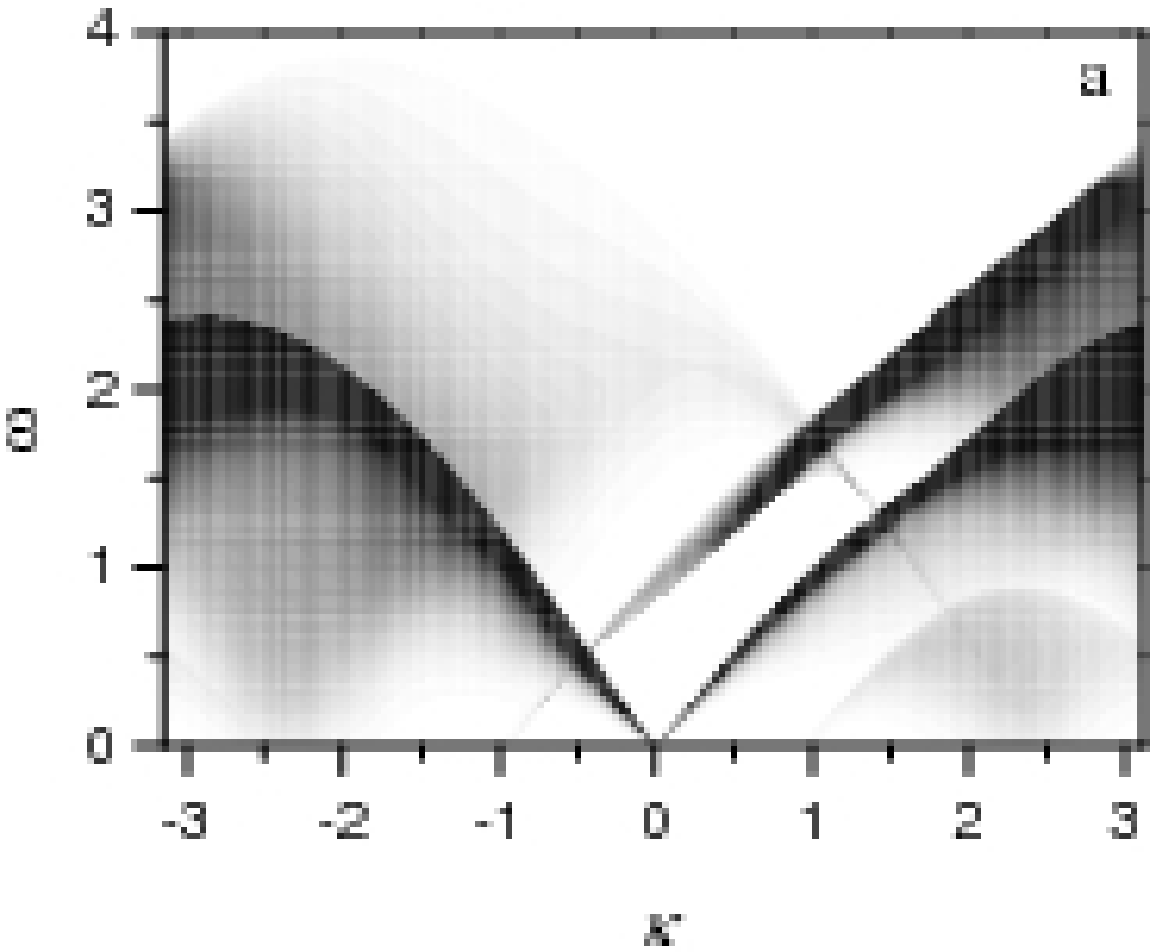, height = 0.25\linewidth}
\epsfig{file = 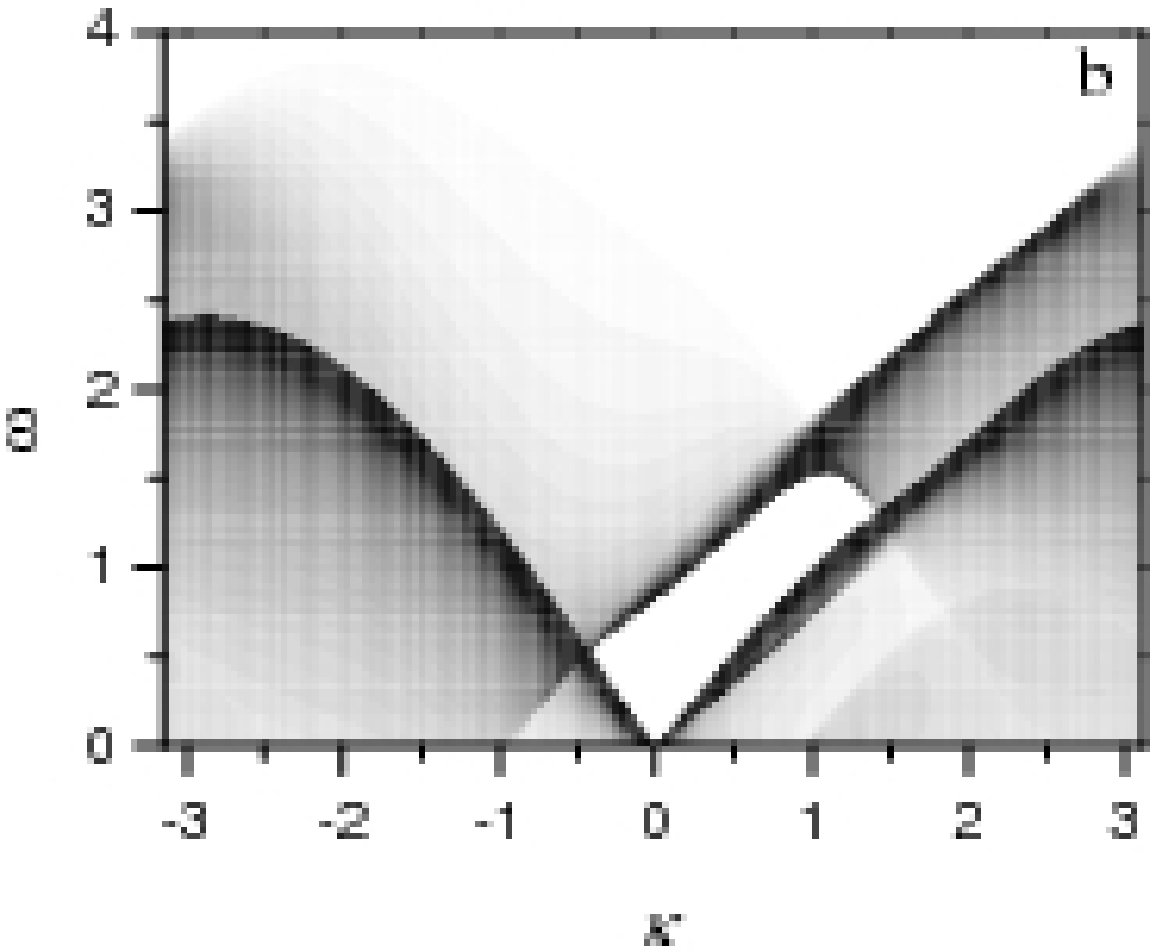, height = 0.25\linewidth}
\epsfig{file = 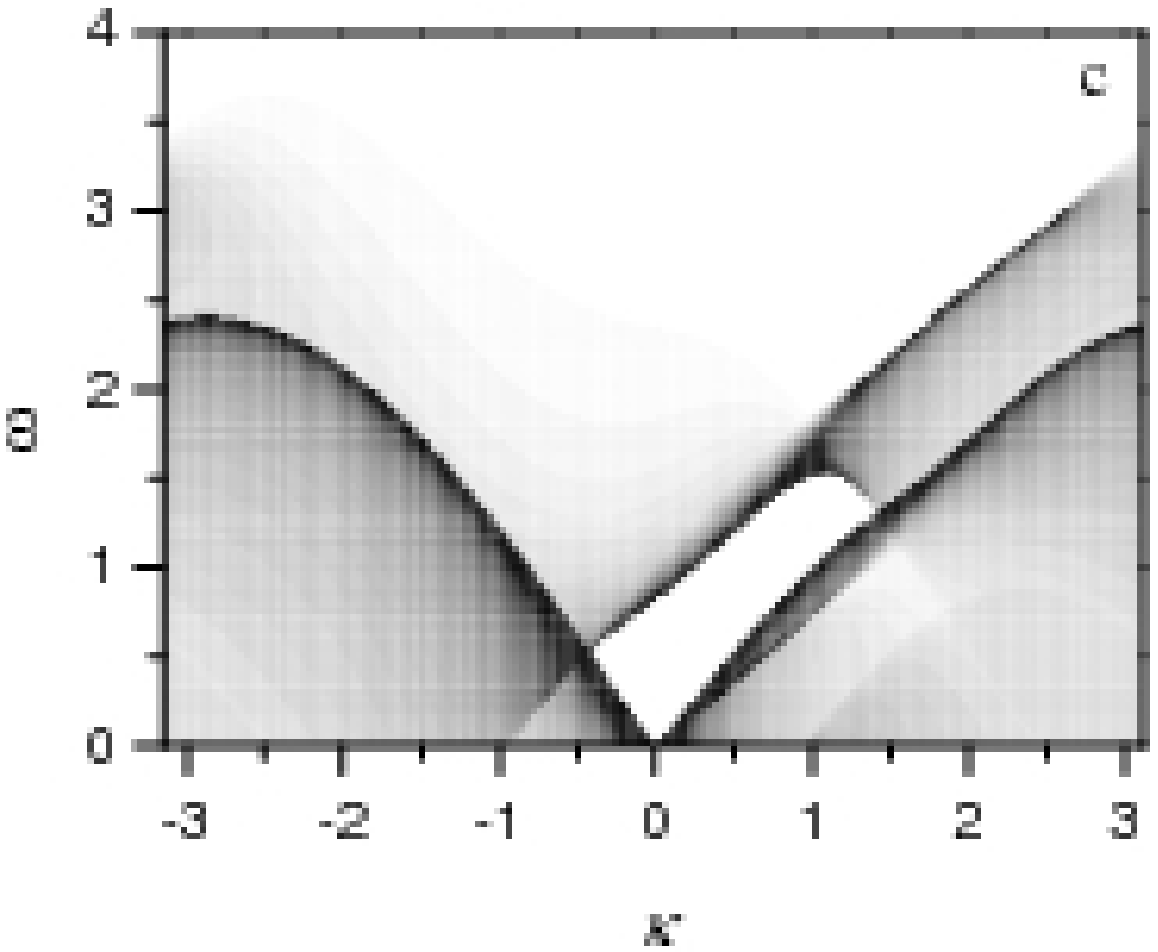, height = 0.25\linewidth}
\caption{}
\end{figure}

\newpage

\begin{figure}
\epsfig{file = 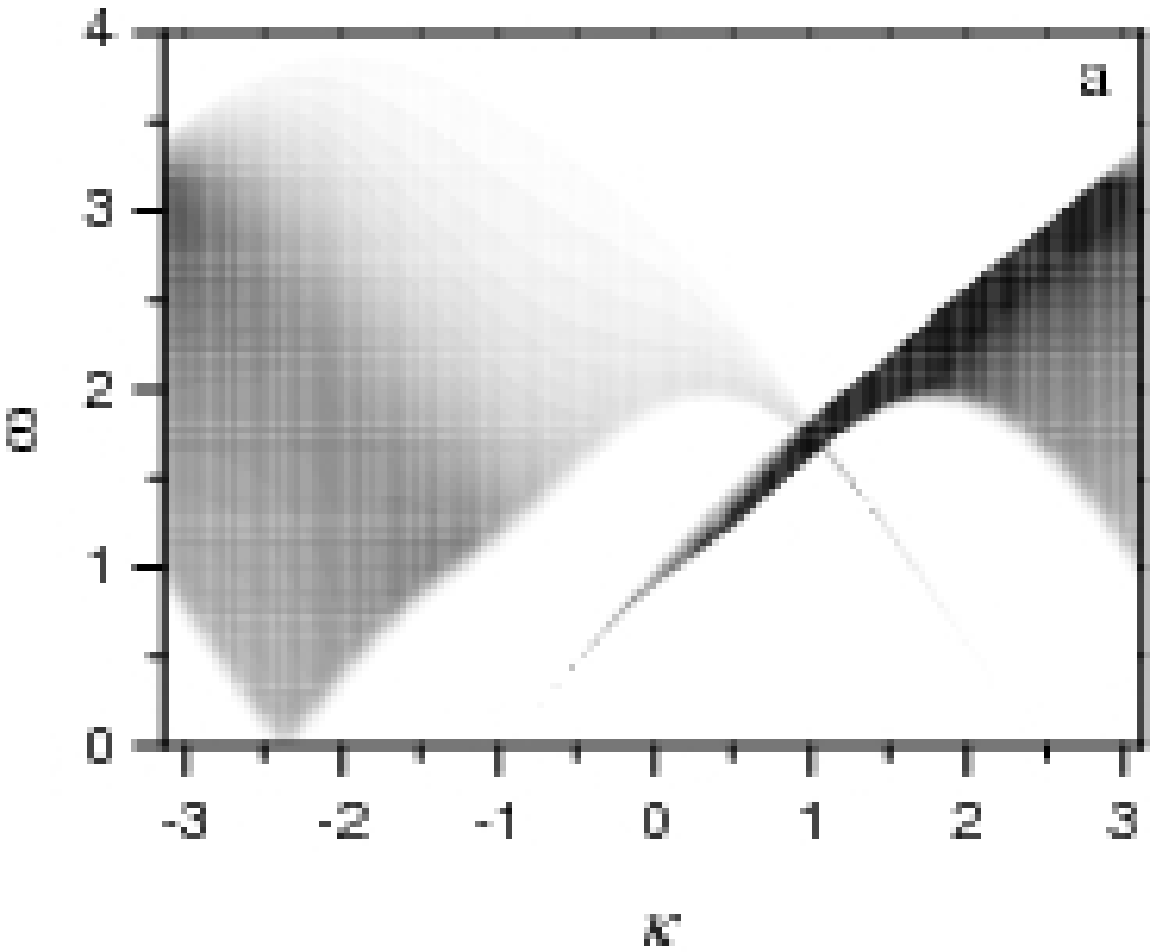, height = 0.25\linewidth}
\epsfig{file = 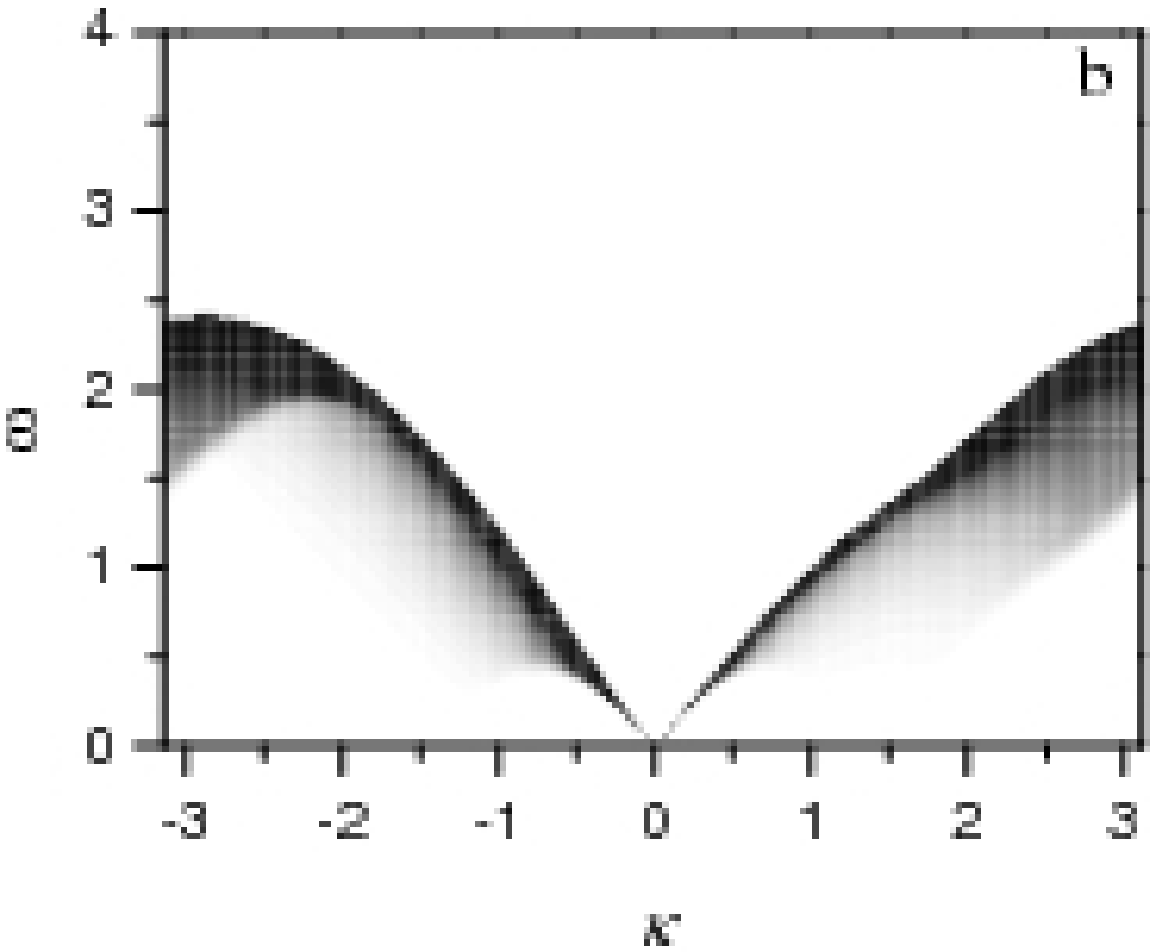, height = 0.25\linewidth}
\epsfig{file = 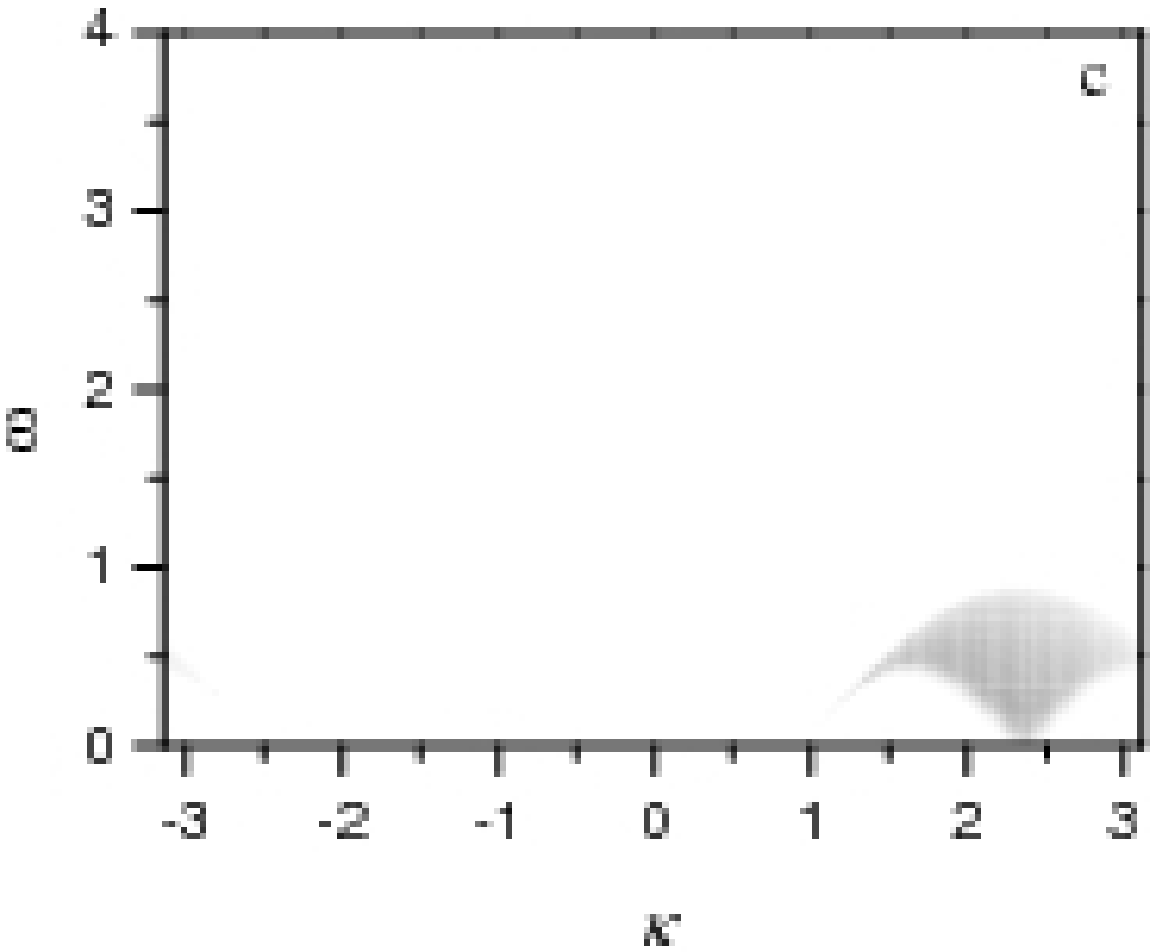, height = 0.25\linewidth}
\caption{}
\end{figure}

\newpage

\begin{figure}
\epsfig{file = 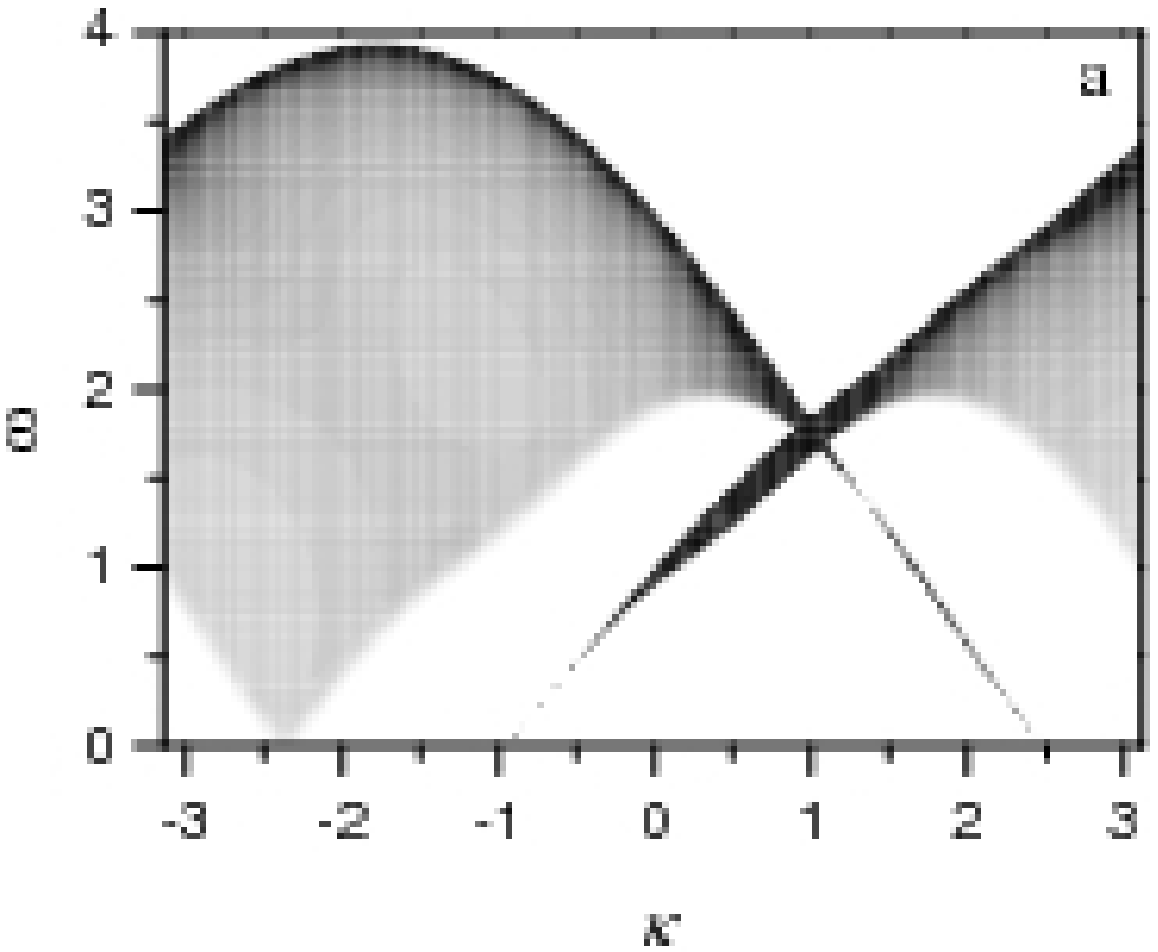, height = 0.25\linewidth}
\epsfig{file = 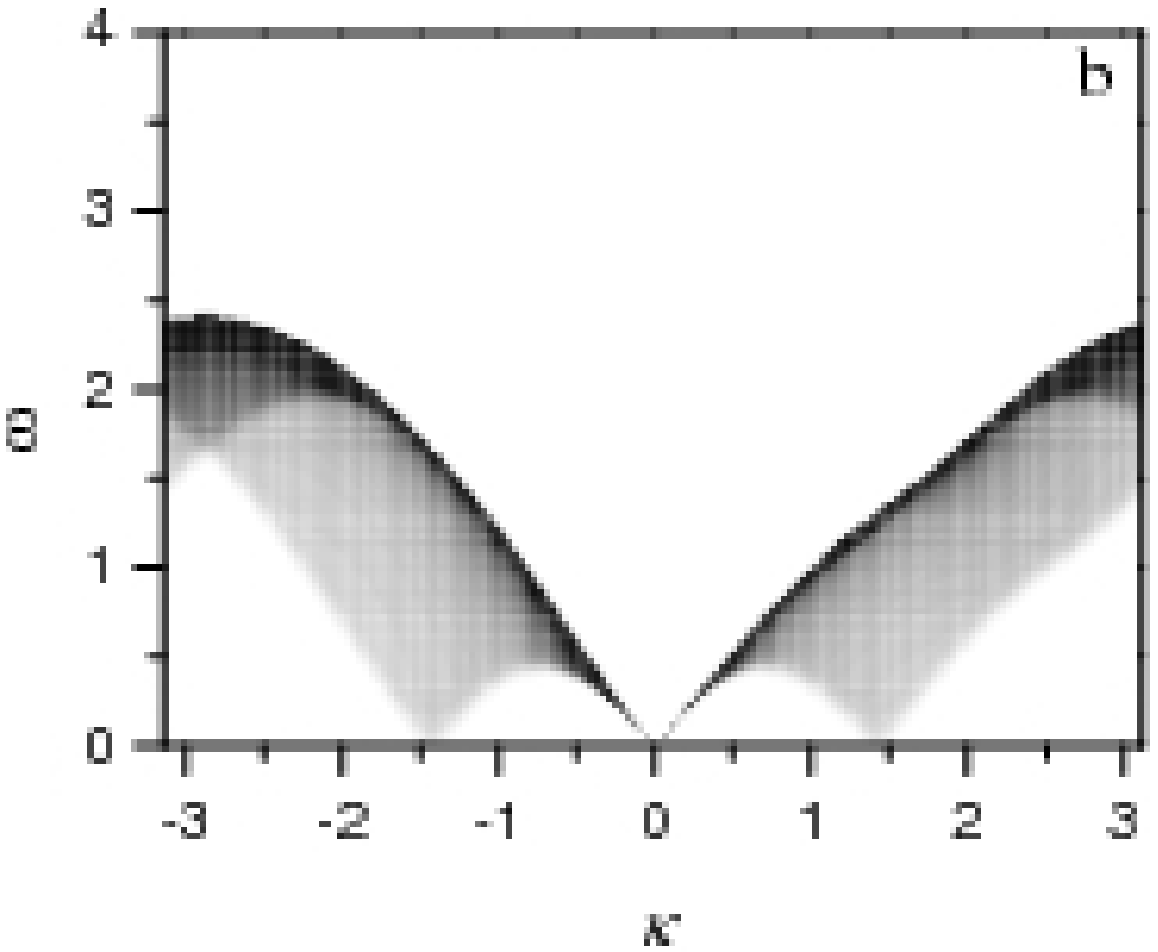, height = 0.25\linewidth}
\epsfig{file = 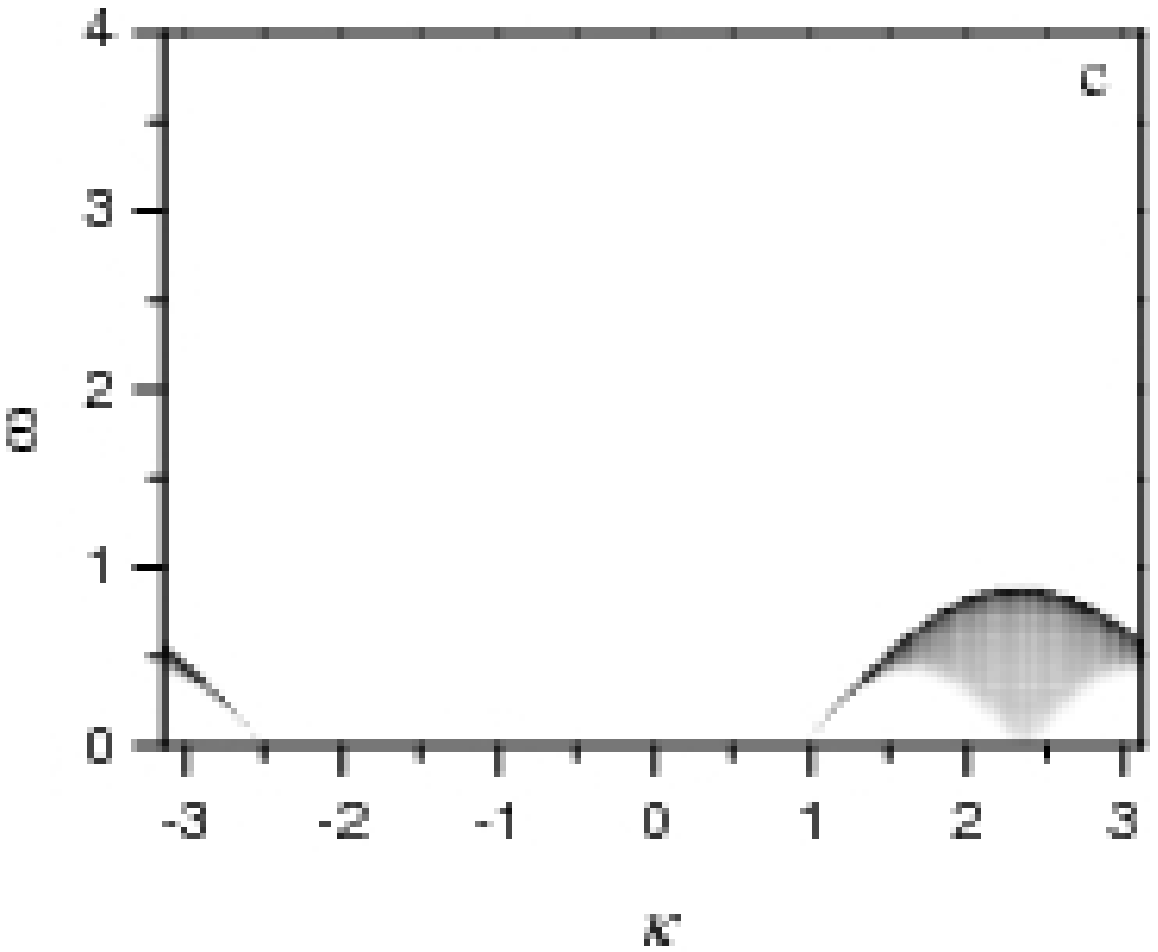, height = 0.25\linewidth}
\caption{}
\end{figure}

\newpage

\begin{figure}
\epsfig{file = 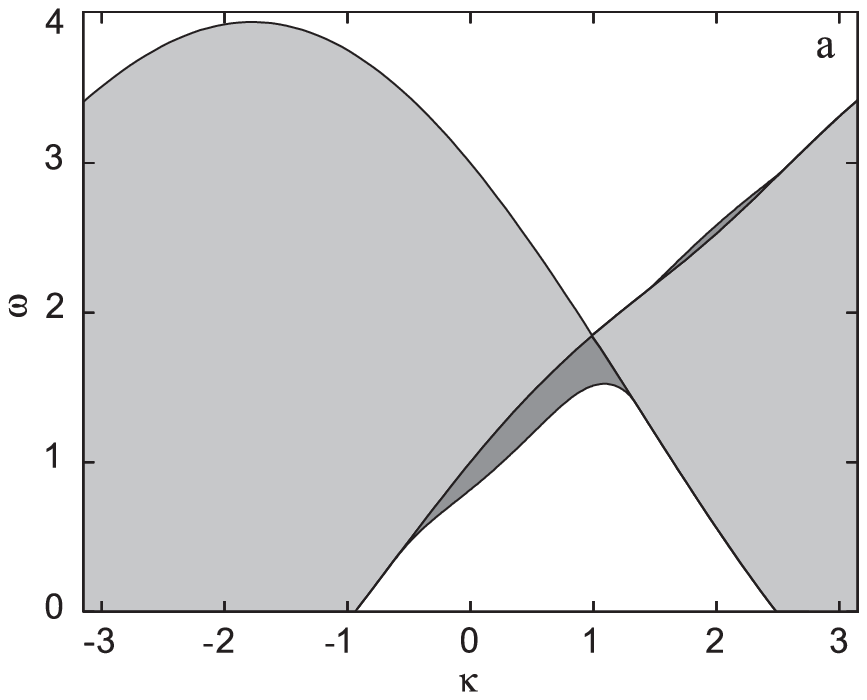, height = 0.25\linewidth}
\epsfig{file = 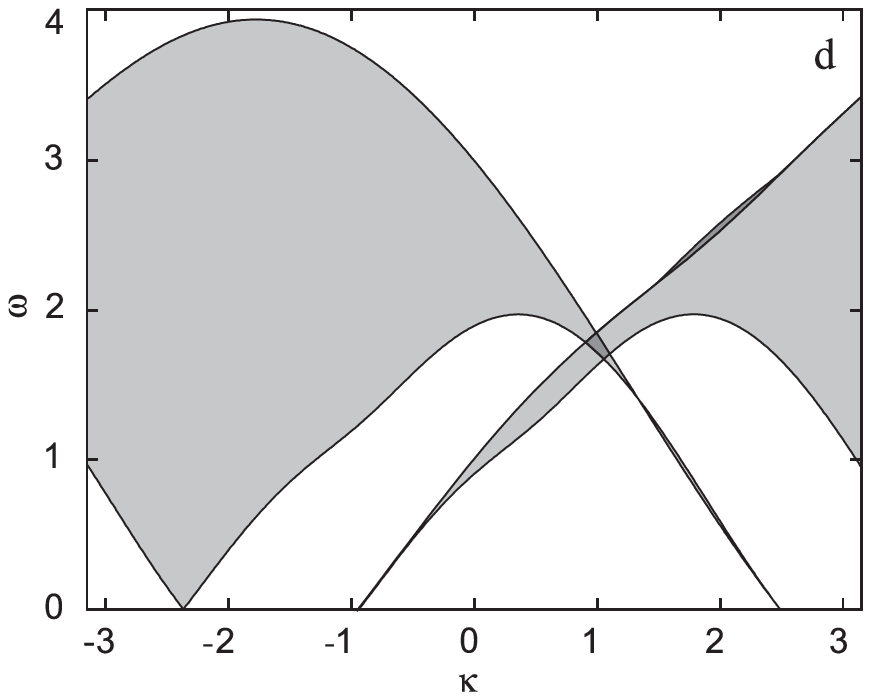, height = 0.25\linewidth}\\
\epsfig{file = 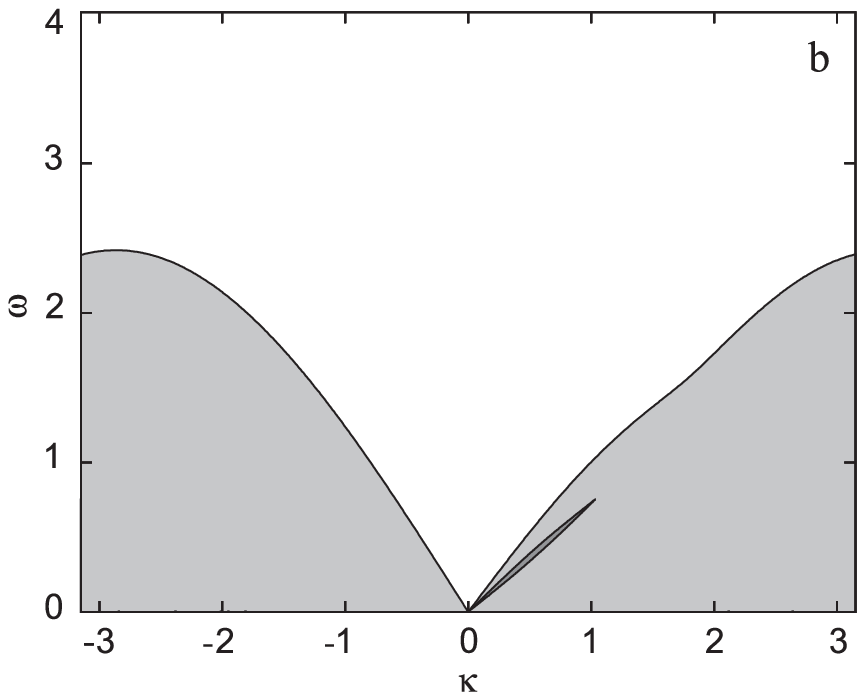, height = 0.25\linewidth}
\epsfig{file = 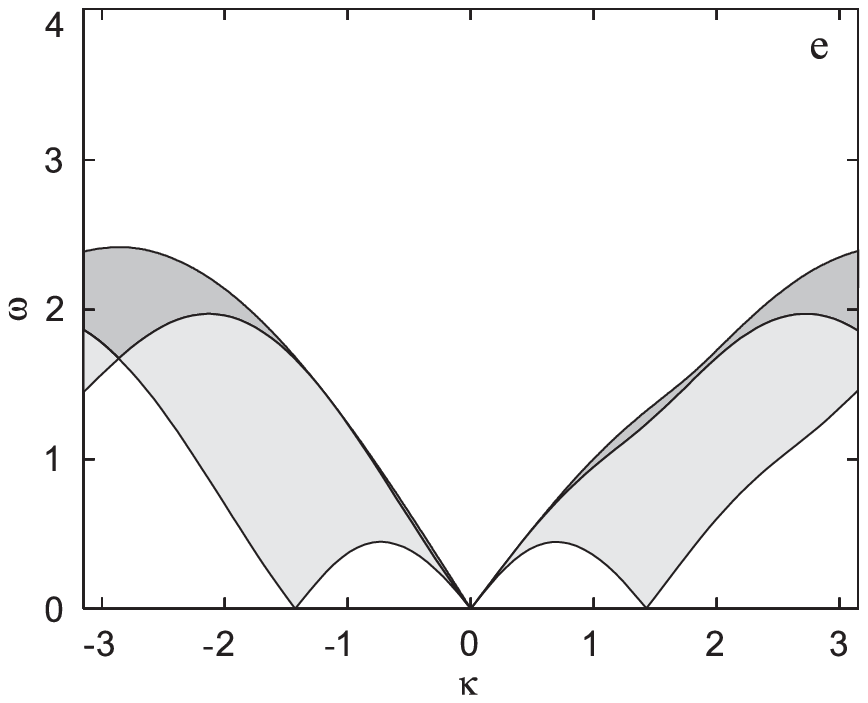, height = 0.25\linewidth}\\
\epsfig{file = 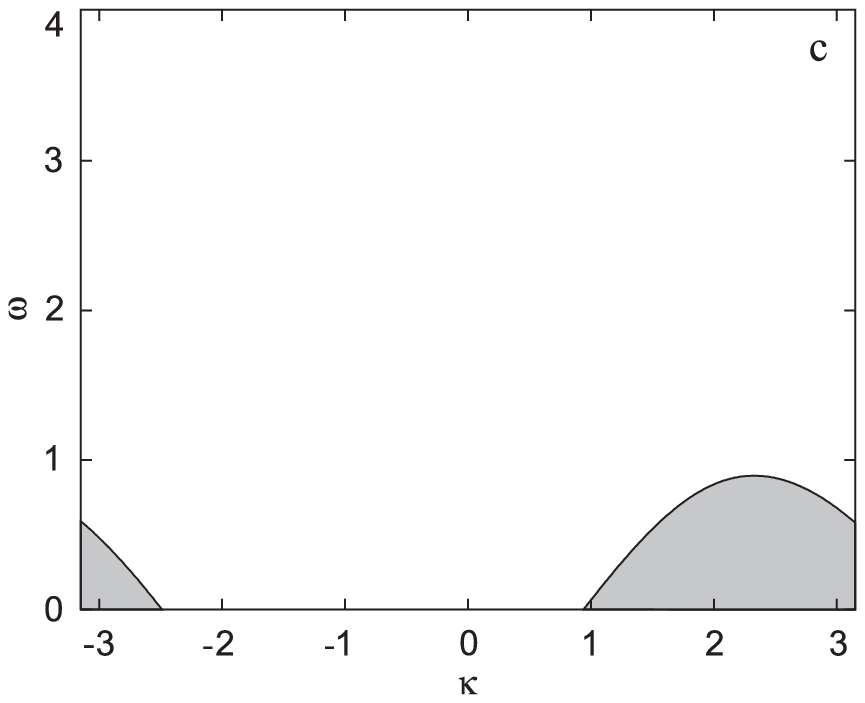, height = 0.25\linewidth}
\epsfig{file = 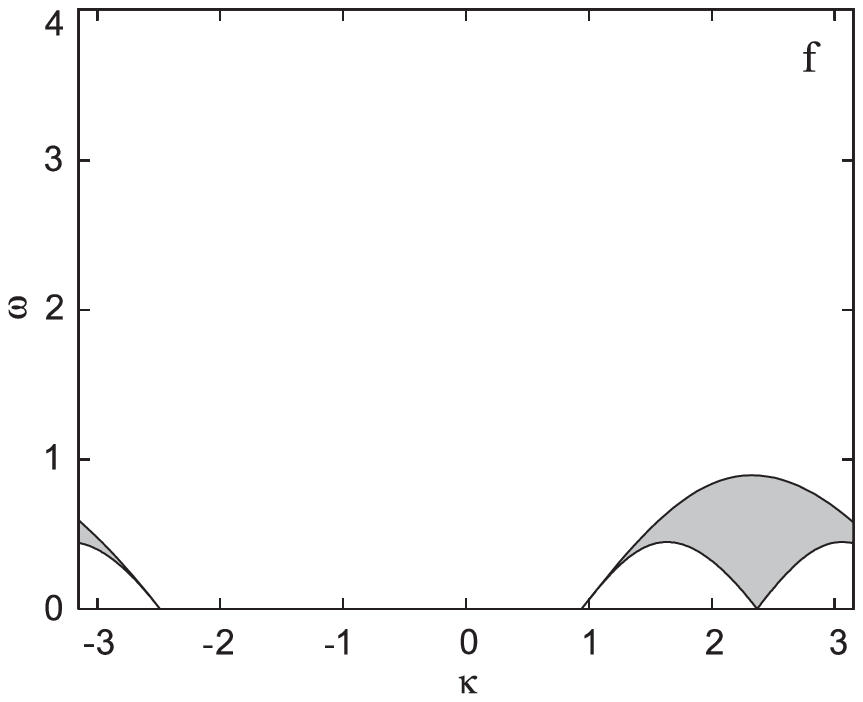, height = 0.25\linewidth}
\caption{}
\end{figure}

\newpage

\begin{figure}
\epsfig{file = 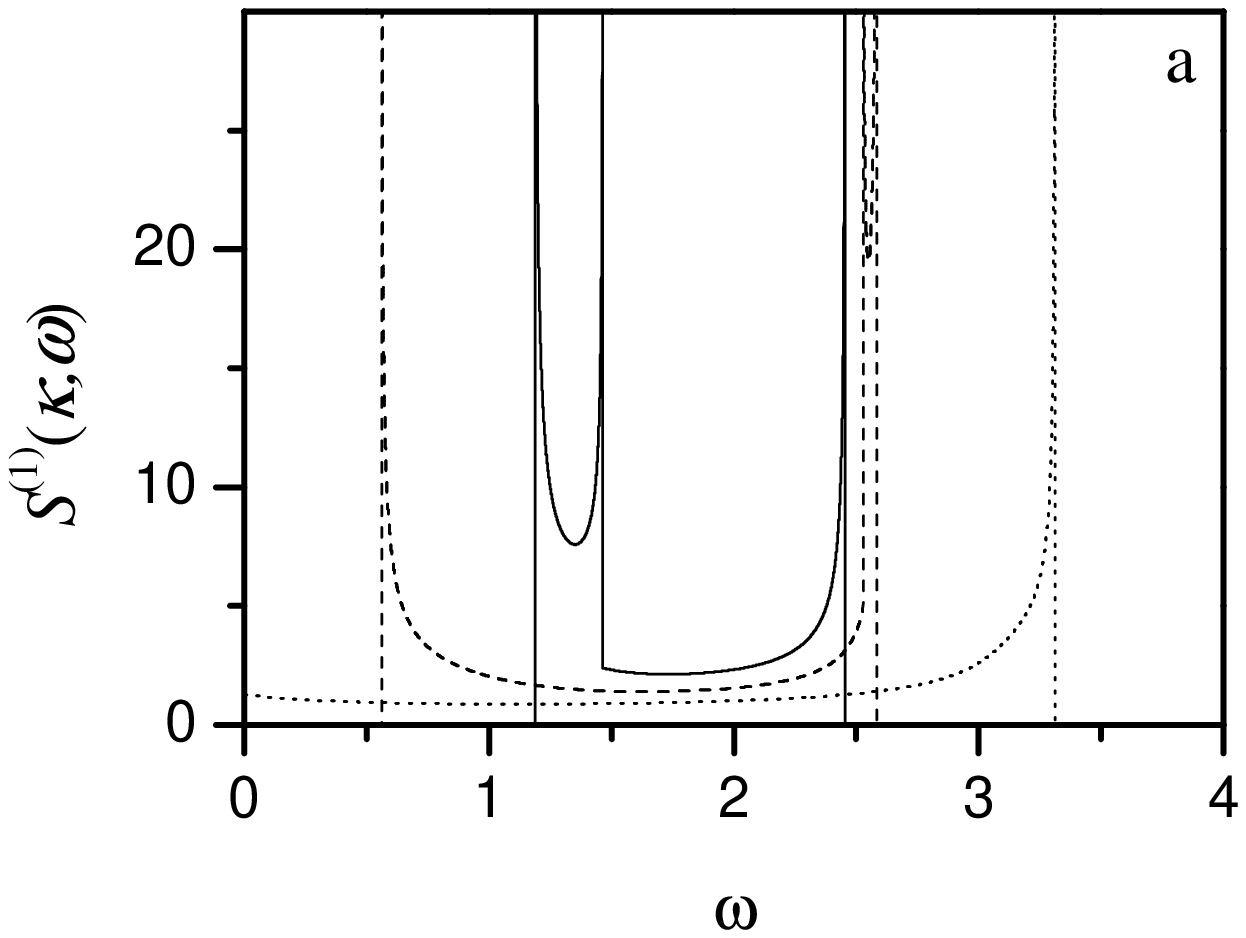, height = 0.3\linewidth}
\epsfig{file = 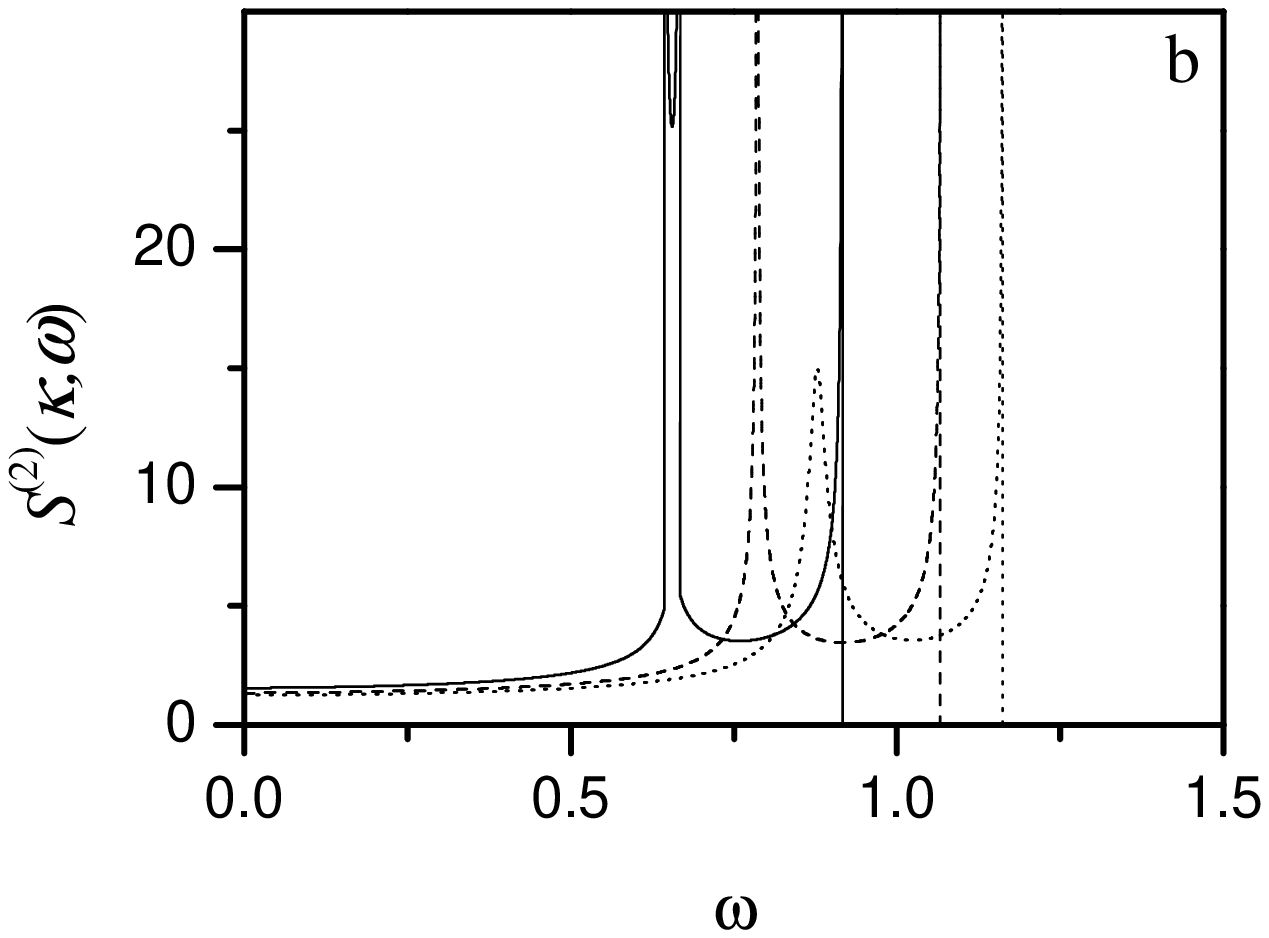, height = 0.3\linewidth}
\caption{}
\end{figure}

\newpage

\begin{figure}
\epsfig{file = 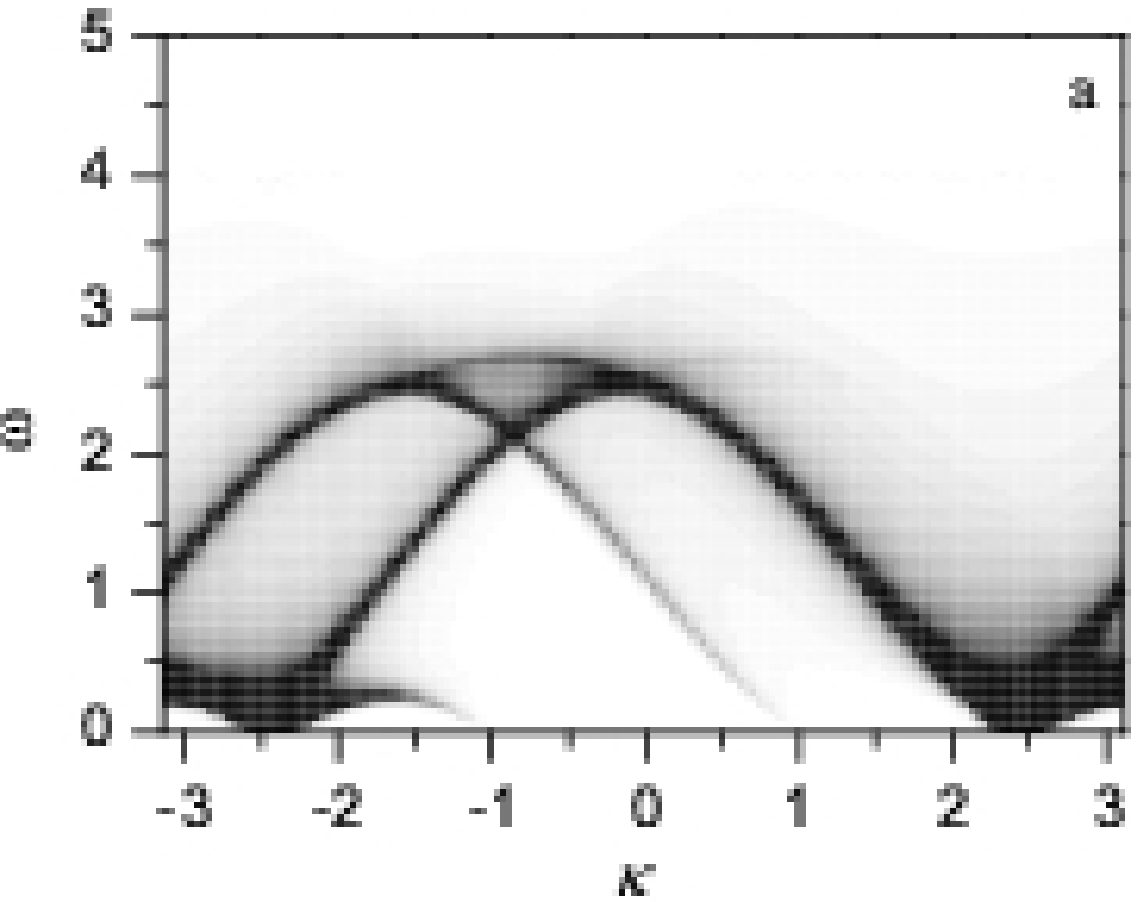, height = 0.25\linewidth}
\epsfig{file = 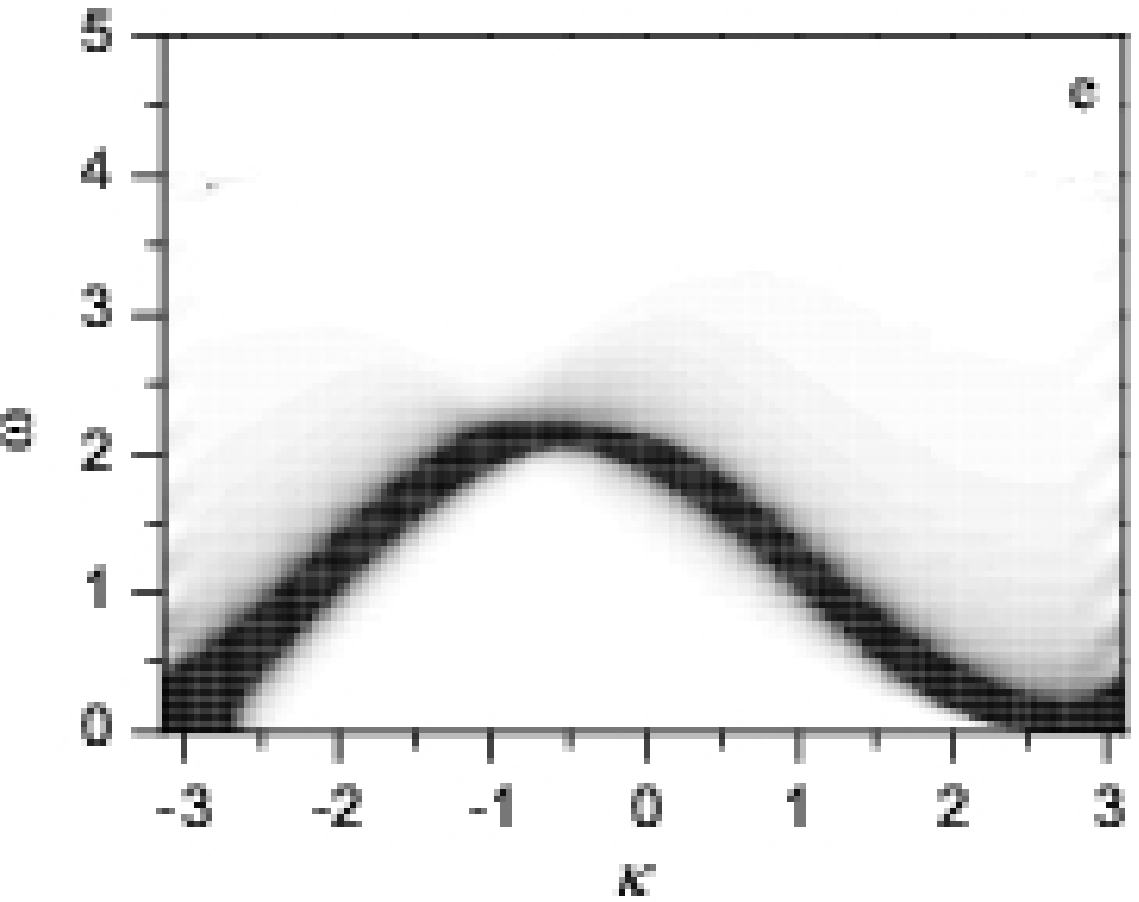, height = 0.25\linewidth}
\epsfig{file = 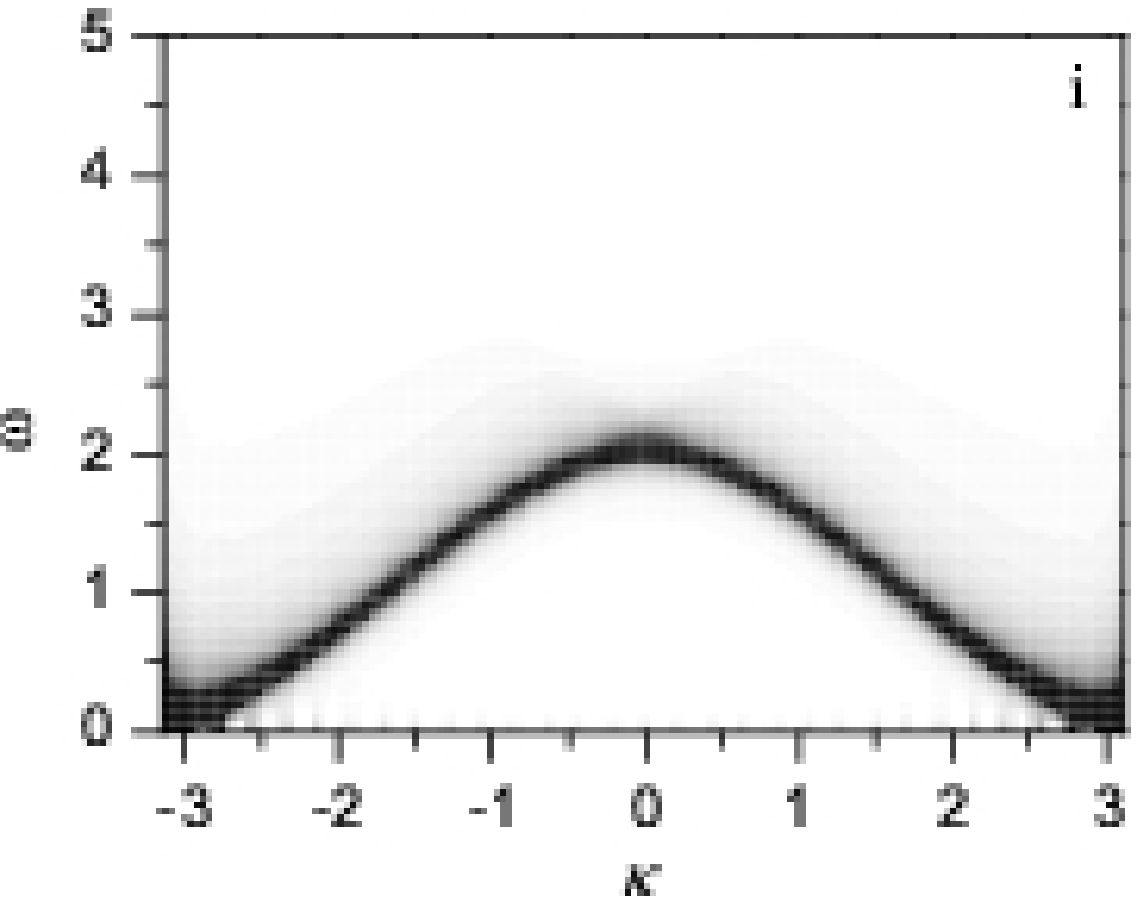, height = 0.25\linewidth}\\
\epsfig{file = 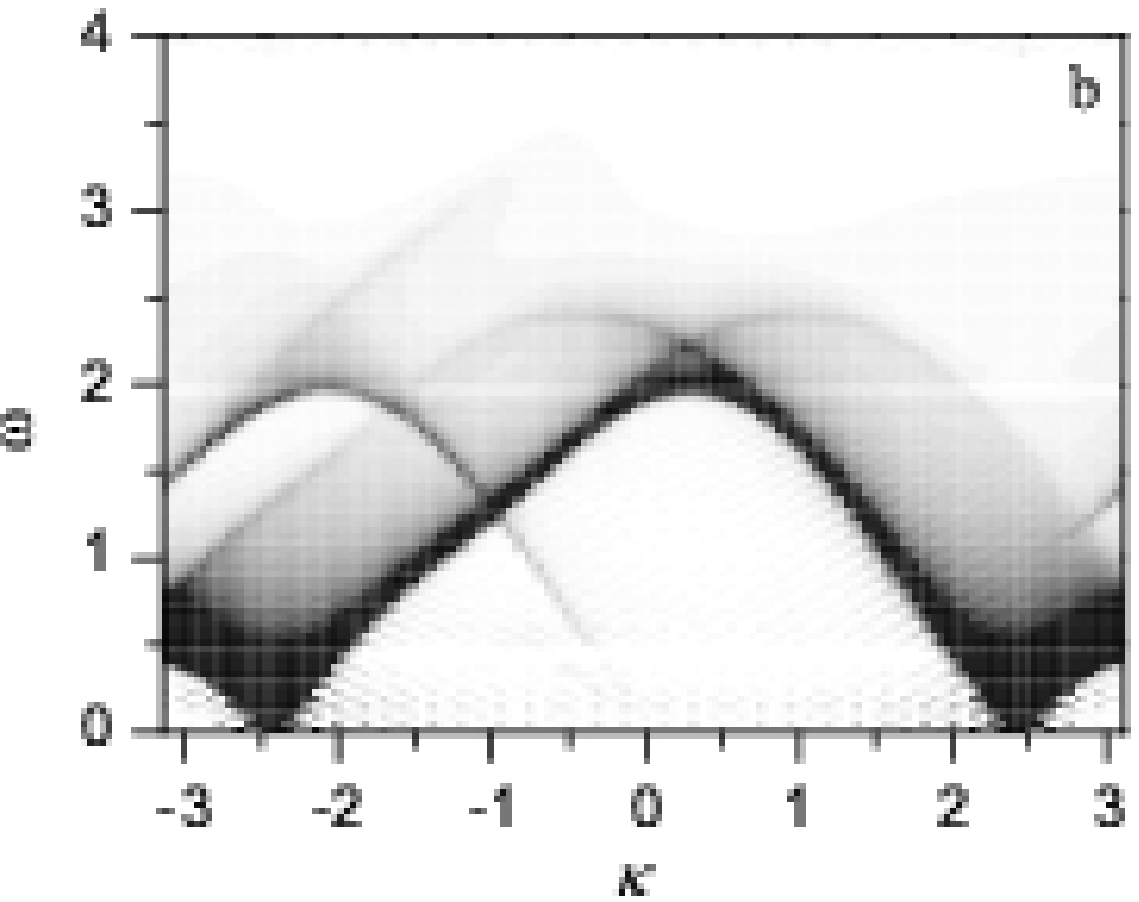, height = 0.25\linewidth}
\epsfig{file = 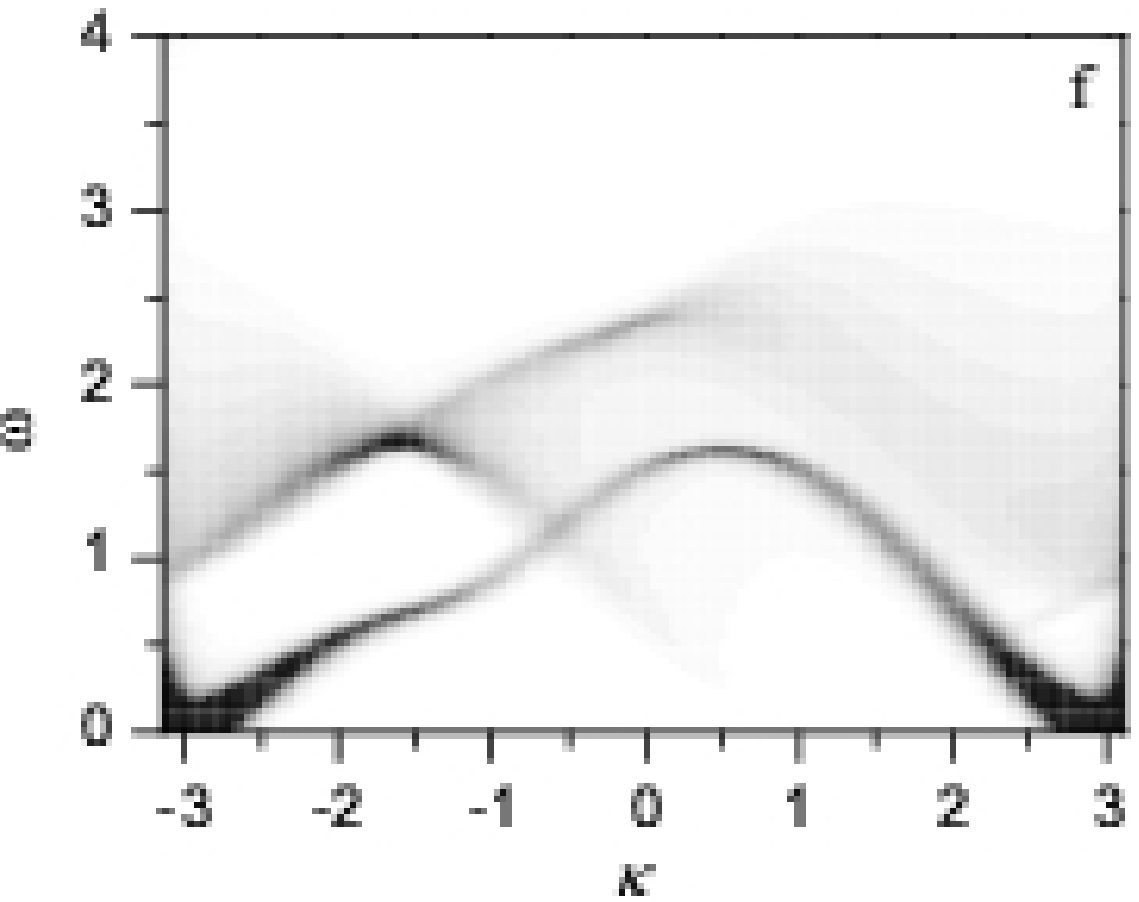, height = 0.25\linewidth}
\epsfig{file = 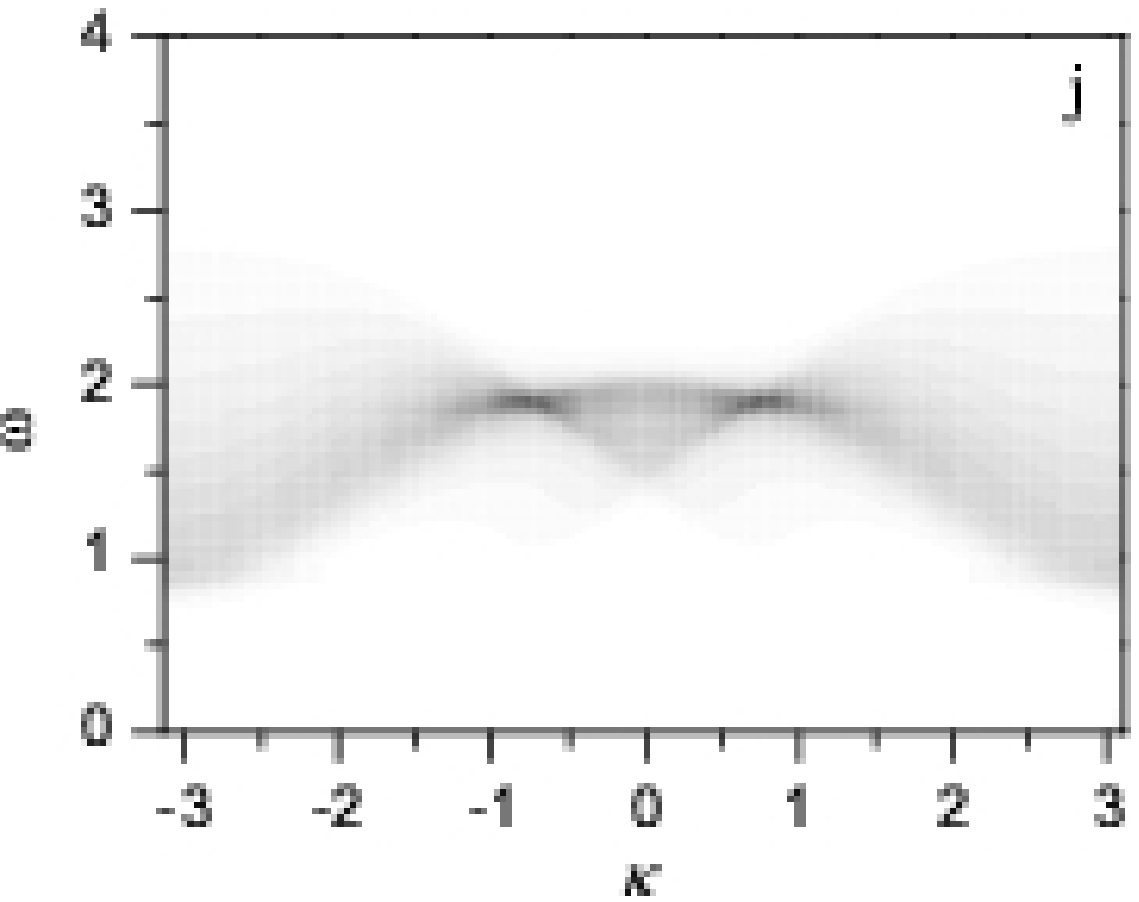, height = 0.25\linewidth}\\
\epsfig{file = 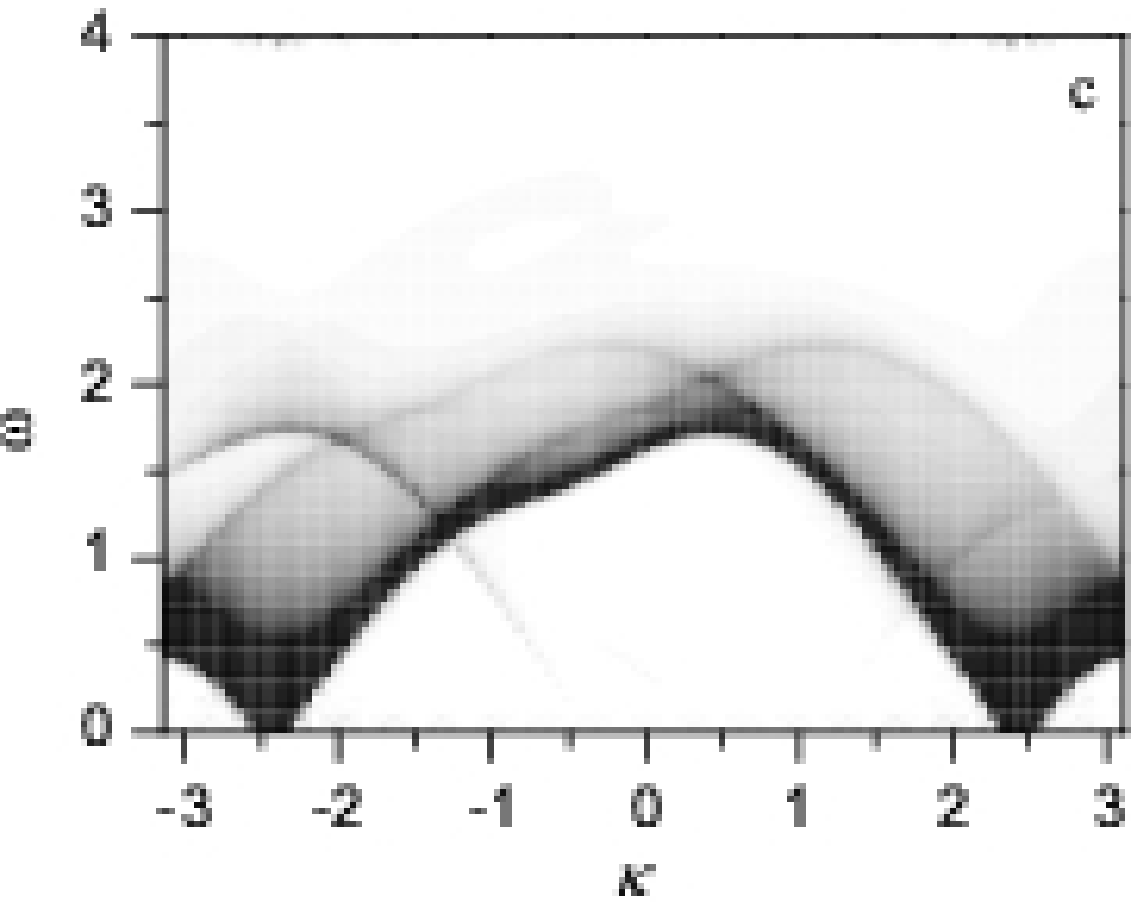, height = 0.25\linewidth}
\epsfig{file = 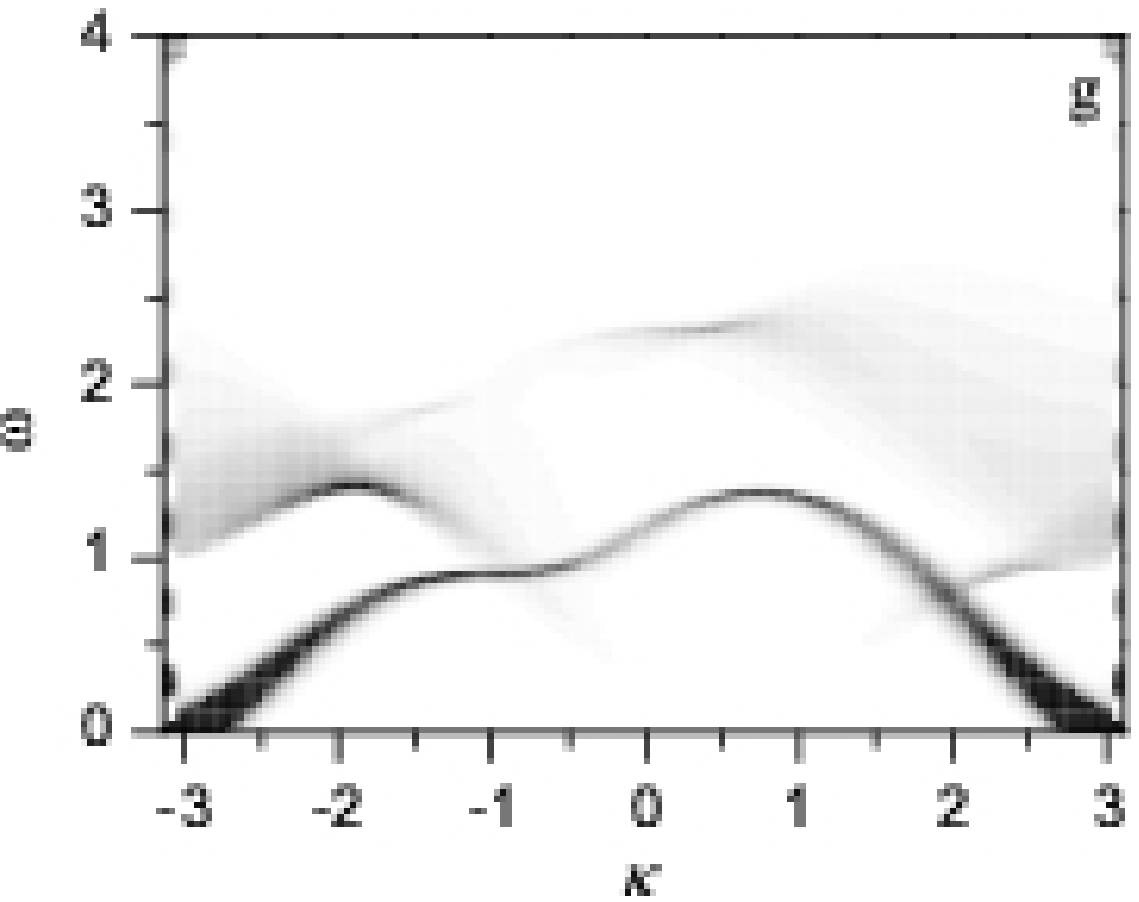, height = 0.25\linewidth}
\epsfig{file = 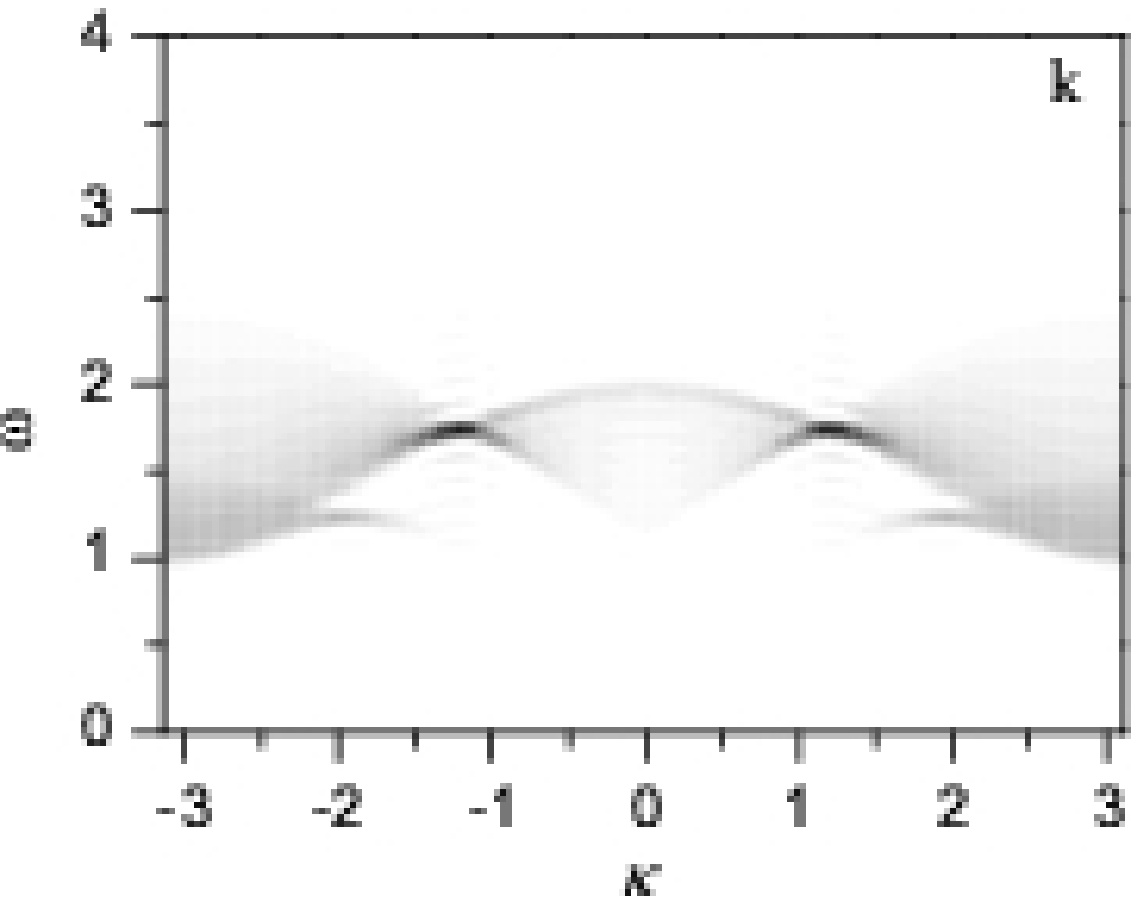, height = 0.25\linewidth}\\
\epsfig{file = 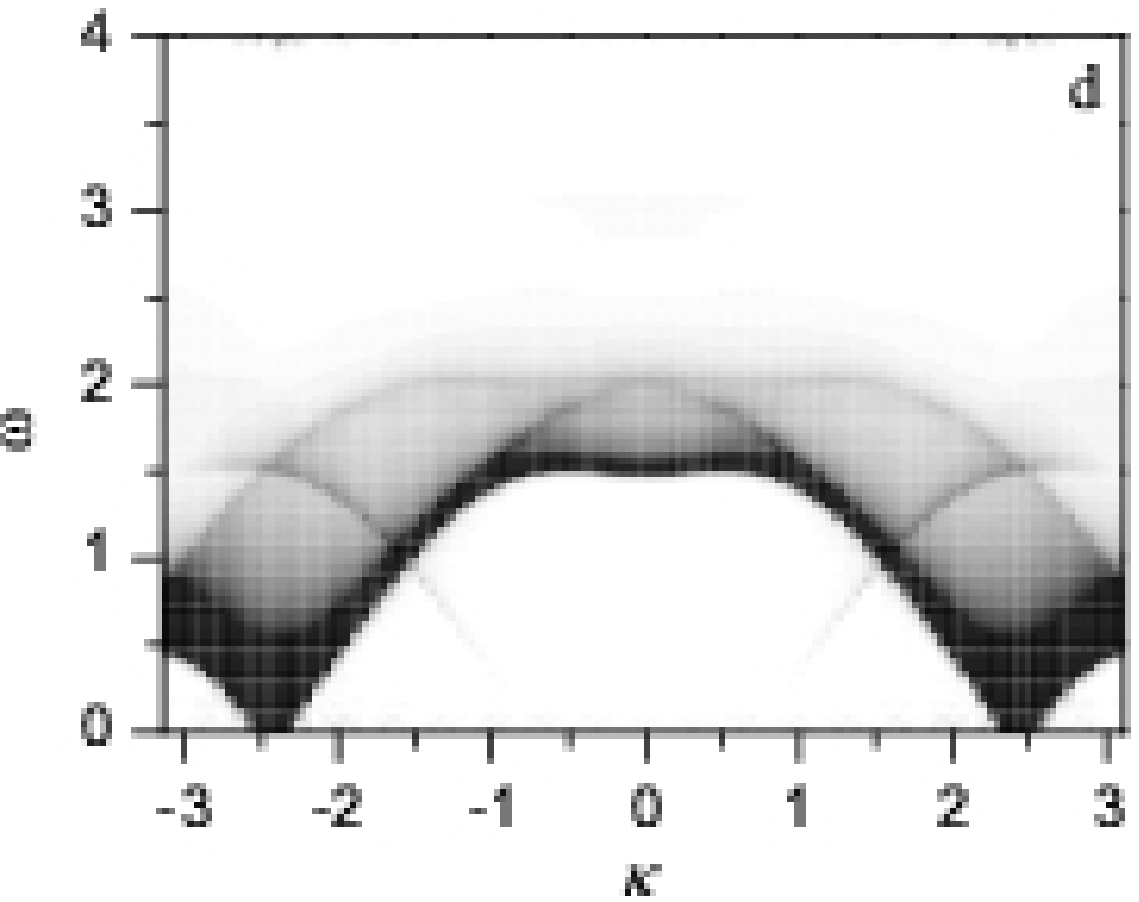, height = 0.25\linewidth}
\epsfig{file = 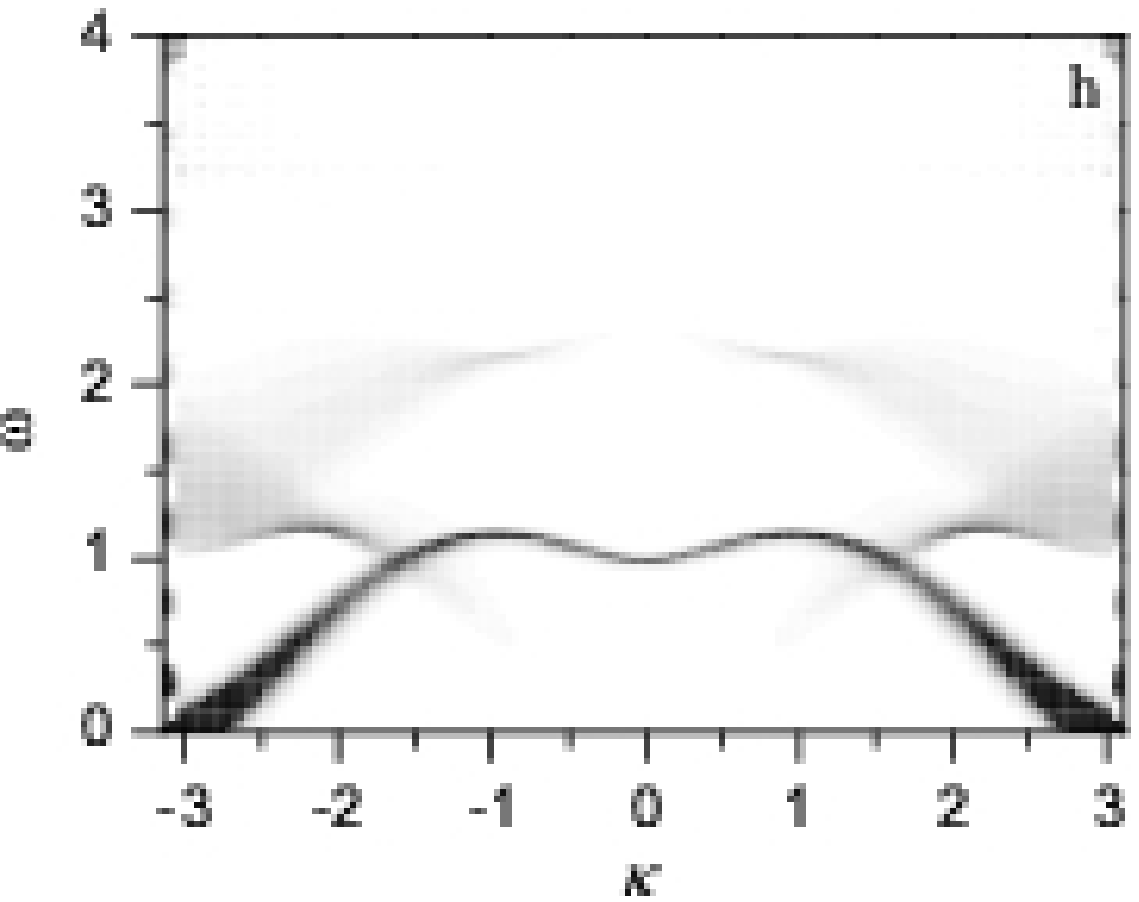, height = 0.25\linewidth}
\epsfig{file = 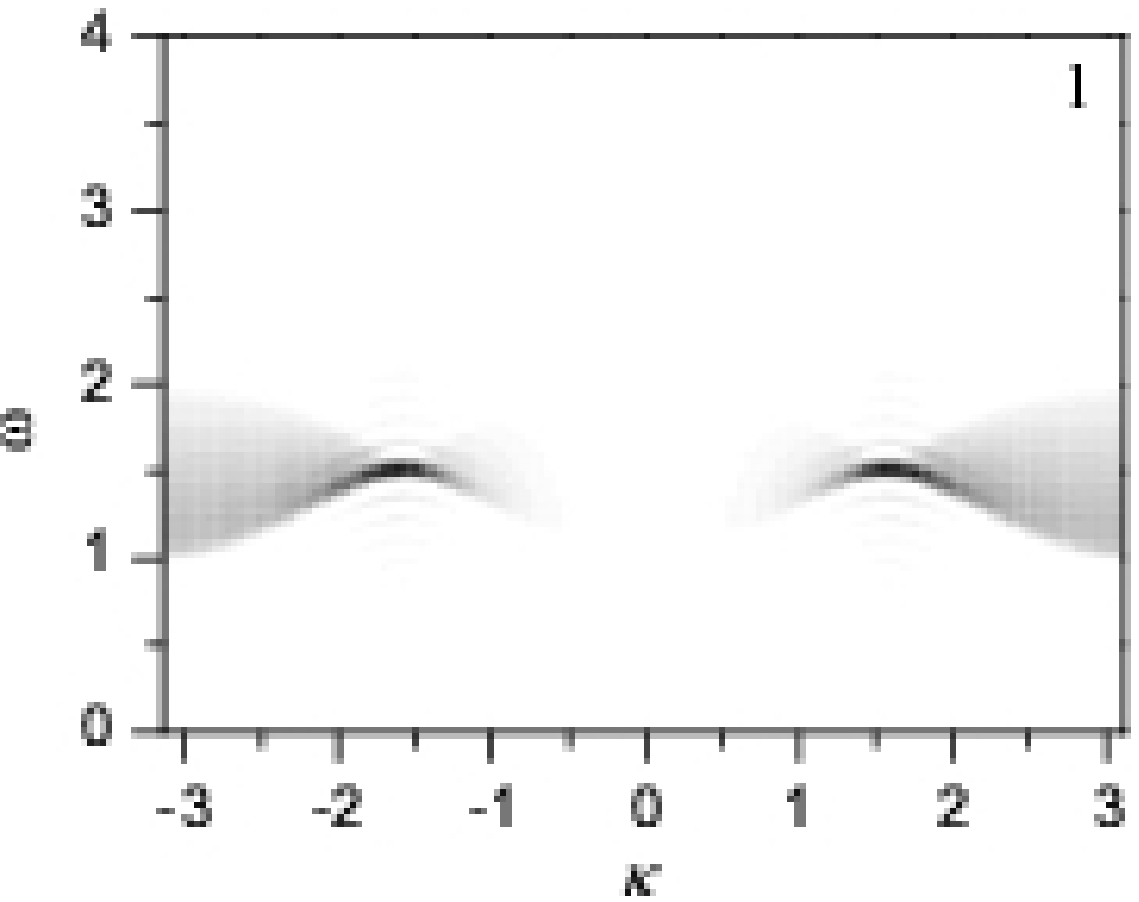, height = 0.25\linewidth}
\caption{}
\end{figure}

\newpage

\begin{figure}
\epsfig{file = 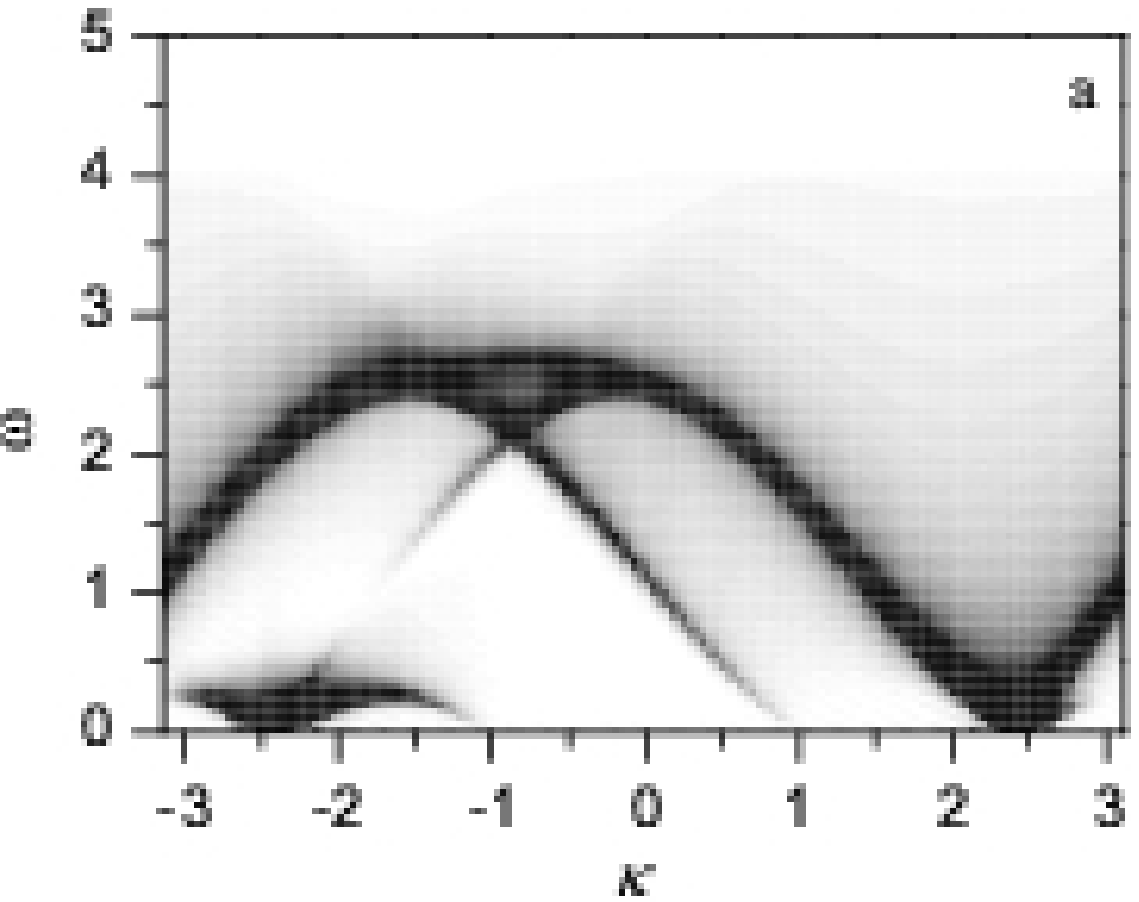, height = 0.25\linewidth}
\epsfig{file = 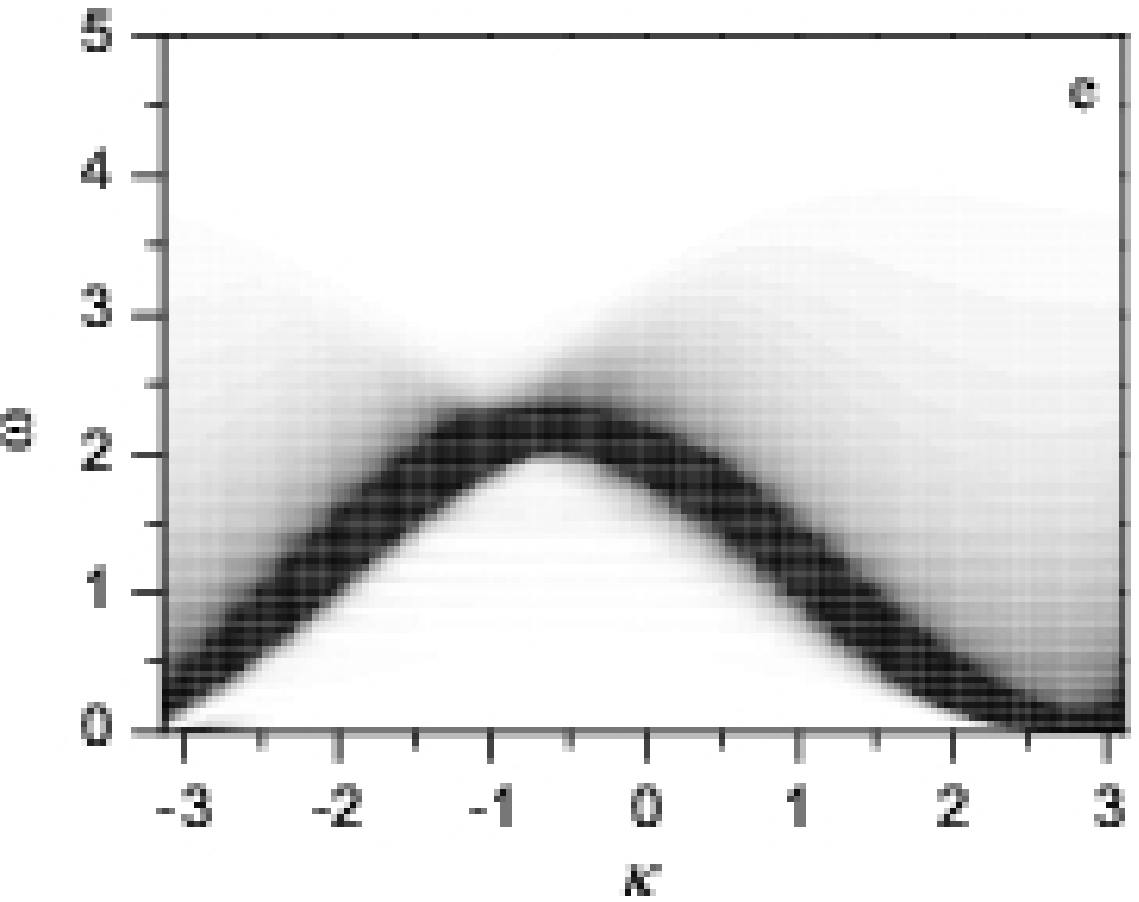, height = 0.25\linewidth}
\epsfig{file = 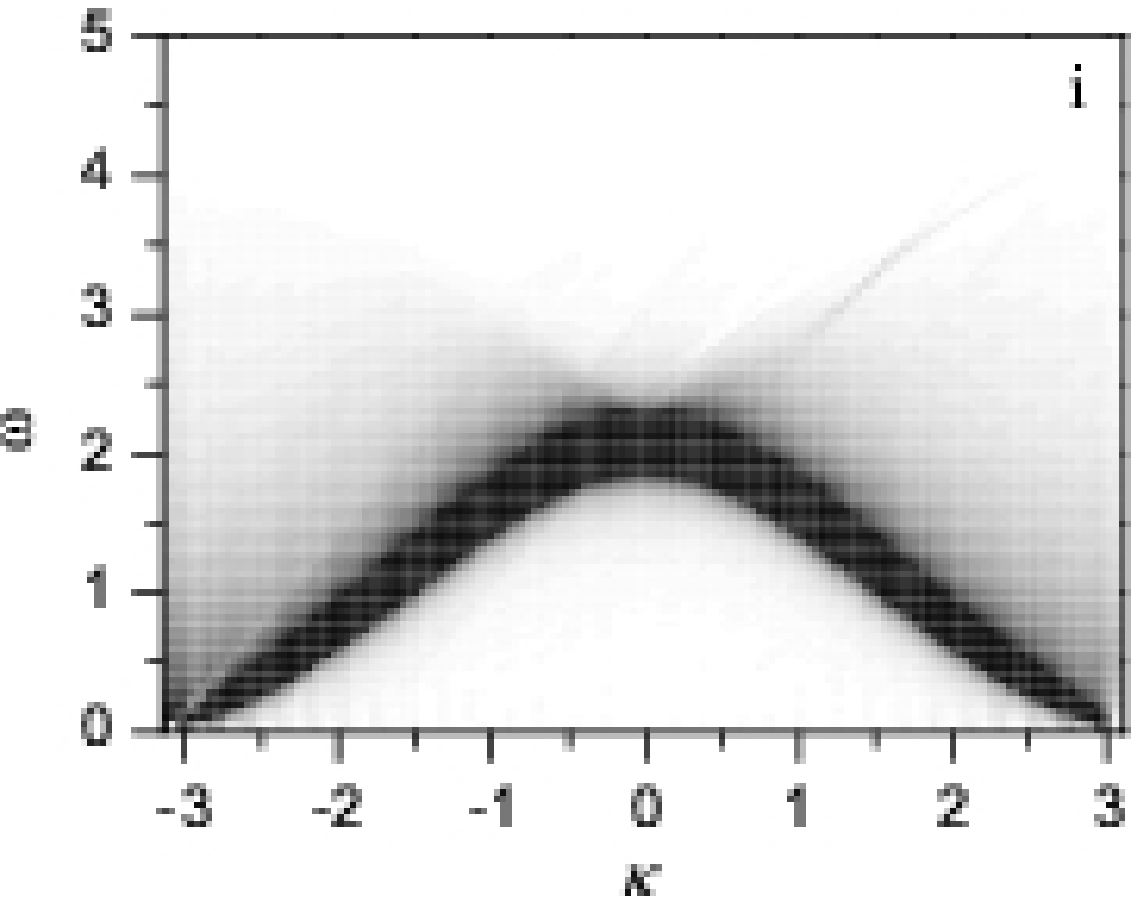, height = 0.25\linewidth}\\
\epsfig{file = 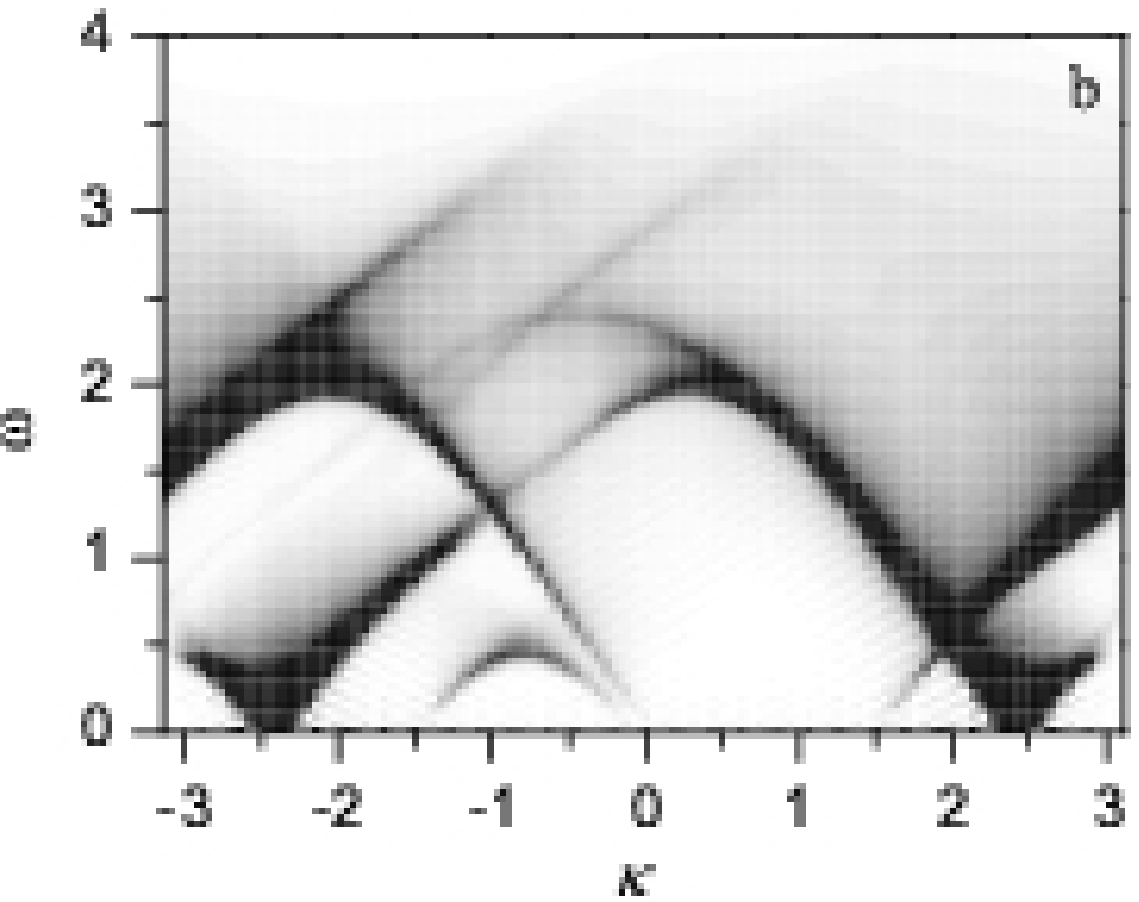, height = 0.25\linewidth}
\epsfig{file = 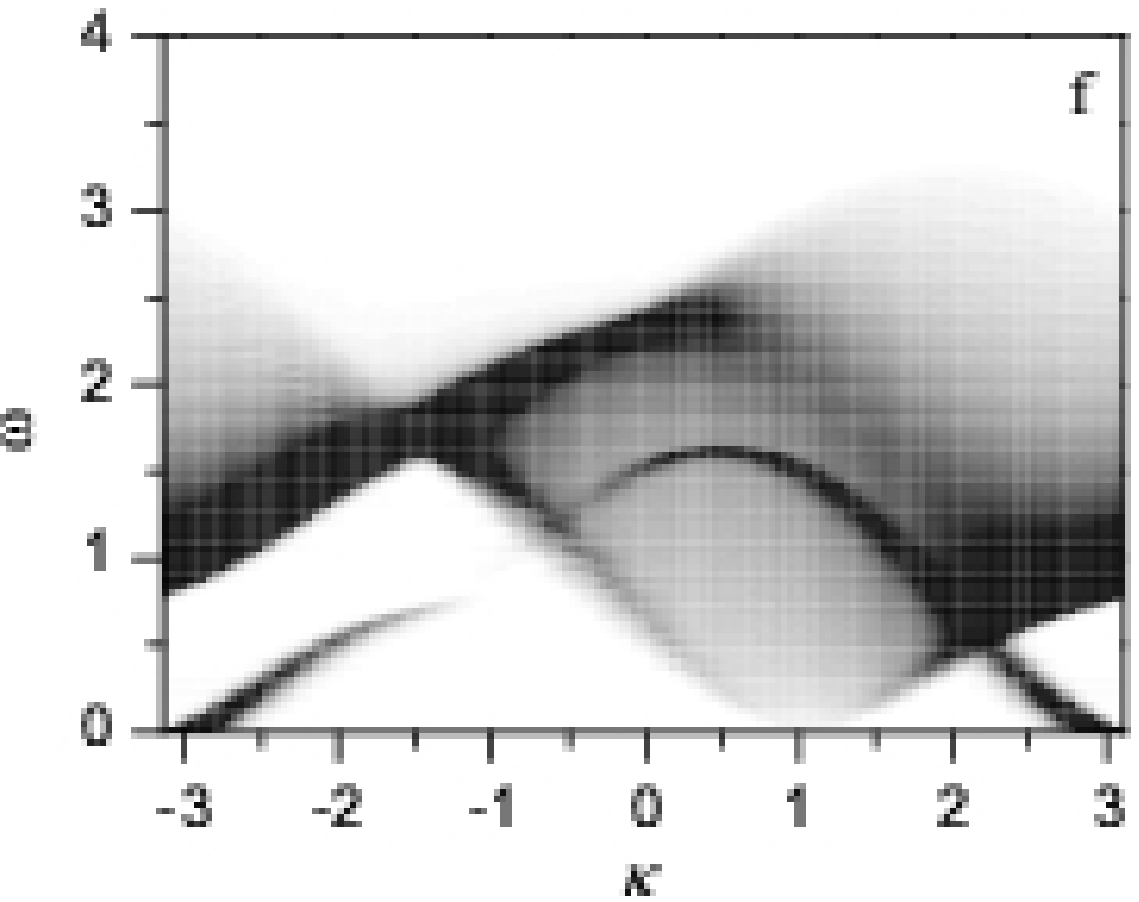, height = 0.25\linewidth}
\epsfig{file = 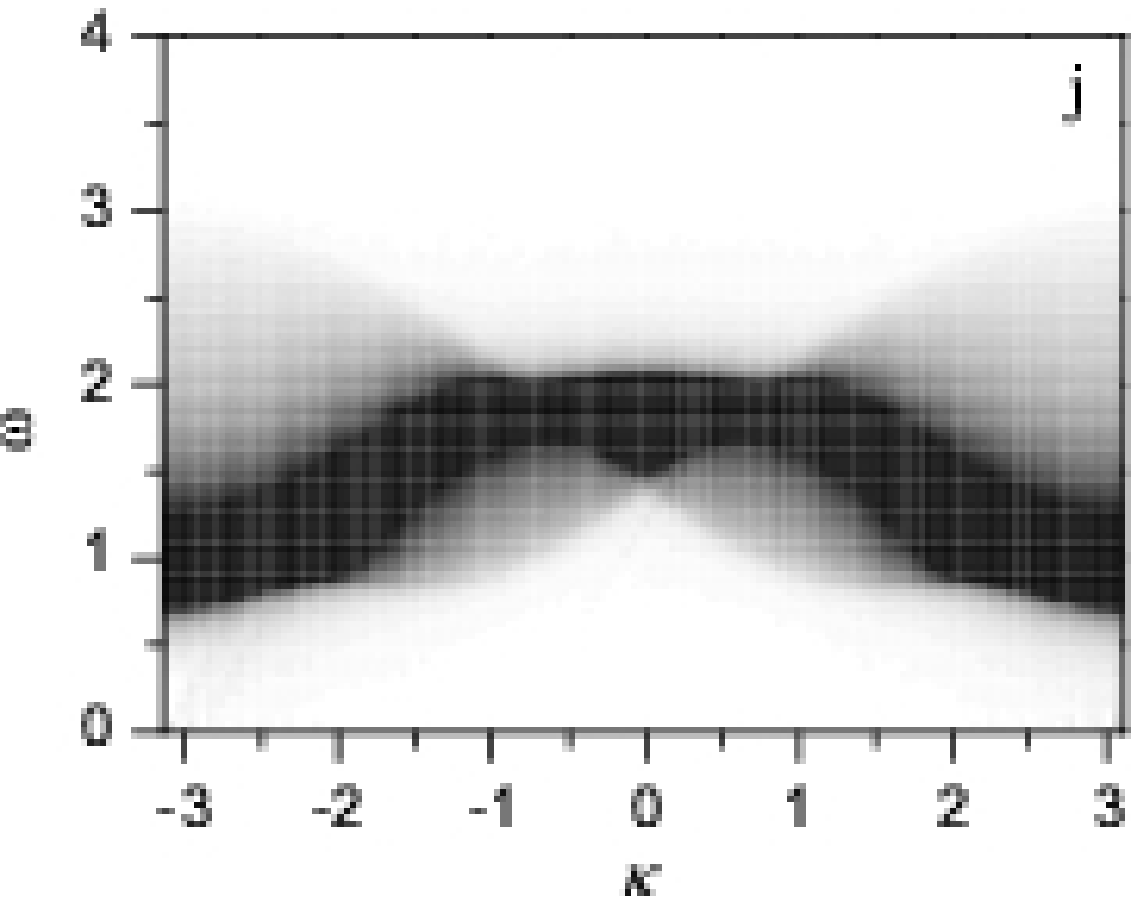, height = 0.25\linewidth}\\
\epsfig{file = 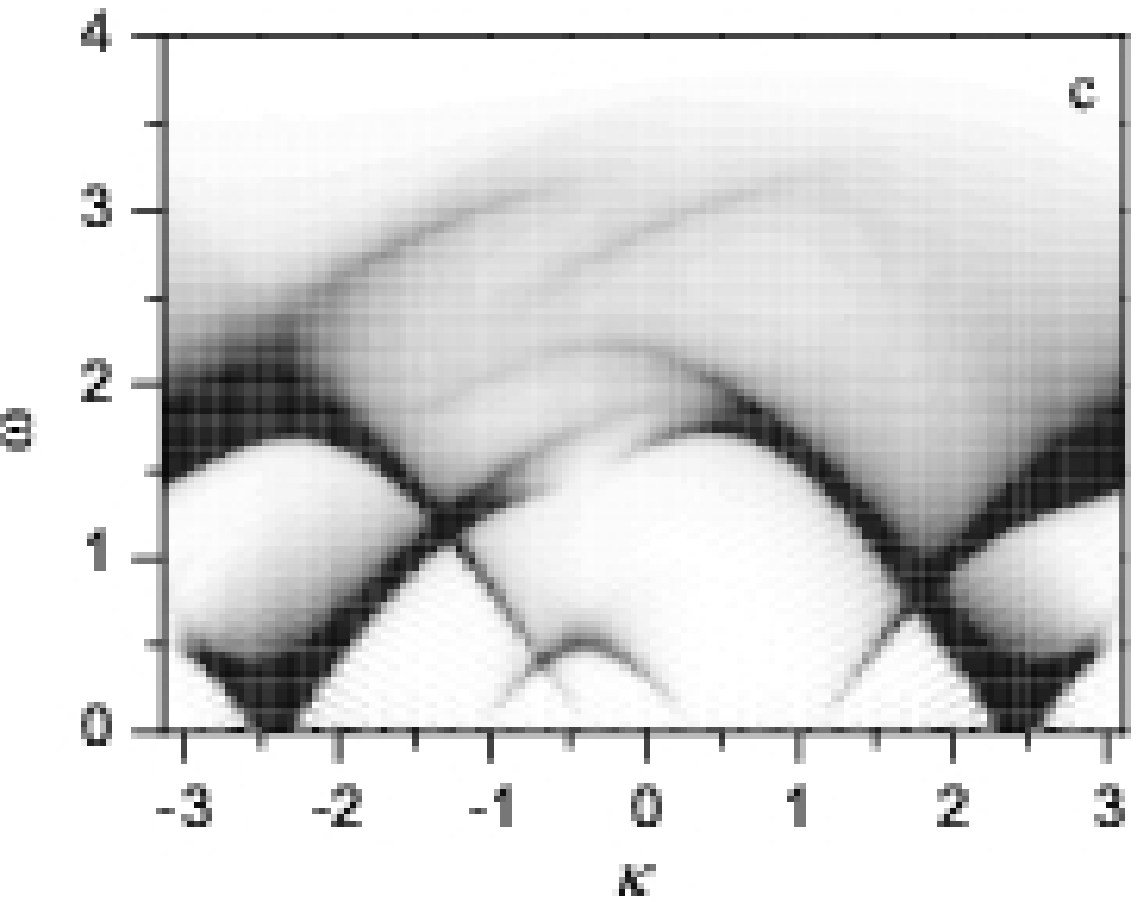, height = 0.25\linewidth}
\epsfig{file = 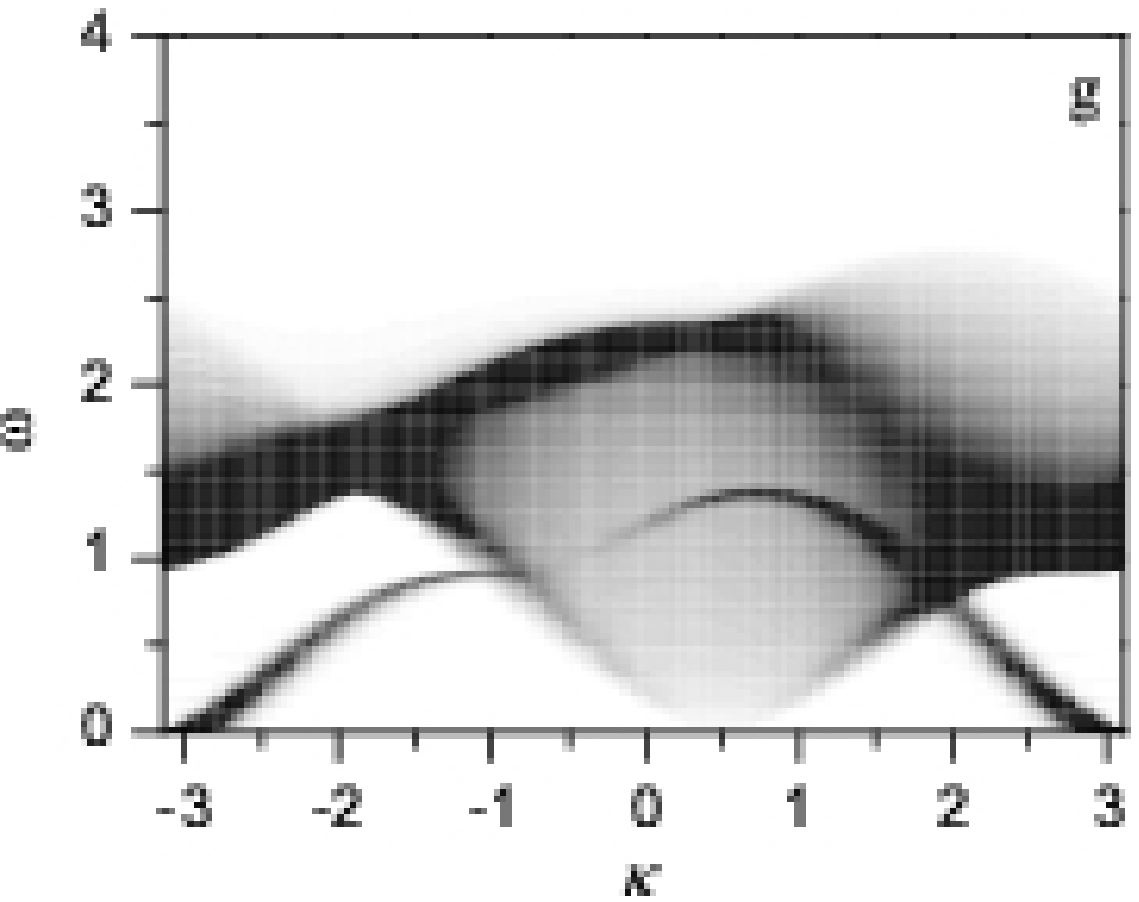, height = 0.25\linewidth}
\epsfig{file = 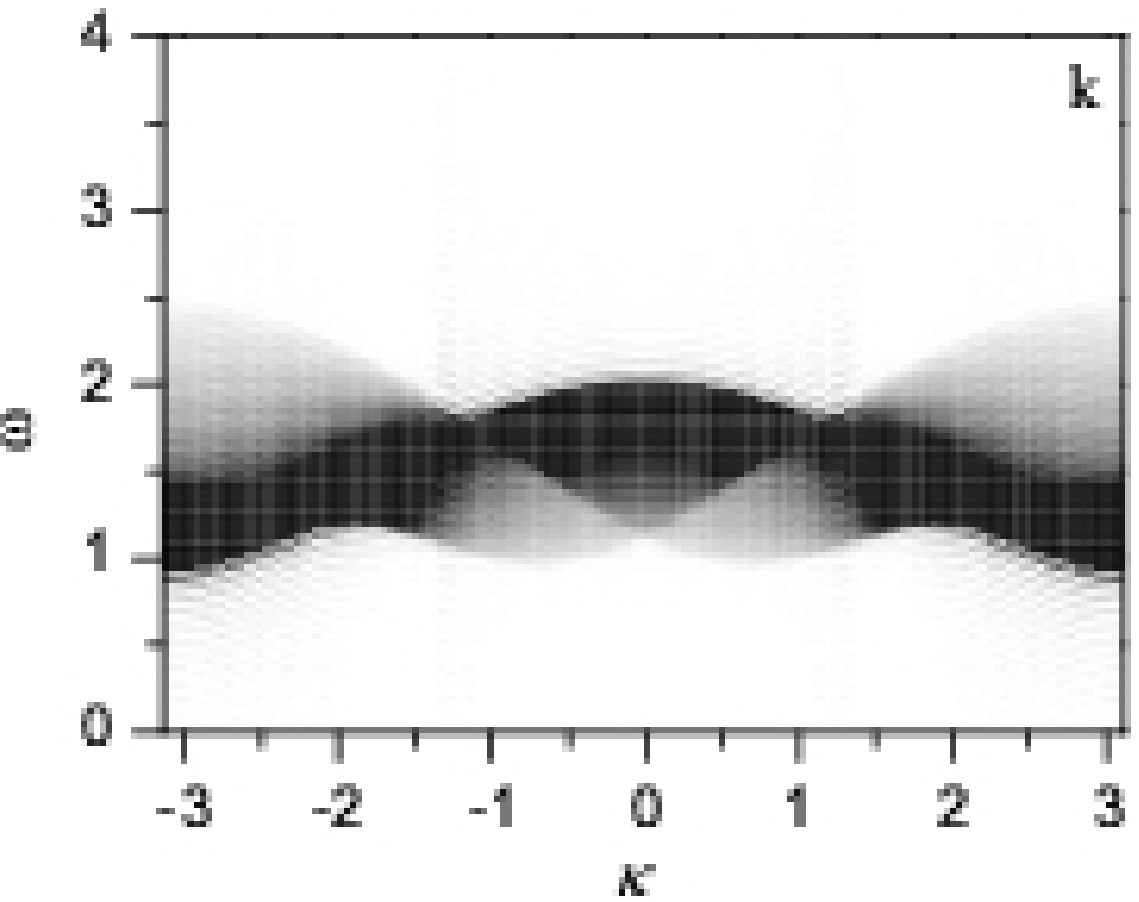, height = 0.25\linewidth}\\
\epsfig{file = 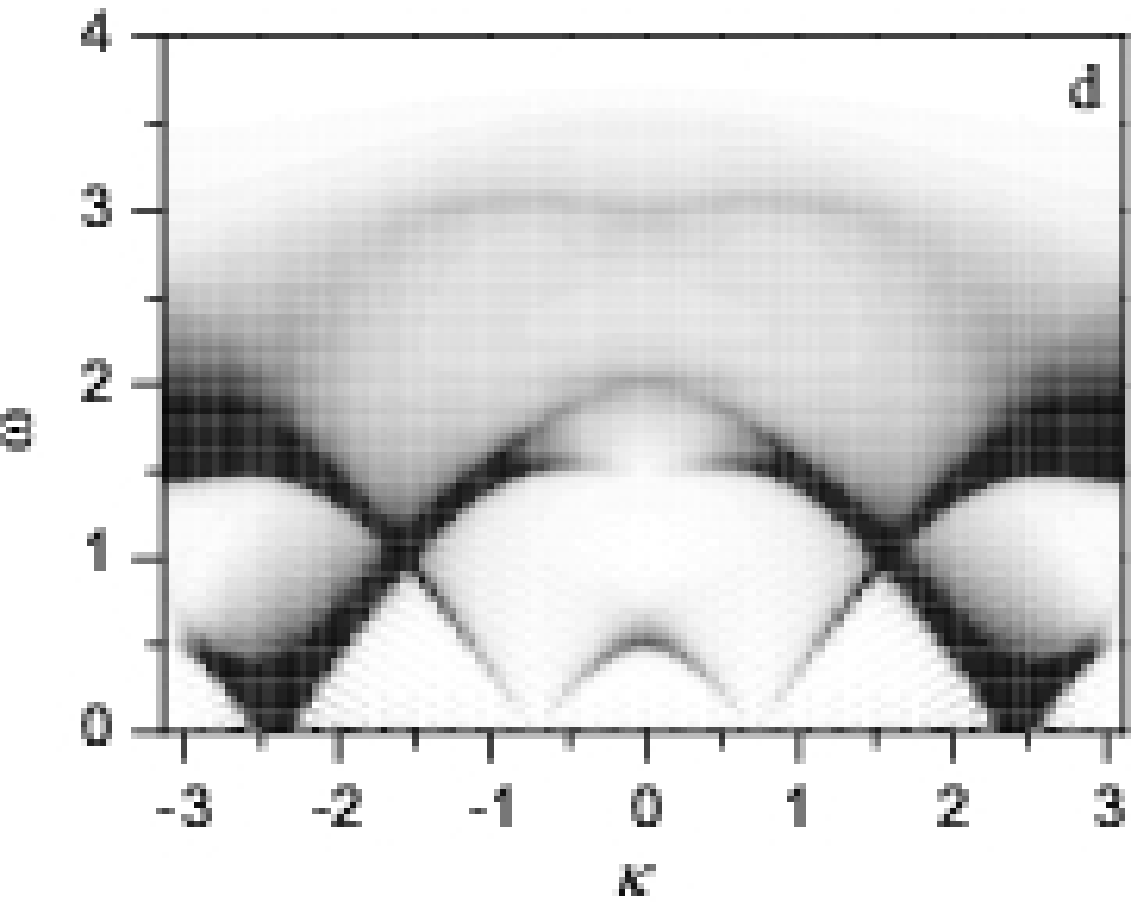, height = 0.25\linewidth}
\epsfig{file = 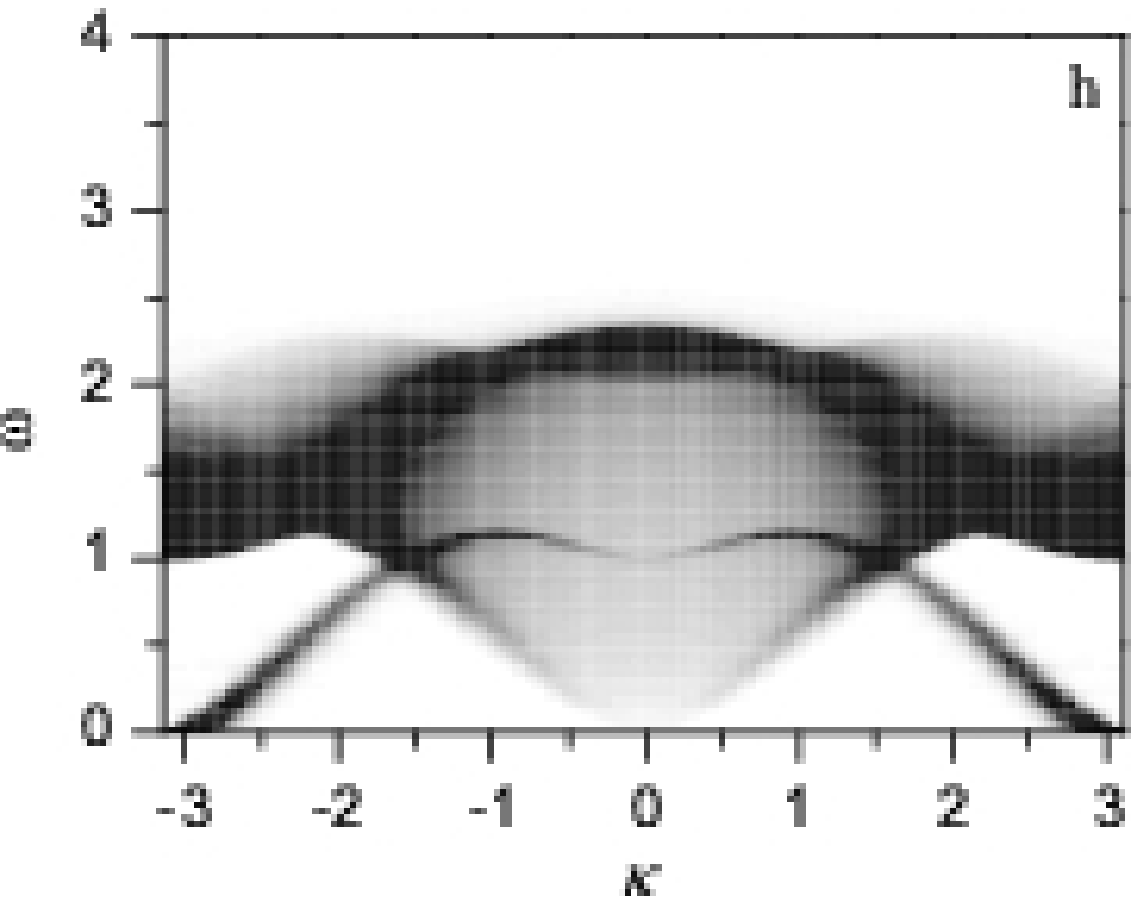, height = 0.25\linewidth}
\epsfig{file = 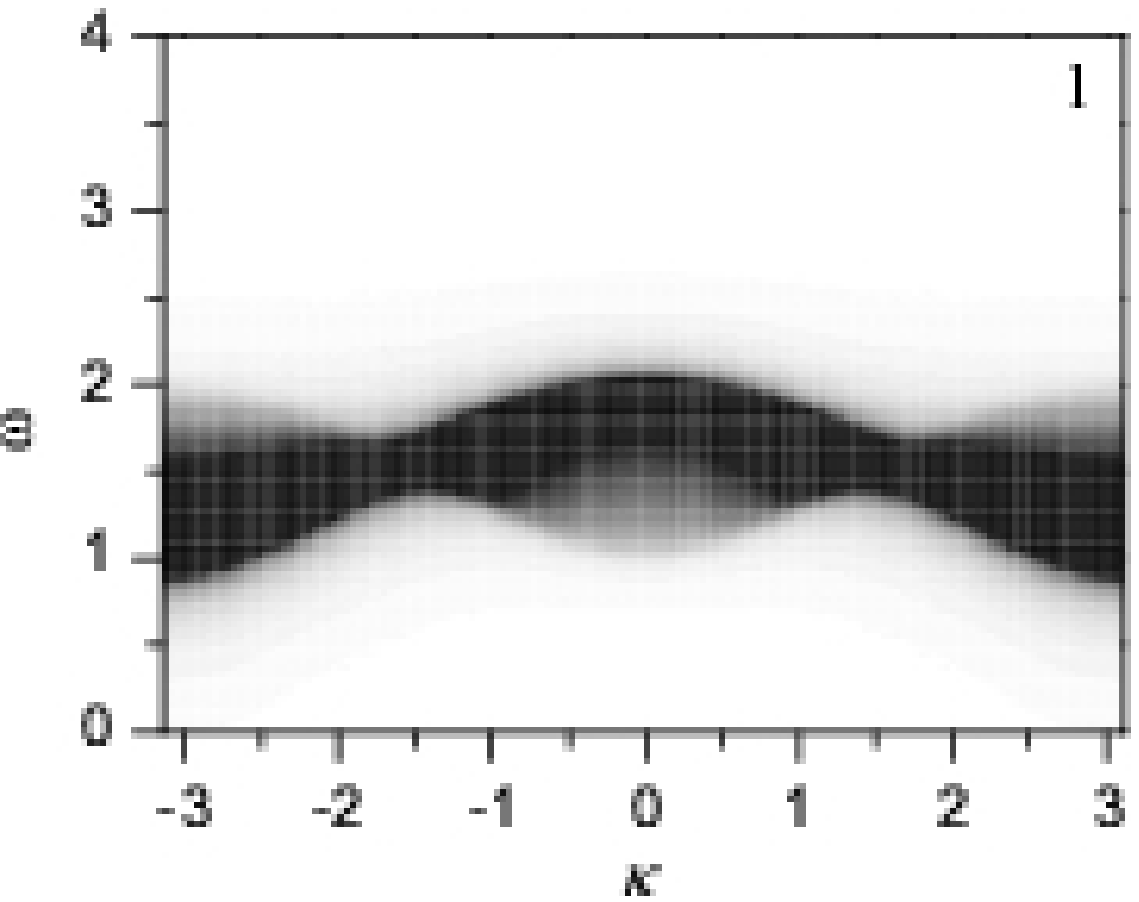, height = 0.25\linewidth}
\caption{}
\end{figure}

\newpage

\begin{figure}
\epsfig{file = 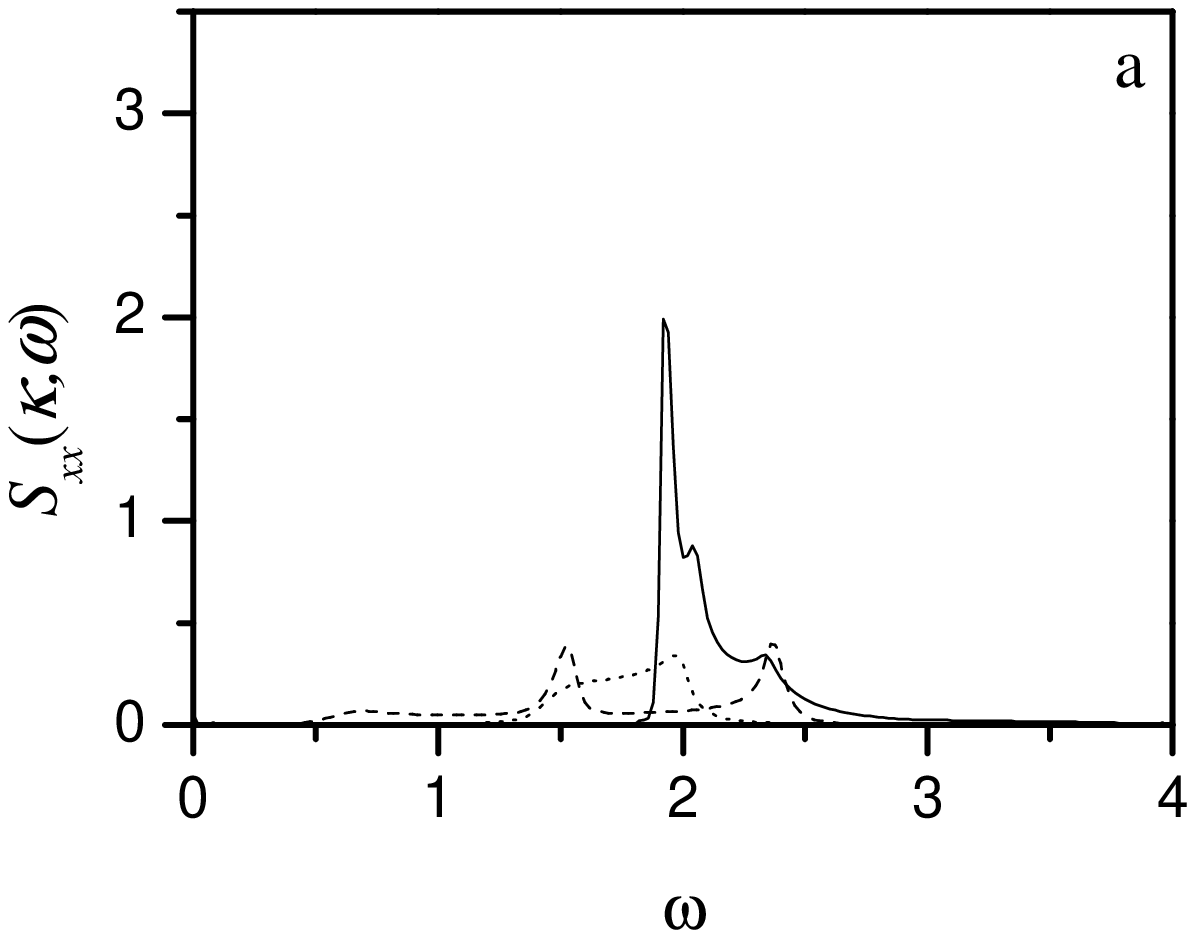, height = 0.23\linewidth}
\epsfig{file = 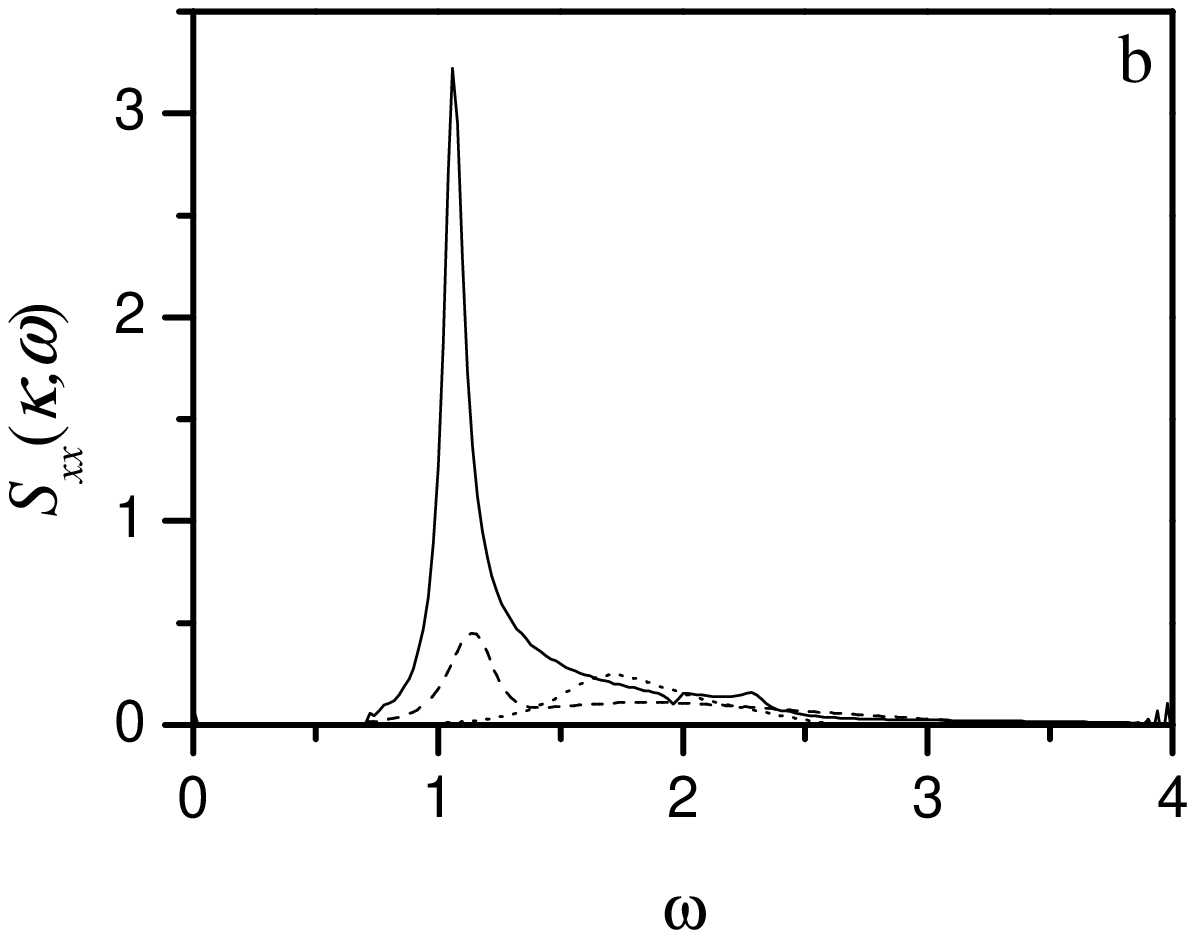, height = 0.23\linewidth}
\epsfig{file = 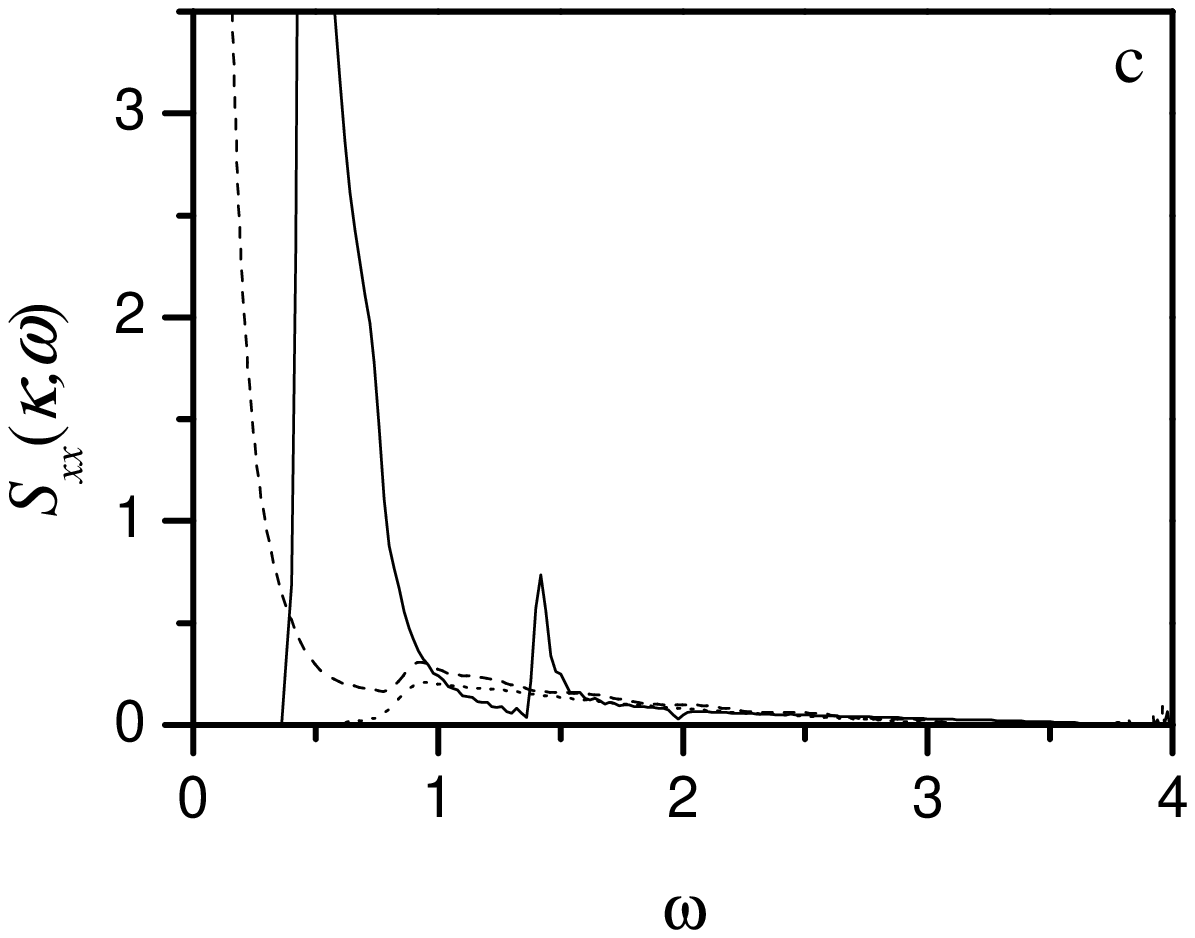, height = 0.23\linewidth}
\caption{}
\end{figure}

\newpage

\begin{figure}
\epsfig{file = 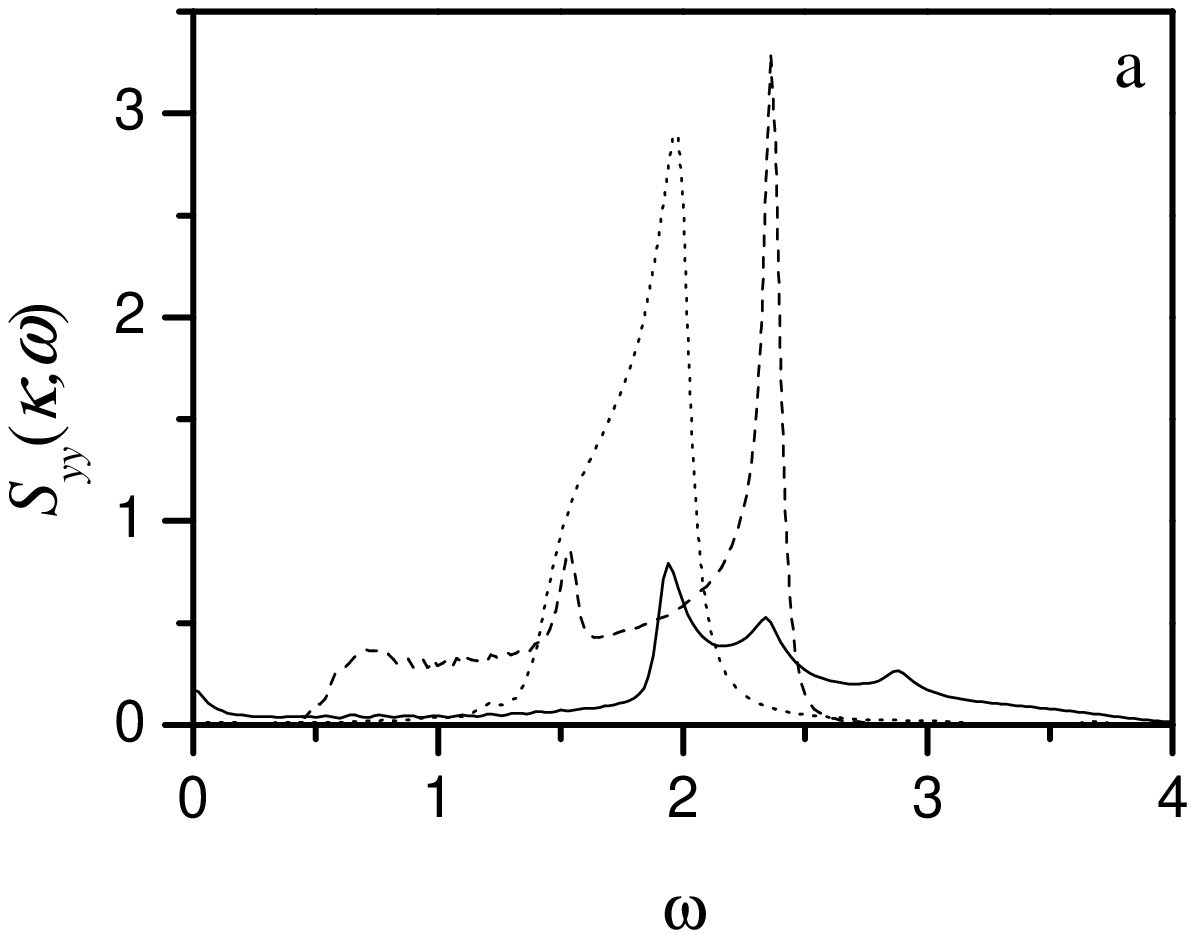, height = 0.23\linewidth}
\epsfig{file = 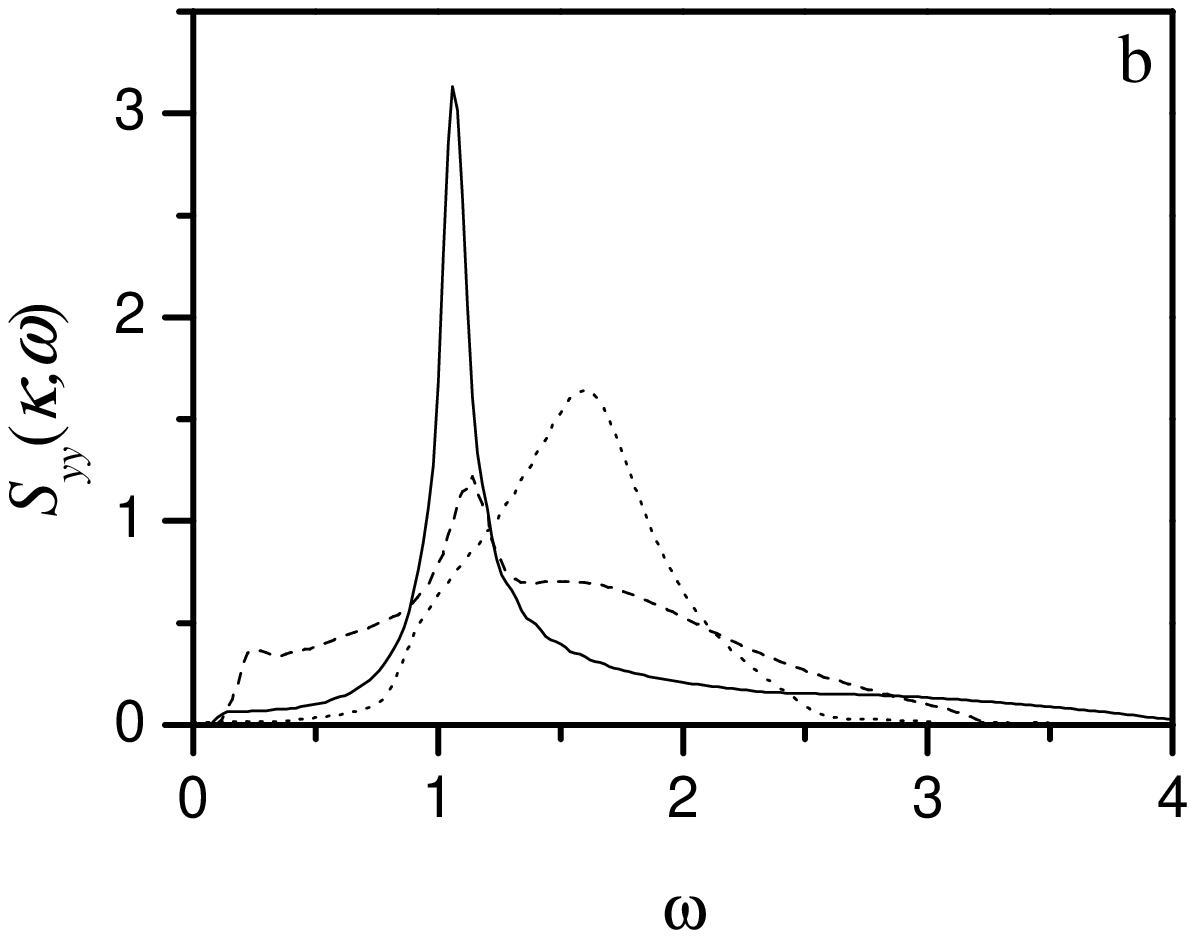, height = 0.23\linewidth}
\epsfig{file = 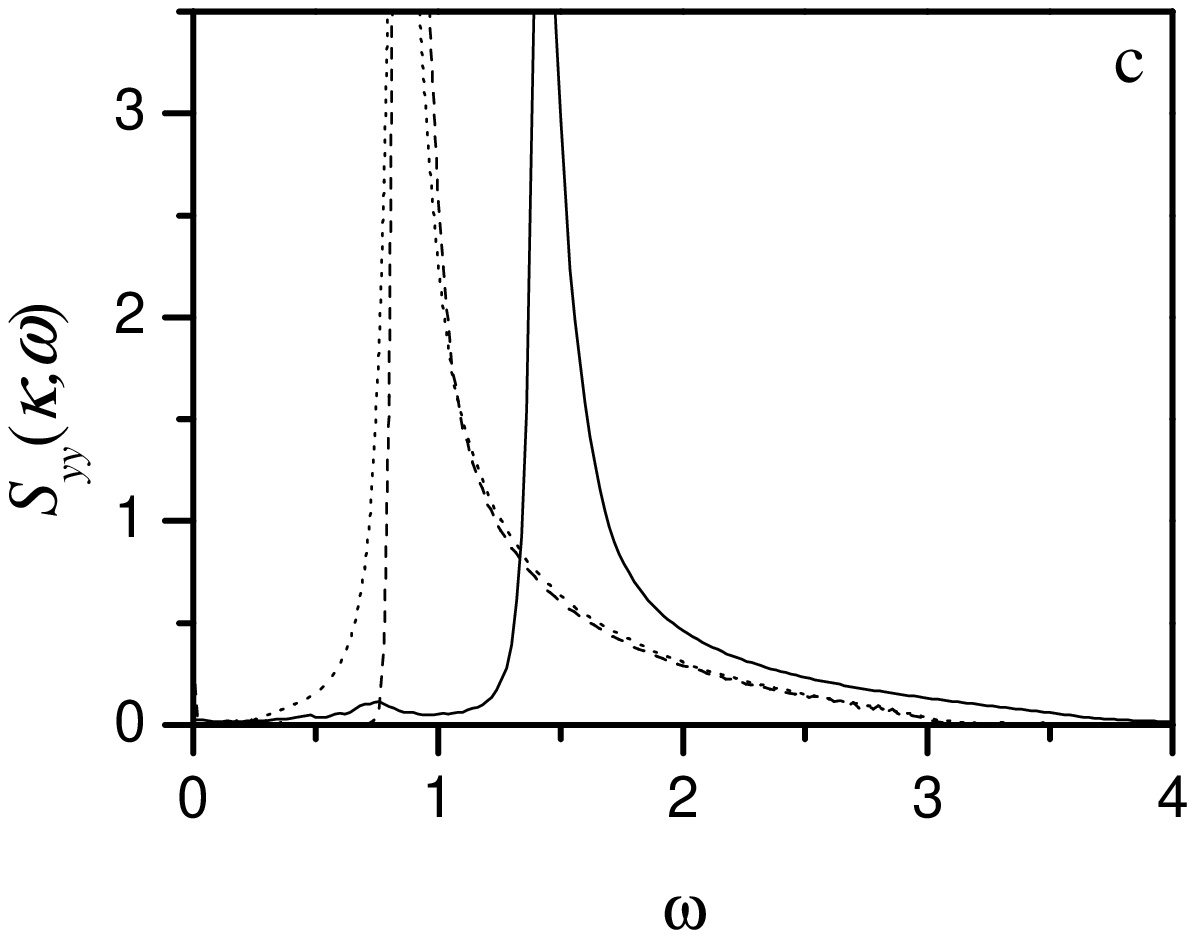, height = 0.23\linewidth}
\caption{}
\end{figure}

\newpage

\begin{figure}
\epsfig{file = 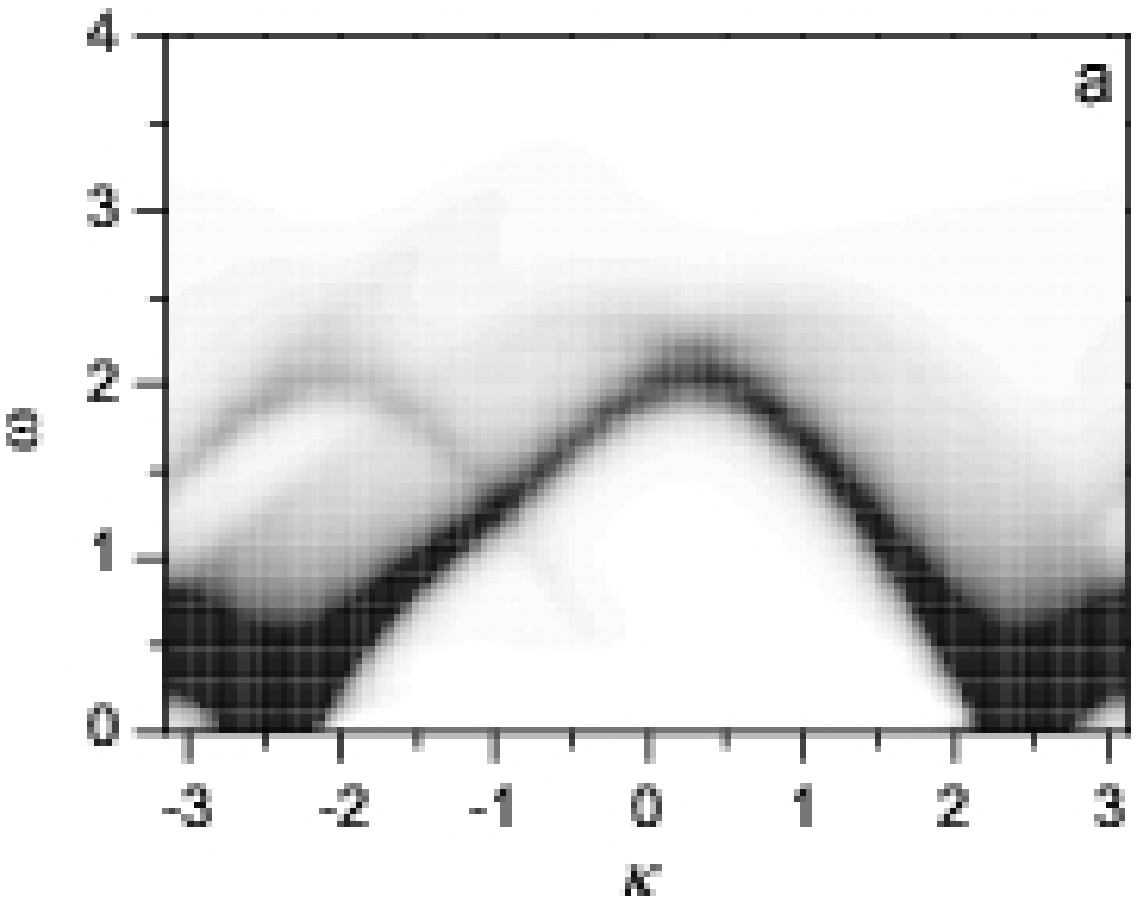, height = 0.25\linewidth}
\epsfig{file = 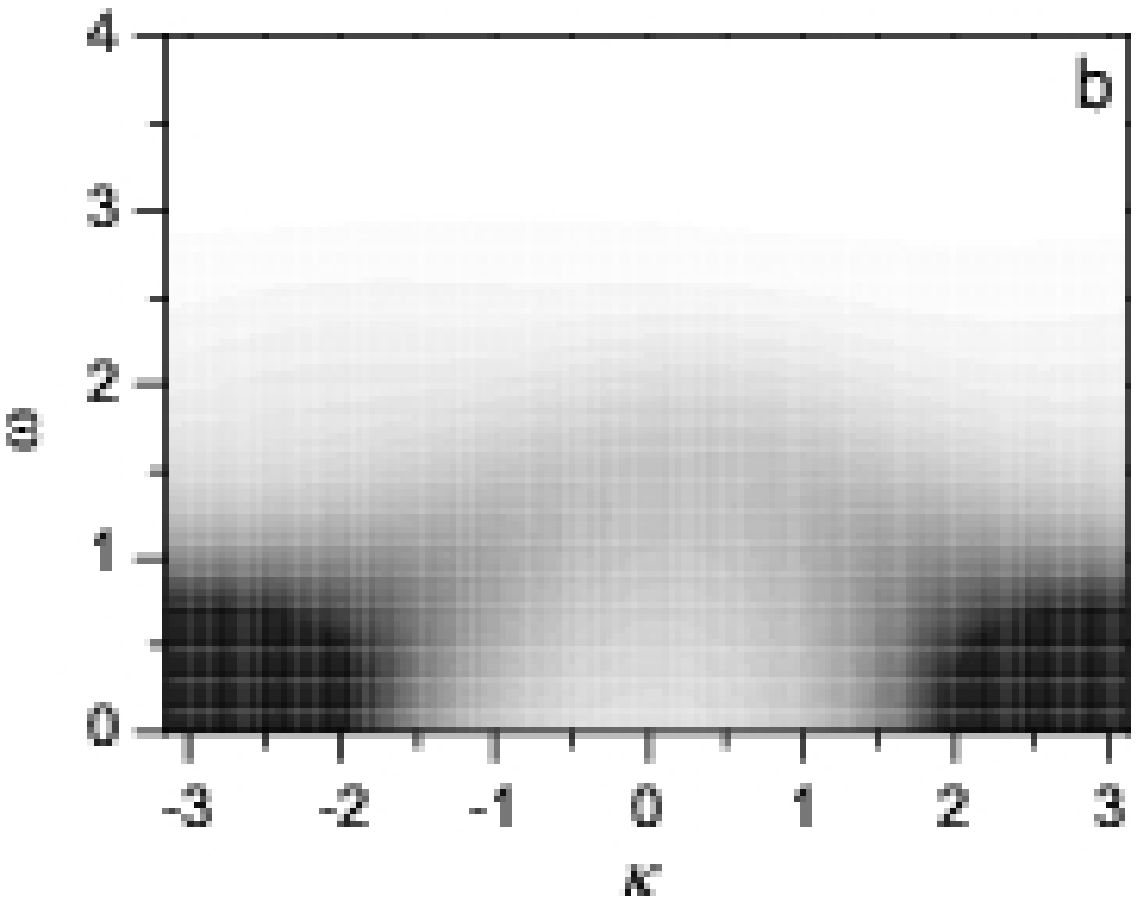, height = 0.25\linewidth}
\epsfig{file = 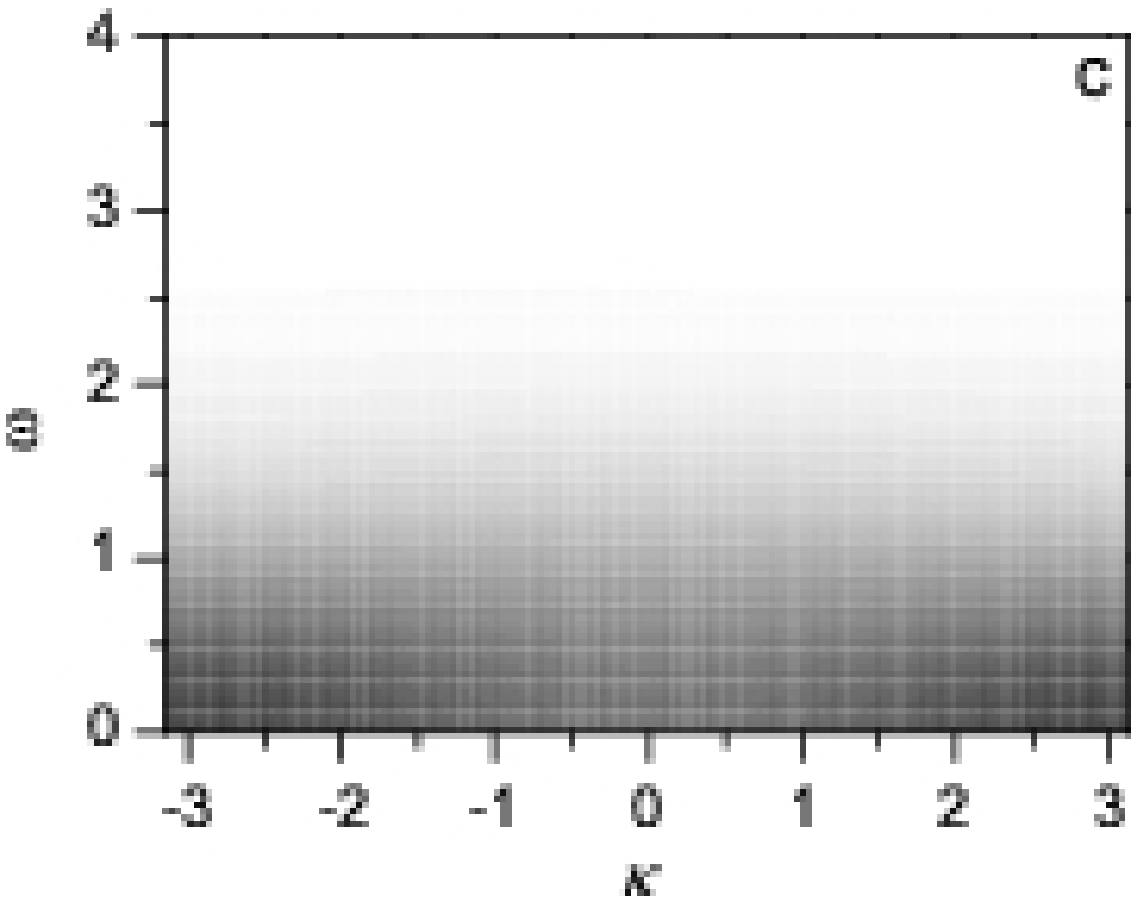, height = 0.25\linewidth}
\caption{}
\end{figure}

\newpage

\begin{figure}
\epsfig{file = 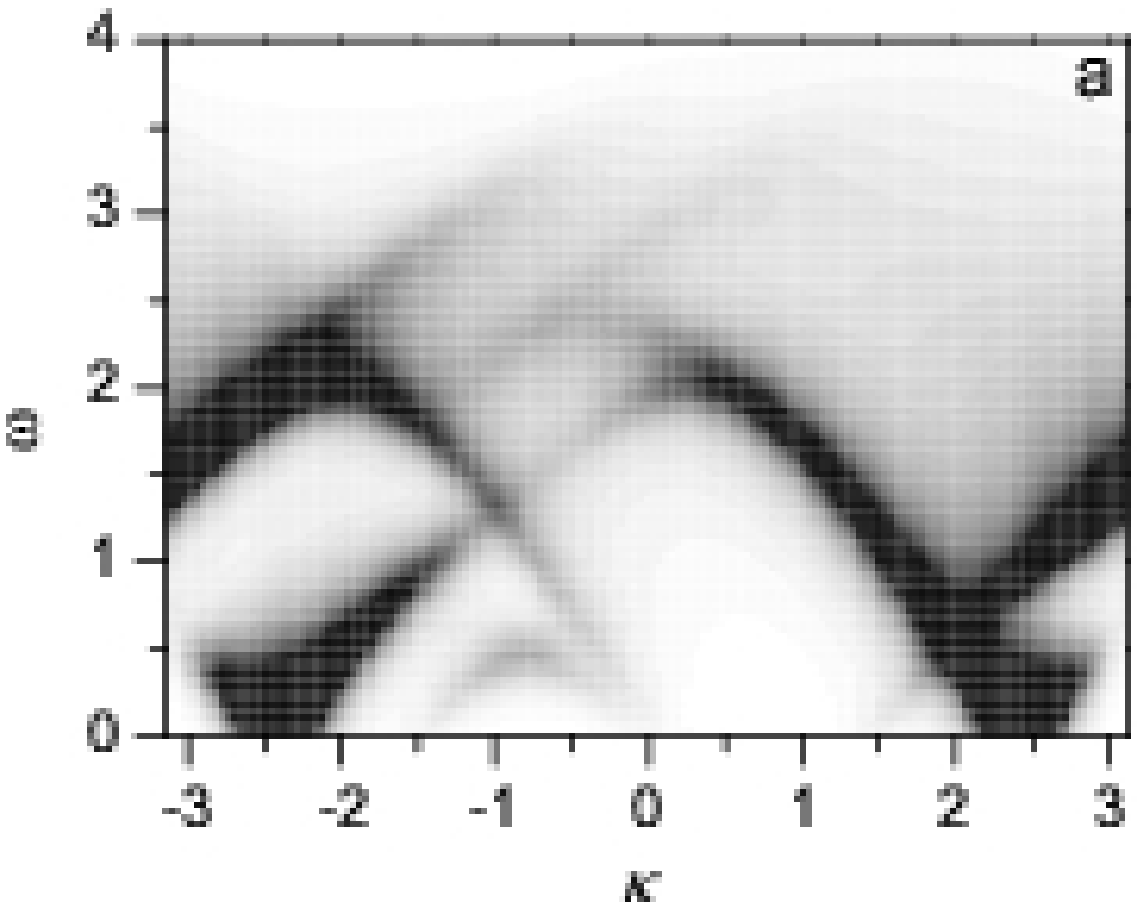, height = 0.25\linewidth}
\epsfig{file = 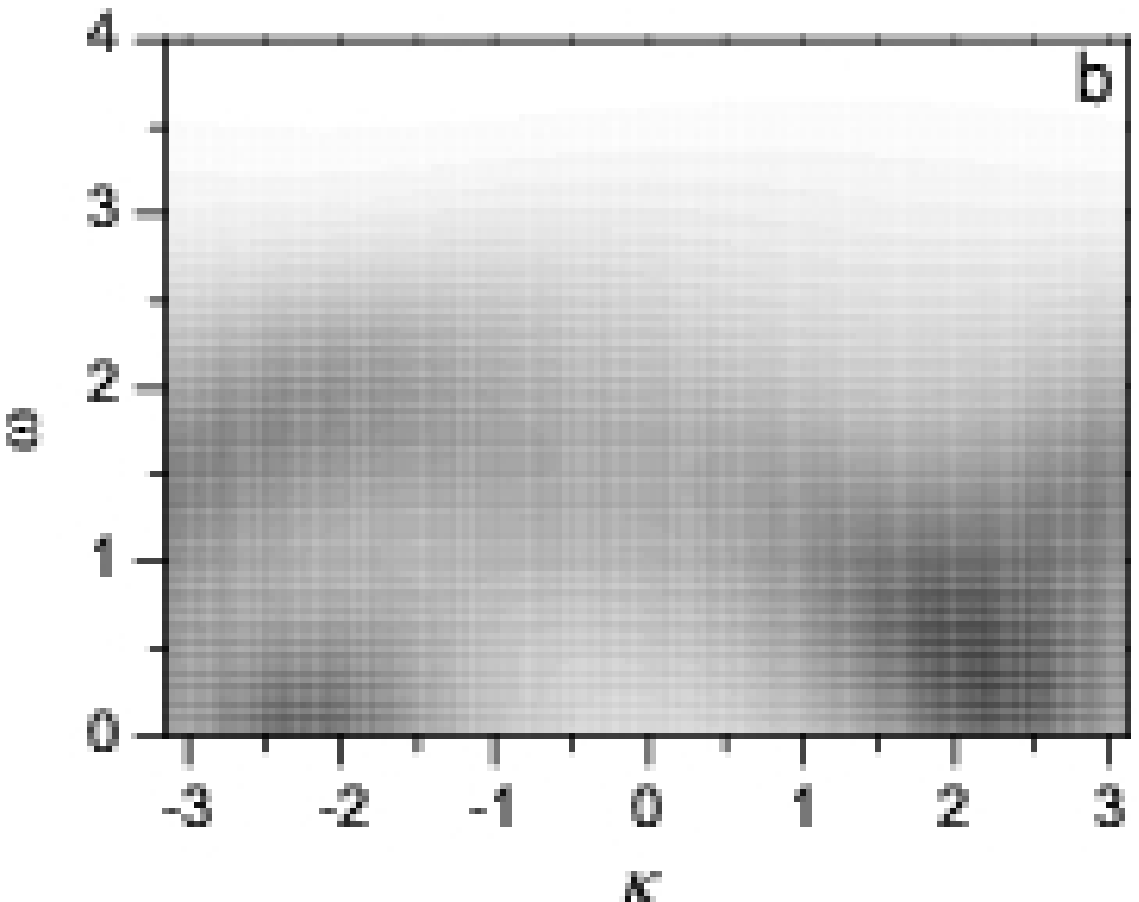, height = 0.25\linewidth}
\epsfig{file = 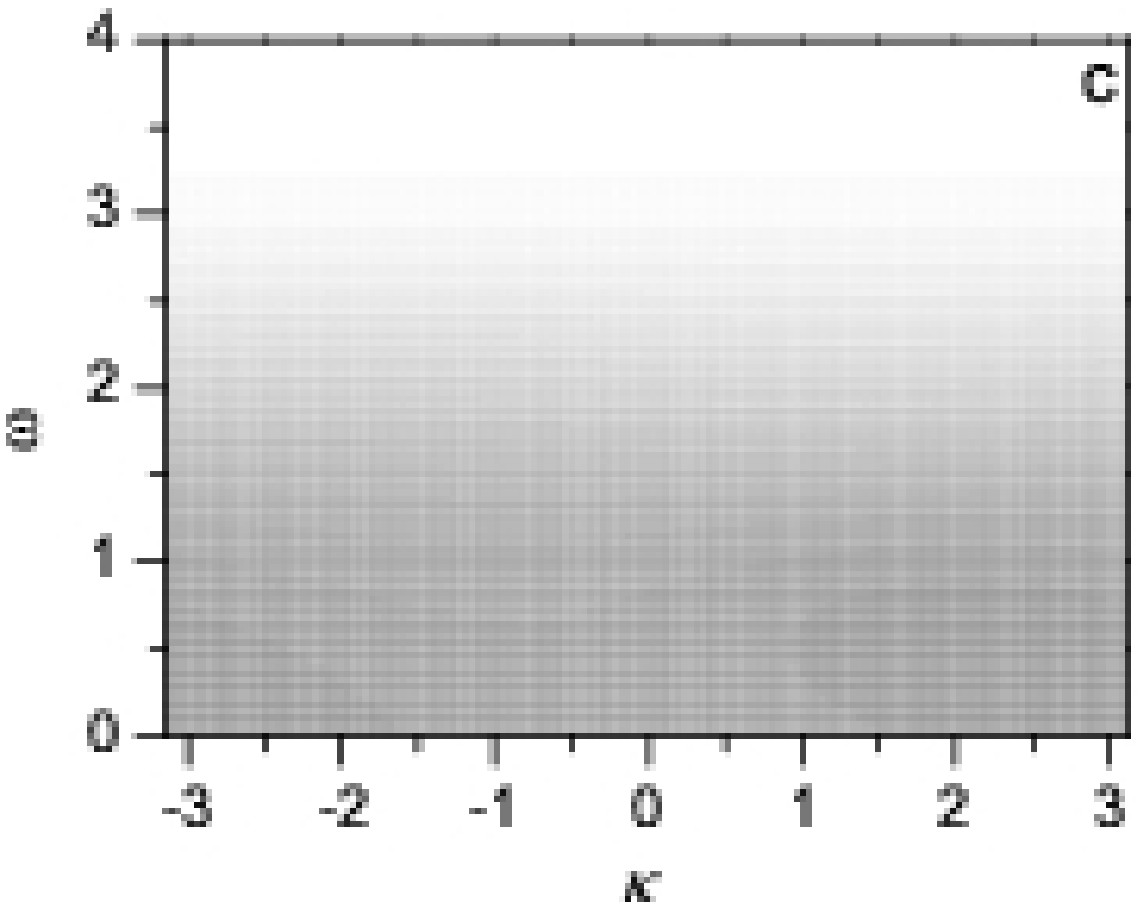, height = 0.25\linewidth}
\caption{}
\end{figure}

\newpage

\begin{figure}
\epsfig{file = 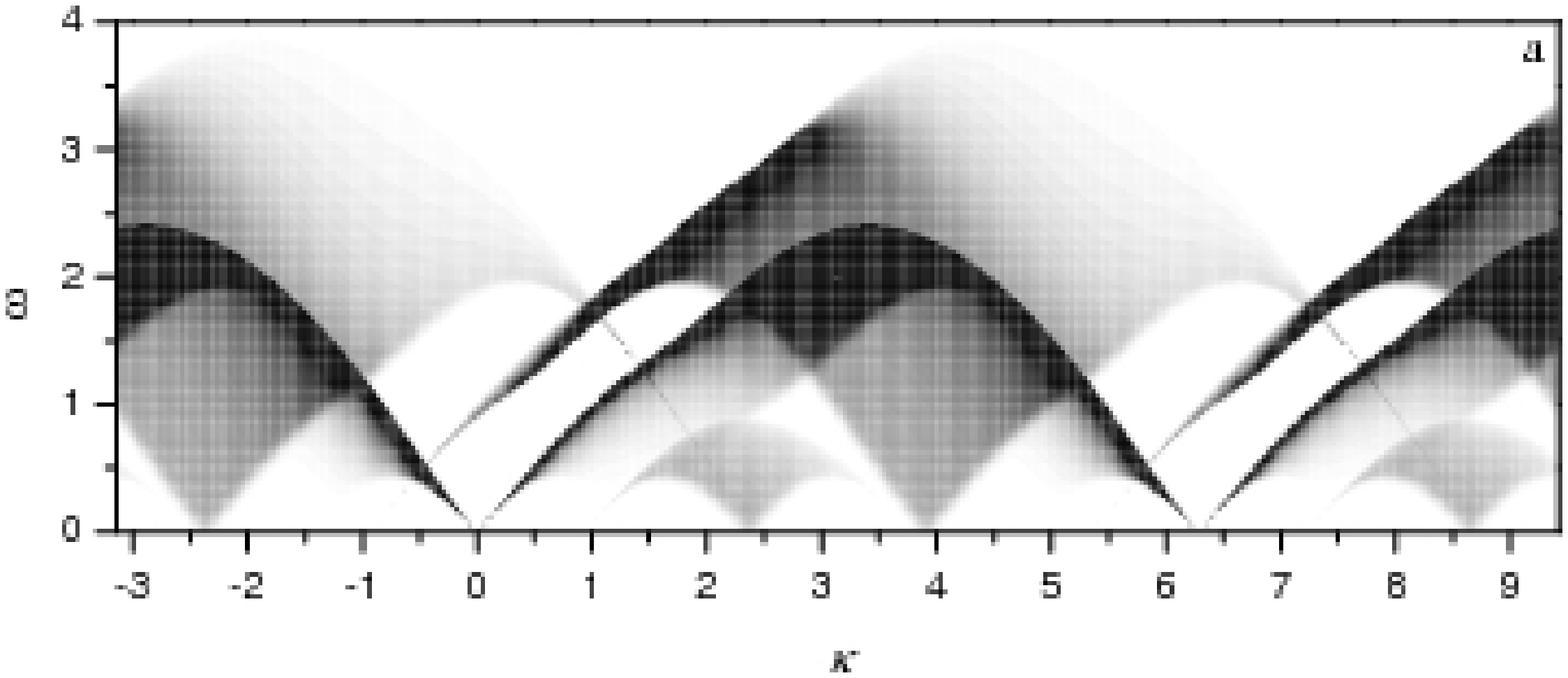, height = 0.25\linewidth}\\
\epsfig{file = 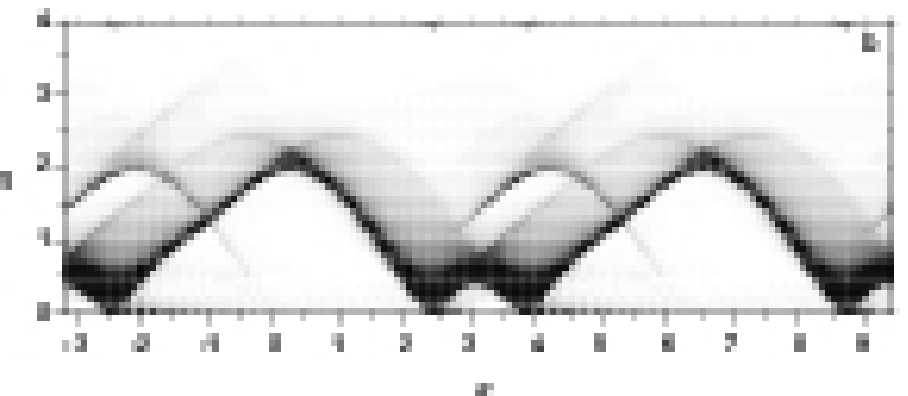, height = 0.25\linewidth}\\
\epsfig{file = 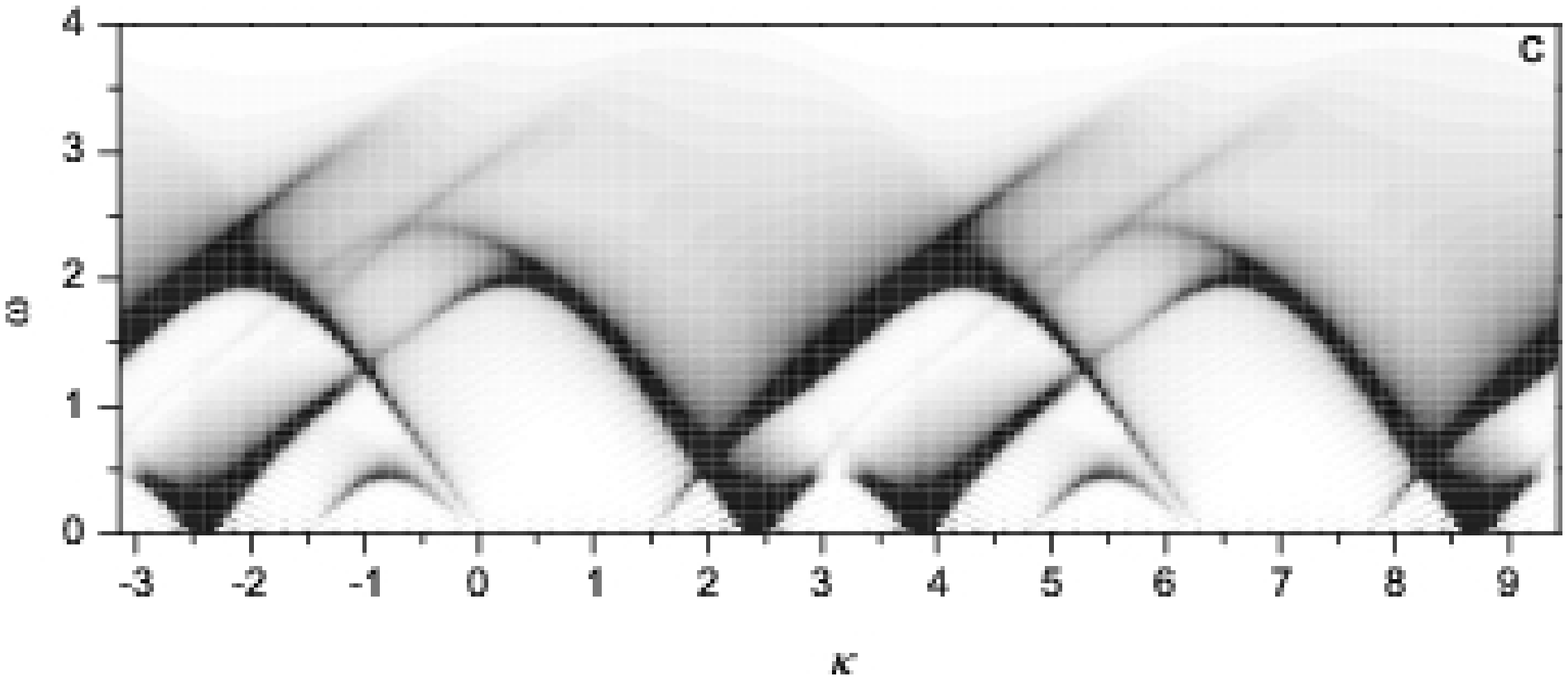, height = 0.25\linewidth}
\caption{}
\end{figure}

\end{document}